\documentclass[twocolumn]{aastex631}
\let\tablenum\relax

\usepackage{url}
\usepackage{threeparttable}
\usepackage{ulem}
\usepackage{natbib}
\usepackage{appendix}
\usepackage{amsmath}
\usepackage{amssymb}
\usepackage{multirow}
\usepackage{float}
\usepackage{physics}
\usepackage{siunitx}
\usepackage{mhchem}

\def\ion#1#2{#1$\;${\sc\@roman{#2}}\relax}
\def\lesssim{\mathrel{\hbox{\rlap{\hbox{\lower4pt\hbox{$\sim$}}}\hbox{$<$}}}}
\def\gtrsim{\mathrel{\hbox{\rlap{\hbox{\lower4pt\hbox{$\sim$}}}\hbox{$>$}}}}
\def\red{\textcolor{black}}

\shorttitle{Fe Abundances of Galaxies at $z=9-12$}
\shortauthors{Nakane et al.}

\begin{document}
\title{
%
Fe Abundances of Early Galaxies at $z=9-12$\\
Derived with Deep JWST Spectra
}

\author[0009-0000-1999-5472]{Minami Nakane}
\affiliation{Institute for Cosmic Ray Research, The University of Tokyo, 5-1-5 Kashiwanoha, Kashiwa, Chiba 277-8582, Japan}
\affiliation{Department of Physics, Graduate School of Science, The University of Tokyo, 7-3-1 Hongo, Bunkyo, Tokyo 113-0033, Japan}

\author[0000-0002-1049-6658]{Masami Ouchi}
\affiliation{National Astronomical Observatory of Japan, 2-21-1 Osawa, Mitaka, Tokyo 181-8588, Japan}
\affiliation{Institute for Cosmic Ray Research, The University of Tokyo, 5-1-5 Kashiwanoha, Kashiwa, Chiba 277-8582, Japan}
\affiliation{Department of Astronomical Science, SOKENDAI (The Graduate University for Advanced Studies), 2-21-1 Osawa, Mitaka, Tokyo, 181-8588, Japan}
\affiliation{Kavli Institute for the Physics and Mathematics of the Universe (WPI), The University of Tokyo, 5-1-5 Kashiwanoha, Kashiwa, Chiba 277-8583, Japan}

\author[0000-0003-2965-5070]{Kimihiko Nakajima}
\affiliation{National Astronomical Observatory of Japan, 2-21-1 Osawa, Mitaka, Tokyo 181-8588, Japan}

\author[0000-0001-9011-7605]{Yoshiaki Ono}
\affiliation{Institute for Cosmic Ray Research, The University of Tokyo, 5-1-5 Kashiwanoha, Kashiwa, Chiba 277-8582, Japan}

\author[0000-0002-6047-430X]{Yuichi Harikane} 
\affiliation{Institute for Cosmic Ray Research, The University of Tokyo, 5-1-5 Kashiwanoha, Kashiwa, Chiba 277-8582, Japan}

\author[0000-0001-7730-8634]{Yuki Isobe}
\affiliation{Kavli Institute for Cosmology, University of Cambridge, Madingley Road, Cambridge, CB3 0HA, UK}
\affiliation{Cavendish Laboratory, University of Cambridge, 19 JJ Thomson Avenue, Cambridge, CB3 0HE, UK}
\affiliation{Waseda Research Institute for Science and Engineering, Faculty of Science and Engineering, Waseda University, 3-4-1, Okubo, Shinjuku, Tokyo 169-8555, Japan}

\author[0000-0001-9553-0685]{Ken'ichi Nomoto}
\affiliation{Kavli Institute for the Physics and Mathematics of the Universe (WPI), The University of Tokyo, 5-1-5 Kashiwanoha, Kashiwa, Chiba 277-8583, Japan}

\author[0000-0003-4656-0241]{Miho N. Ishigaki}
\affiliation{National Astronomical Observatory of Japan, 2-21-1 Osawa, Mitaka, Tokyo 181-8588, Japan}

\author[0009-0006-6763-4245]{Hiroto Yanagisawa}
\affiliation{Institute for Cosmic Ray Research, The University of Tokyo, 5-1-5 Kashiwanoha, Kashiwa, Chiba 277-8582, Japan}
\affiliation{Department of Physics, Graduate School of Science, The University of Tokyo, 7-3-1 Hongo, Bunkyo, Tokyo 113-0033, Japan}

\author[0000-0001-9044-1747]{Daichi Kashino}
\affiliation{National Astronomical Observatory of Japan, 2-21-1 Osawa, Mitaka, Tokyo 181-8588, Japan}

\author[0000-0001-8537-3153]{Nozomu Tominaga}
\affiliation{National Astronomical Observatory of Japan, 2-21-1 Osawa, Mitaka, Tokyo 181-8588, Japan}
\affiliation{Astronomical Science Program, Graduate Institute for Advanced Studies, SOKENDAI, 2-21-1 Osawa, Mitaka, Tokyo 181-8588, Japan}
\affiliation{Department of Physics, Faculty of Science and Engineering, Konan University, 8-9-1 Okamoto, Kobe, Hyogo 658-8501, Japan}

\author[0000-0002-6705-6303]{Koh Takahashi}
\affiliation{National Astronomical Observatory of Japan, 2-21-1 Osawa, Mitaka, Tokyo 181-8588, Japan}

\author[0000-0003-4321-0975]{Moka Nishigaki}
\affiliation{Department of Astronomical Science, SOKENDAI (The Graduate University for Advanced Studies), 2-21-1 Osawa, Mitaka, Tokyo, 181-8588, Japan}
\affiliation{National Astronomical Observatory of Japan, 2-21-1 Osawa, Mitaka, Tokyo 181-8588, Japan}

\author[0009-0005-2897-002X]{Yui Takeda}
\affiliation{Department of Astronomical Science, SOKENDAI (The Graduate University for Advanced Studies), 2-21-1 Osawa, Mitaka, Tokyo, 181-8588, Japan}
\affiliation{National Astronomical Observatory of Japan, 2-21-1 Osawa, Mitaka, Tokyo 181-8588, Japan}

\author[0000-0002-2740-3403]{Kuria Watanabe}
\affiliation{Department of Astronomical Science, SOKENDAI (The Graduate University for Advanced Studies), 2-21-1 Osawa, Mitaka, Tokyo, 181-8588, Japan}
\affiliation{National Astronomical Observatory of Japan, 2-21-1 Osawa, Mitaka, Tokyo 181-8588, Japan}



\begin{abstract}

We derive Fe-abundance ratios of $7$ galaxies at $z=9-12$ with $-22<M_{\mathrm{UV}}<-19$ whose JWST/NIRSpec spectra achieve very high signal-to-noise ratios, $\mathrm{SNR}=60-320$, at the rest-frame UV wavelength. We fit stellar population synthesis model spectra to these JWST spectra, masking out nebular emission lines, and obtain Fe-abundance ratios of $\mathrm{[Fe/H]}=-1-0$ \red{dex} for $5$ galaxies and upper limits of $\mathrm{[Fe/H]}\sim-2-0$ \red{dex} for $2$ galaxies.
We compare these [Fe/H] values with the oxygen abundances of these galaxies ($7.4<12+\log{\mathrm{(O/H)}}<8.4$) in the same manner as previous studies of $z\sim2-6$ galaxies, and derive oxygen-to-iron abundance ratios [O/Fe]. 
We find that $2$ out of $7$ galaxies, GS-z11-0 and GN-z11, show Fe enhancements relative to O ($\mathrm{[O/Fe]}<0$ \red{dex}), especially GS-z11-0 ($z=11.12$) with a Fe enhancement ($\mathrm{[O/Fe]}=-0.68_{-0.55}^{+0.37}$ \red{dex}) beyond the solar-abundance ratio at $\sim2\sigma$. 
Because, unlike GS-z11-0, GN-z11 ($z=10.60$) may be an AGN, we constrain [O/Fe] via Fe{\sc ii} emission under the assumption of AGN and confirm that the Fe enhancement is consistent even in the case of AGN. 
While [O/Fe] values of most galaxies are comparable to those of core-collapse supernovae (CCSNe) yields, the Fe enhancements of GS-z11-0 and GN-z11 are puzzling. We develop chemical evolution models, and find that the Fe enhancements in GS-z11-0 and GN-z11 can be explained by 1) pair-instability supernovae/bright hypernovae with little contribution of CCSNe or 2) Type-Ia supernovae with short delay time ($\sim30-50$ Myr) with a top-light initial mass function.

\end{abstract}

\keywords{Galaxy chemical evolution (580); Galaxy evolution (594); Galaxy formation (595); High-redshift galaxies (734); Star formation (1596)}


\section{Introduction}
\label{sec:introduction}

The chemical enrichment of galaxies caused by stellar nucleosynthesis and supernovae provides the insight into star formation of the galaxies. The high sensitivity of the Near Infrared Spectrograph (NIRSpec; \citealt{Jakobsen2022}) of the James Webb Space Telescope (JWST; \citealt{Gardner2023}) allows us to measure the various chemical abundance ratios of galaxies at $z\sim4-14$ from emission lines: O/H (e.g., \citealt{Curti2023,Nakajima2023,Sanders2024}), C/O (e.g., \citealt{Arellano-Cordova2022,Isobe2023b,D'Eugenio2024a}), N/O (e.g., \citealt{Cameron2023,Isobe2023b,Castellano2024,Topping2024,Topping2025a,Naidu2025}), Ne/O (e.g., \citealt{Arellano-Cordova2022,Isobe2023b}), and Ar/O (e.g., \citealt{Bhattacharya2025,Stanton2025}). 
In massive stars, these elements are produced in the outer to intermediate layers via the nuclear fusion reactions (e.g., \citealt{Nomoto2013}). However, abundances of Fe, which is produced in the innermost regions of the massive stars, are investigated for only a few high-$z$ galaxies, such as GS\_3073 ($z=5.55$; \citealt{Ji2024a}), GS 9422 ($z=5.94$; \citealt{Tacchella2024}), and GN-z11 ($z=10.60$; \citealt{Ji2024b,Nakane2024}) due to weak Fe emission lines. 

For star-forming galaxies at $z=2-6$, iron abundances are measured from spectral fitting including UV stellar absorption lines of iron (e.g., \citealt{Rix2004}) with stellar population synthesis model spectra \citep{Steidel2016,Chisholm2019,Cullen2019,Harikane2020,Topping2020a,Topping2020b,Cullen2021,Kashino2022,Chartab2024,Stanton2024}. \citet{Nakane2024} obtain the abundance ratio of [O/Fe] for GN-z11 at $z=10.60$, with the iron abundance derived with the same spectral fitting method and the oxygen abundance measured from the emission lines. The resulting ratio of $\mathrm{[O/Fe]}=-0.37^{+0.43}_{-0.22}$ \red{dex} is lower than the ones of the star-forming galaxies at $z=2-6$ ($\mathrm{[O/Fe]\sim0.3-0.6}$ \red{dex}; \citealt{Steidel2016,Cullen2019,Harikane2020,Cullen2021,Kashino2022}) and Milky Way (MW) stars ($\mathrm{[O/Fe]\sim-0.2-0.7}$ \red{dex}; e.g., \citealt{Bensby2013}), implying iron enhancement relative to oxygen ($\mathrm{[O/Fe]<0}$ \red{dex}), which we hereafter refer to simply as ``Fe enhancement", in the early epoch only $\SI{430}{Myr}$ after the Big Bang. 

In the MW stars, [O/Fe] ratios are high at low [Fe/H] and low at high [Fe/H], which is explained with core-collapse supernovae (CCSNe) and Type-Ia supernovae (SNe Ia) (e.g., \citealt{Suzuki2018}). This is because iron enrichment is normally caused by SNe Ia, which occur later than instantaneous CCSNe due to the delay time (typically $\sim0.1-1.0\ \mathrm{Gyr}$) for the white dwarf formation and gas accretion/white dwarf merger. In the case of GN-z11, however, the early Fe enhancement by SNe Ia is difficult unless the delay times is very short, and other possibilities of bright hypernovae (BrHNe; e.g., \citealt{Umeda2008,Leung2024}) and/or theoretical pair-instability supernovae (PISNe; e.g, \citealt{Takahashi2018}), which eject a lot of iron due to the high explosion energies and high mass cut, are also suggested \citep{Nakane2024}. In similar cases, the observed low [O/Fe] ratios of extremely metal poor galaxies (EMPGs) in the local Universe, which are very young ($\sim$ a few tens of Myr), suggest the Fe enhancement by BrHNe/PISNe or SNe Ia with short delay time \citep{Kojima2021,Isobe2022,Watanabe2024}.

In this study, we investigate iron abundances of $7$ galaxies at $z\sim9-12$, using NIRSpec deep spectra observed in the multiple programs to explore the processes of iron enrichment of high-$z$ galaxies with a larger sample. This paper is organized as follows. Section \ref{sec:data} describes the NIRSpec data obtained from the multiple programs and defines our sample of galaxies at $z\sim9-12$. In Section \ref{sec:chemical}, we measure iron abundances with the spectral fitting and oxygen abundances with the nebular emission lines, and present the resulting [O/Fe] ratios of our sample galaxies. In Section \ref{sec:discussion}, we construct the chemical evolution models including CCSNe, SNeIa, and PISNe, and discuss the origins of the iron enrichment in the early epoch of the Universe. Section \ref{sec:summary} summarizes our findings. We assume a standard $\Lambda$CDM cosmology with $\Omega_\Lambda=0.7$, $\Omega_m=0.3$, and $H_0=70$ km $\mathrm{s}^{-1}$ $\mathrm{Mpc}^{-1}$. All magnitudes are in the AB system \citep{Oke&Gunn1983}. Throughout this paper, we utilize the solar metallicity values of $12+\log{\mathrm{(O/H)}_\odot}=8.69$, $12+\log{\mathrm{(Fe/H)}_\odot}=7.50$, and $Z_\odot=0.0142$ \citep{Asplund2009}. The notation [X/Y] (dex) is defined as log(X/Y) relative to the solar abundance ratio, i.e., $[\mathrm{X/Y}]=\log(\mathrm{X/Y})-\log(\mathrm{X/Y})_\odot$.

\section{Data and Sample} \label{sec:data}
\subsection{NIRSpec Spectra}
\label{subsec:spectra}

The spectroscopic data used in this study were obtained with NIRSpec in multiple programs of the public observations; the Cosmic Evolution Early Release Science (CEERS; ERS-1345, PI: S. Finkelstein; \citealt{Finkelstein2023}, \citealt{Haro2023a}), the Director’s Discretionary Time (DDT) observations (DDT-2750, PI: P. Arrabal Haro; \citealt{Haro2023b}, DDT-2767, PI: P. Kelly; \citealt{Williams2023}), the General Observer (GO) observations (GO-1433, PI: D. Coe; \citealt{Hsiao2024a}, GO-2198, PI: L. Barrufet \& P. Oesch; \citealt{Barrufet2025}, and GO-3073, PI: M. Castellano; \citealt{Castellano2024}), the Ultra-deep NIRCam and NIRSpec Observations Before the Epoch of Reionization (UNCOVER; GO-2561, PI: I. Labbe \& R. Bezanson; \citealt{Bezanson2024}), and the \red{Guaranteed} Time Observations (GTO) of the JWST Advanced Deep Extragalactic Survey (JADES; GTO-1180, GTO-1181, PI: D. Eisenstein, GTO-1210, GTO-1286, PI: N. L\"{u}tzgendorf, GO-3215, PI: D. Eisenstein \& R. Maiolino; \citealt{Eisenstein2023a,Eisenstein2023b,Bunker2024,D'Eugenio2024}). All of the NIRSpec observations were conducted with the MOS mode using MSA. We use low-resolution ($R\sim100$) prism data that cover $0.6-5.3\ \mu\mathrm{m}$ because continua, which are crucial for measuring Fe abundances (see Section \ref{subsec:Stellar_FeH}), are detected clearly in the prism spectra compared to the medium-resolution ($R\sim1000$) and high-resolution ($R\sim2700$) grating spectra.

\begin{figure}[th!]
    \centering
    \includegraphics[width=0.45\textwidth]{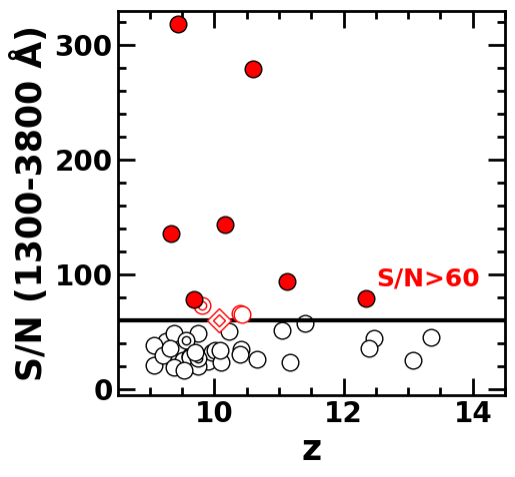}
    \includegraphics[width=0.45\textwidth]{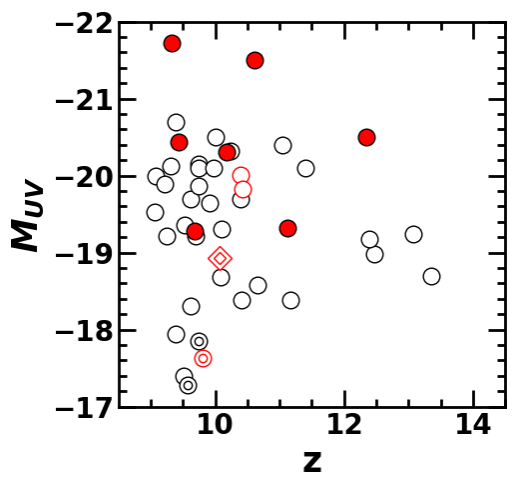}
    \caption{Top: Signal-to-noise ratio around $1300-3800$ \AA\ as a function of redshift. The red-filled circles represent our sample galaxies with $\mathrm{S/N}(1300-3800\ \mathrm{\AA})>60$. The red open circles (diamond) denote the galaxies with $\mathrm{S/N}(1300-3800\ \mathrm{\AA})>60$ which we do not select due to the lack of sufficient emission lines for oxygen abundance measurements (X-ray detection as a strong signature of AGN; \citealt{Goulding2023,Bogdan2024}). The black open circles show the galaxies with $\mathrm{S/N}(1300-3800\ \mathrm{\AA})\leq60$. The double circles (diamond) indicate the lensed galaxies whose magnification factors are $\mu>5$. The black solid line presents the criteria of $\mathrm{S/N}(1300-3800\ \mathrm{\AA})=60$ for selecting the our sample galaxies. Bottom: Absolute UV magnitude distribution as a function of redshift. The symbols are the same as in the top panel.}
    \label{fig:sample_OFe}
\end{figure}

\subsection{Data Reduction}
\label{subsec:reduction}

We have reduced the GO-3073 data in the same way as \citet{Nakajima2023} and \citet{Harikane2024}. We have extracted the raw data from the Mikulski Archive for Space Telescopes (MAST) archive and performed level-2 and -3 calibrations using the JWST pipelines (ver.1.16.0) with the Calibration Reference Data System (CRDS) context file of \texttt{jwst\_1298.pmap}. We have obtained noise spectra from readout noise and Poisson noise. We use the spectroscopic redshifts $z_\mathrm{spec}$ of the objects measured from the nebular emission lines and Lyman break features by \citet{Castellano2024} and \citet{Napolitano2025}.

The data of CEERS, DDT-2750, and GO-1433 have been reduced by \citet{Nakajima2023} and \citet{Harikane2024} with the JWST pipeline version 1.8.5 with the CRDS context file of \texttt{jwst\_1028.pmap} or \texttt{jwst\_1027.pmap} with additional processes improving the flux calibration, noise estimate, and the composition. \citet{Nakajima2023} and \citet{Harikane2024} have measured spectroscopic redshifts with optical emission lines (H$\beta$ and [O {\sc iii}] $\lambda\lambda4959,5007$). The JADES data are publicly available \footnote{\url{https://archive.stsci.edu/hlsp/jades}}, and have already been reduced with the pipeline developed by the ESA NIRSpec Science Operations Team and the NIRSpec GTO Team. 
The redshifts are spectroscopically measured by \citet{Bunker2024} and \citet{D'Eugenio2024} with the emission lines (e.g., H$\beta$ and [O {\sc iii}] $\lambda\lambda4959,5007$) or Lyman break features. The UNCOVER data are publicly available\footnote{\url{https://jwst-uncover.github.io/\#}} and have been reduced by \citet{Price2024}. The redshifts of the galaxies at $z>9$ are spectroscopically measured by \citet{Fujimoto2023} based on the emission lines and Lyman break features. We also utilize a subset of the data from DDT-2767, GO-1433, GO-2198, and GO-1181 that \red{has been} reduced as part of the DAWN JWST Archive (DJA) \footnote{\url{https://dawn-cph.github.io/dja/}} with \texttt{msaexp} \citep{Brammer2023}. For the DJA data, we only use the spectra with grade $3$, whose redshifts are robustly estimated \citep{de_Graaff2025}. See \citet{de_Graaff2025} and \citet{Heintz2025} for details of the data reduction.

\subsection{Sample} 
\label{subsec:sample}

\begin{table*}
    \caption{Sample in This Study}
    \begin{center}
    \begin{tabular}{ccccc}
    \hline
    \hline
    Name&$z_\mathrm{spec}$&$M_\mathrm{UV}$ (mag)&log $(M_*/M_\odot)$&Ref.\\
    (1)&(2)&(3)&(4)&(5)\\
    \hline
    GHZ2&$12.342$&$-20.53$&$9.05_{-0.25}^{+0.10}$&\citet{Castellano2024}\\
    GS-z11-0&$11.122$&$-19.32$&$8.3_{-0.1}^{+0.1}$&\citet{Hainline2024}\\
    GN-z11&$10.603$&$-21.50$&$9.1_{-0.3}^{+0.4}$&\citet{Bunker2023,Tacchella2023}\\
    MACS0647-JD&$10.17$&$-20.3$&$7.6\pm0.1$&\citet{Hsiao2024a}\\
    JADES6438&$9.689$&$-19.28$&-&\citet{Bunker2024,Kageura2025}\\
    GS-z9-0&$9.433$&$-20.43$&$8.18_{-0.06}^{+0.06}$&\citet{Curti2024}\\
    Gz9p3&$9.325$&$-21.66$&$9.2_{-0.2}^{+0.1}$&\citet{Fujimoto2023,Boyett2024}\\
    \hline
    \end{tabular}
    \par
    \vspace{0.05\hsize}
    \footnotesize{(1) Name. (2) Spectroscopic redshift. (3) Absolute UV magnitude. (4) Stellar mass. (5) References for spectroscopic redshift, absolute UV magnitude, and stellar mass.}
    \label{tab:sample_chemistry}
    \end{center}
\end{table*}

For iron abundance measurements, we select high-redshift galaxies whose continua are detected at high signal-to-noise (S/N) ratios with the prism. We first have collected a total of $44$ galaxies at $z>9$ from the data described in Sections \ref{subsec:spectra} and \ref{subsec:reduction}. We then evaluate S/N ratios in the rest-frame wavelength range of $1300-3800$ \AA, S/N($1300-3800$ \AA), which is used to derive [Fe/H] with stellar population synthesis models (see Section \ref{subsubsec:procedure}). We calculate S/N($1300-3800$ \AA) by taking the root sum of squares of the S/N ratios at each spectral pixel over the rest-frame wavelengths of $1300-3800$ \AA. In the top panel of Figure \ref{fig:sample_OFe}, we present S/N($1300-3800$ \AA) as a function of redshift for the $44$ galaxies. The lensed galaxies with magnification factors of $\mu>5$ are shown by the double-outline symbols. We select individual galaxies with the following criteria: $\mathrm{S/N}(1300-3800\ \mathrm{\AA})>60$ (corresponding to $\mathrm{S/N}\gtrsim3$ per spectral pixel), detections of sufficient emission lines to measure the abundance ratios of O/H (see Section \ref{subsec:OH}), and no detections of clear signatures of active galactic nuclei (AGN; e.g., X-ray and broad component of Balmer emission). Using the selection criteria, we select a total of $7$ galaxies, GHZ2 (e.g., \citealt{Castellano2024,Zavala2024a,Zavala2024b}), GS-z11-0 (e.g., \citealt{Hainline2024}), GN-z11 \red{(e.g., \citealt{Bunker2023,Tacchella2023,Maiolino2024})}, MACS0647-JD (e.g., \citealt{Hsiao2024a,Hsiao2024b}), JADES6438 (e.g., \citealt{Bunker2024,Jones2024}), GS-z9-0 (e.g., \citealt{Curti2024}), and Gz9p3 (e.g., \citealt{Fujimoto2023,Boyett2024}), which \red{is} summarized in Table \ref{tab:sample_chemistry}. In the bottom panel of Figure \ref{fig:sample_OFe}, we show the redshift and absolute UV magnitude ($M_\mathrm{UV}$) distribution of the $44$ galaxies. For galaxies observed in CEERS, DDT-2750, DDT-2767, GO-1433, GO-3073, UNCOVER, and JADES, we use the $M_\mathrm{UV}$ values reported in the literature \citep{Haro2023a,Fujimoto2023,Castellano2024,Curti2024,Hainline2024,Harikane2024,Hsiao2024a,Tang2024,Yanagisawa2024,Kageura2025,Napolitano2025}. We derive the $M_\mathrm{UV}$ values of galaxies observed in GO-2198 and JADES, which are not reported in the literature, from the prism spectra by taking the average fluxes over the rest-frame wavelength range of $1400-1600$ \AA. The $M_\mathrm{UV}$ values are corrected for magnification factor, but not for dust extinction. Our sample galaxies at $z=9.3-12.3$ have wide range of UV magnitude, $-22<M_\mathrm{UV}<-19$. We show the prism spectra of our sample galaxies in Figure \ref{fig:individual_spectra}. For GN-z11, which is observed in JADES GTO-1181, we also utilize the data observed with the medium-resolution filter-grating pairs of F070LP-G140M, F170LP-G235M, and F290LP-G395M covering the wavelength of \red{$0.7-1.3$, $1.7-3.1$}, and $2.9-5.1$ $\mu\mathrm{m}$, respectively. This is because the grating spectra allow us to accurately measure the Mg \textsc{ii} $\lambda\lambda2796,2803$ lines, which are necessary to derive Fe abundances with AGN models (see Section \ref{subsec:AGN}).

\section{Chemical Abundances} \label{sec:chemical}
\subsection{Iron Abundances from Stellar Population Synthesis Spectrum Fitting}
\label{subsec:Stellar_FeH}

\begin{figure*}
    \centering
    \includegraphics[width=0.8\linewidth]{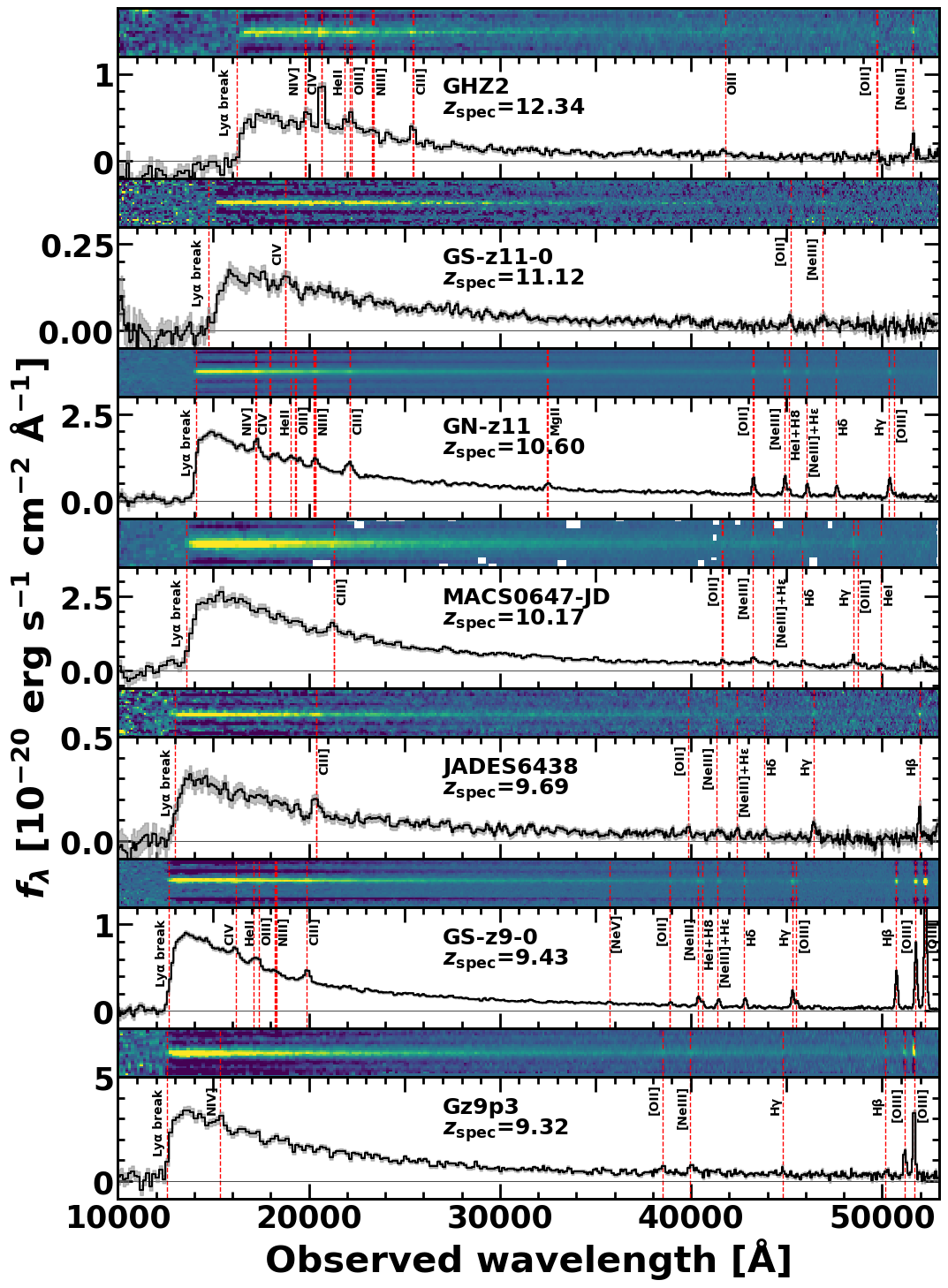}
    \caption{Spectra of the $7$ galaxies in our sample. Each panel shows the two-dimensional spectrum (top) and the one-dimensional spectrum (bottom). The black-solid lines and shaded regions represent the observed spectra and their $1\sigma$ uncertainties, respectively. The red-dashed lines indicate Lyman break features and other emission lines.}
    \label{fig:individual_spectra}
\end{figure*}

In this section, we estimate iron abundances of [Fe/H], assuming that the UV radiation of the galaxies are dominated by stellar components. The shape of the stellar continuum in the rest-frame far-ultraviolet (FUV) is sensitive to photospheric line blanketing, which is primarily caused by transitions of highly ionized iron (Fe \textsc{iii}, Fe \textsc{iv}, and Fe \textsc{v}) from massive stars \citep{Dean&Bruhweiler1985,Brandt1998}. The stellar metallicity derived from the FUV spectrum of a galaxy is thus, to a first approximation, the iron abundance in the photosphere of massive stars in the galaxy (e.g., \citealt{Steidel2016,Cullen2019,Cullen2021,Kashino2022,Chartab2024,Stanton2024}). We conduct spectral fitting with stellar population synthesis models in the similar way as \citet{Nakane2024}, following the studies of galaxies at $z\sim2-6$ (\citealt{Steidel2016,Cullen2019,Harikane2020,Cullen2021,Kashino2022}).

\subsubsection{Stellar Population Synthesis Models}
\label{subsubsec:stellar_model}

\begin{figure*}
    \centering
    \includegraphics[width=0.7\linewidth]{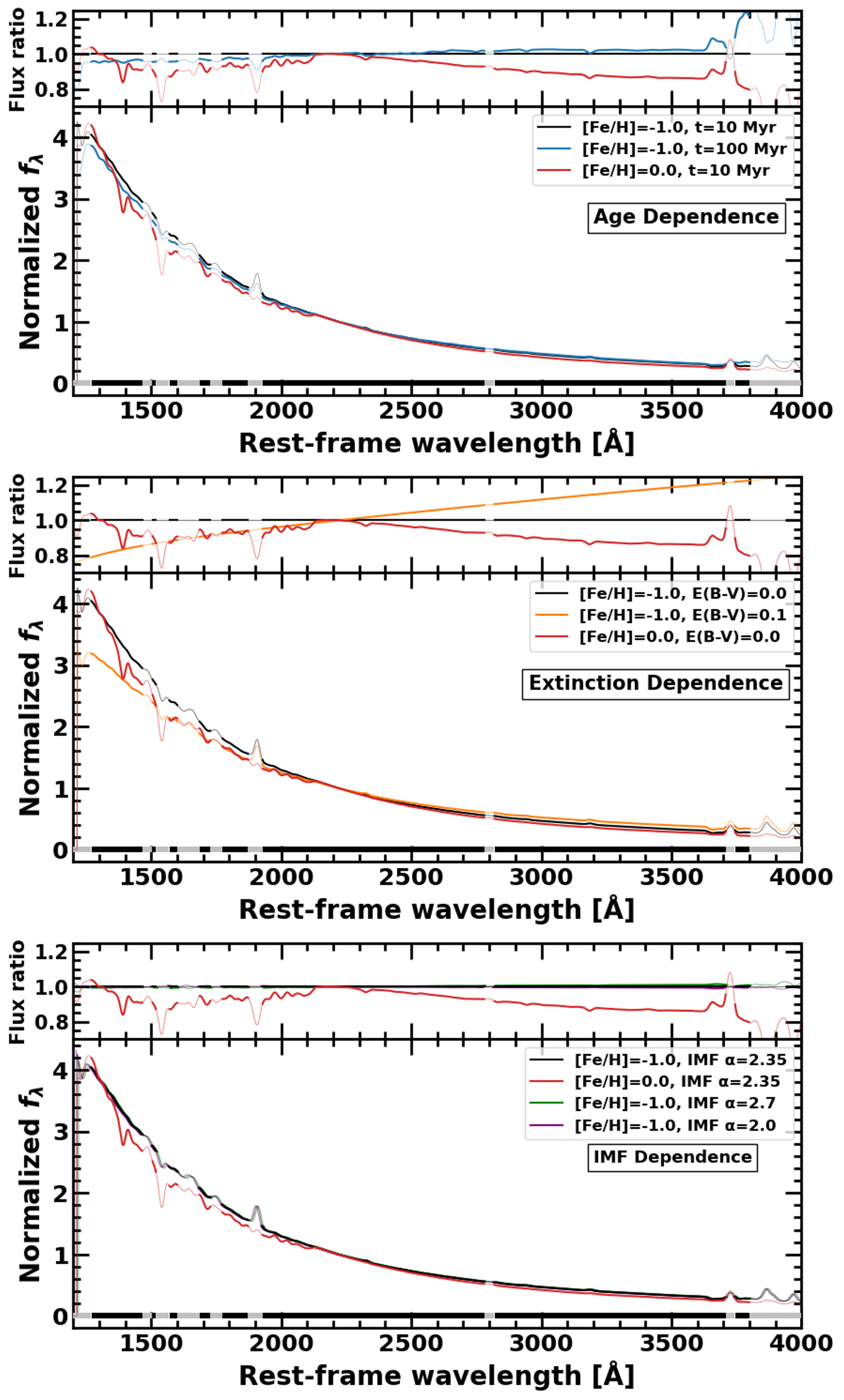}
    \caption{Comparison of the stellar population synthesis model spectra. Three panels show the differences in the spectra with respect to [Fe/H] and the following factors: stellar age (top), dust extinction (middle) and IMF (bottom). Each panel shows the flux ratio relative to the reference model with $\mathrm{[Fe/H]}=-1.0$ \red{dex}, $t=\SI{10}{Myr}$, $E(B-V)=0.0$, and a \citet{Salpeter1955} IMF ($\alpha=2.35$) (upper) and normalized model spectra at $2200$ \AA\ (lower). The black lines represent the reference models. The red, blue, orange, green, and purple lines indicate the same models as the black lines but for $\mathrm{[Fe/H]}=0.0$ \red{dex}, $t=\SI{100}{Myr}$, $E(B-V)=0.1$, an IMF with $\alpha=2.7$, and an IMF with $\alpha=2.0$, respectively. For display purposes, the purple and green lines are shifted by $-5$ and $+5$ \AA\ in the $x$-axis, respectively. The faint lines show the wavelength regions excluded from the spectral fitting to avoid the impacts of nebular emission lines. The age change cannot make the spectral shape similar to the one of [Fe/H] change (top). The extinction change flattens the spectra, which are totally different from the ones of [Fe/H] change (middle). The IMF change cannot make the differences as large as the [Fe/H] change (bottom).}
    \label{fig:model_spectra}
\end{figure*}

The stellar population synthesis spectra consist of the stellar and nebular continua, reddened by the dust extinction and attenuated by the intergalactic medium (IGM) absorption. For the stellar continuum, we use the population synthesis code BPASS v2.2.1 \citep{Eldridge2017,Stanway2018}. We adopt the \citet{Salpeter1955} initial mass function (IMF), of which slope is $\alpha=2.35$, with a high-mass cutoff of $\SI{100}{M_\odot}$ including binary stars, and constant star formation history, varying stellar metallicities ($Z_*=0.00001$, $0.0001$, $0.001$,  $0.002$, $0.003$, $0.004$, $0.006$, $0.008$, $0.010$, $0.014$, $0.020$, $0.030$, and $0.040$) and ages ($\log(t/\mathrm{yr})=6.0-11.0$ in increments of $0.1$ dex). We again note that although the BPASS models adopt a solar abundance pattern, the FUV spectrum is mostly influenced by the iron abundances of massive stars. We can thus translate the inferred stellar metallicity into the iron abundances (i.e., $\log(Z_*/Z_\odot)=\mathrm{[Fe/H]}$).
We also use the IMF with slope $\alpha=2.0$ and $2.7$ in the mass range of $M>0.5M_\odot$ for comparison. We calculate the nebular continuum by the photoionization code \textsc{Cloudy} v23.01 \citep{Ferland1998,Gunasekera2023}, adopting the BPASS spectrum as the incident spectrum, a plane-parallel geometry, unity for a covering factor, an ionization parameter of $\log(\mathrm{U})=-2.0$ (e.g., \citealt{Bunker2023,Castellano2024}), and a density of $n_\mathrm{H}=\SI{300}{cm^{-2}}$ (e.g., \citealt{Steidel2016}). Although recent JWST studies report the high electron densities of $n_e\sim\SI{1000}{cm^{-3}}$ for galaxies at $z\gtrsim9$ (e.g., \citealt{Isobe2023a,Abdurro'uf2024,Topping2025b}), \citet{Nakane2024} confirm that changing the density to $n_\mathrm{H}=\SI{1000}{cm^{-3}}$ has little effect on [Fe/H] measurements. We utilize \citet{Calzetti2000} extinction law parametrized by a color excess of $E(B-V)$ and the IGM absorption models of \citet{Inoue2014}. The stellar population synthesis spectra are smoothed to match the observed resolutions of $R=100$ for the prism spectra. 

In Figure \ref{fig:model_spectra}, we compare the smoothed stellar population synthesis model spectra for different values of stellar metallicity ($=\mathrm{[Fe/H]}$), stellar age, dust extinction, and IMF. In the top panel, both of the spectra with higher [Fe/H] (red line) and older stellar age (blue line) are reddened compared to the reference spectrum (black line). However, the spectrum with higher [Fe/H] have a different spectral shape, primarily due to photospheric line blanketing, which is not caused by increasing stellar age. In addition, the spectra at wavelengths longer the Balmer break show significant difference driven by stellar age. In the middle panel, the spectrum with larger dust extinction (orange line) flattened over wide range of wavelengths, resulting in the different spectral shape from that with higher [Fe/H]. In the bottom panel, the spectra with different IMF slopes of $\alpha=2.7$ (green line) and $\alpha=2.0$ (purple line) show much less differences in spectral shape compared to the change of [Fe/H], stellar age, and dust extinction.

\subsubsection{Fitting Procedure}
\label{subsubsec:procedure}

We fit the stellar population synthesis models to the prism spectra with $4$ free parameters of the iron abundance [Fe/H], stellar age $t$, color excess $E(B-V)$, and normalization factor of the model spectra $f$. To obtain the posterior probability distributions functions (PDFs) of the free parameters, we perform Markov Chain Monte Carlo (MCMC) simulations with \texttt{emcee} \citep{Foreman2013}. We utilize flat priors of $-3.15\leq\mathrm{[Fe/H]}\leq0.45$ \red{dex}, $6.00\leq\log(t/\mathrm{yr})\leq t_\mathrm{C}$, $0.0\leq E(B-V)\leq1.0$, and $-2.0\leq\log(f)\leq2.0$, where $t_\mathrm{C}$ is the cosmic age corresponding to the redshift of the galaxy. We determine the best-fit parameter from the mode (i.e., a peak of the posterior distribution) and its $1\sigma$ uncertainty from the $68$\% highest posterior density interval (HPDI; i.e., the narrowest interval containing $68$\%) of the posterior distribution. 
We conduct the spectral fitting over the rest-frame wavelength range of $1300-3800$ \AA\ in our fiducial analysis. The lower limit of the fitting range is chosen to avoid contamination from Ly$\alpha$ damping wing absorption. Based on careful inspection of the individual spectra, we adjust the lower limits to $1270$ \AA\ for GN-z11, and to $1400$ \AA\ for GS-z11-0 and MACS0647-JD. For the upper limit of the fitting range, it is crucial to choose which wavelengths to include for accurately deriving the free parameters. As shown in Figure \ref{fig:model_spectra}, while the FUV spectra ($\sim1000-2000$ \AA) are primarily sensitive to stellar metallicity (particularly iron abundance), the optical spectra ($\gtrsim3600$ \AA) are strongly impacted by the stellar age. Although incorporating the optical spectra into the fitting is essential for constraining stellar age, it may in turn affect the inferred metallicities due to the potentially different metallicities of old stellar populations.
We thus conservatively set the upper limit of the fitting range to $3800$ \AA. Additionally, we exclude some wavelength regions where the spectra \red{are} contaminated by non-stellar features. In \citet{Nakane2024}, we carefully mask out the wavelength regions around the non-stellar features (nebular emission and interstellar absorption lines), and furthermore mask non-iron stellar absorption lines to ensure that the inferred stellar metallicity primarily reflects the iron abundance. However, some non-stellar features and non-iron stellar absorption lines would have little effects on estimating the iron abundance due to the low resolution of the prism spectra. In this work, we mask the following emission lines, which are detected for high-$z$ galaxies: N \textsc{iv}] $\lambda\lambda1483,1486$, C \textsc{iv} $\lambda\lambda1548,1550$, He \textsc{ii} $\lambda1640$, O \textsc{iii}] $\lambda\lambda1661,1666$, N \textsc{iii}] $\lambda1747-1752$, C \textsc{iii}] $\lambda\lambda1907,1909$, Mg \textsc{ii} $\lambda\lambda2796,2803$, and [O \textsc{ii}] $\lambda\lambda3726,3729$. We also exclude the emission lines of the individual galaxies, O \textsc{iii} $\lambda3133$ of GHZ2 \citep{Castellano2024} and [Ne \textsc{v}] $\lambda3422$ of GS-z9-0 \citep{Curti2024}. We note that extending the fitting range and reducing the masked regions do not introduce a significant bias in the [Fe/H] estimates (see Section \ref{subsubsec:fit_res}).


\subsubsection{Fitting Results}
\label{subsubsec:fit_res}

\begin{figure*}
    \centering
    \includegraphics[width=0.95\linewidth]{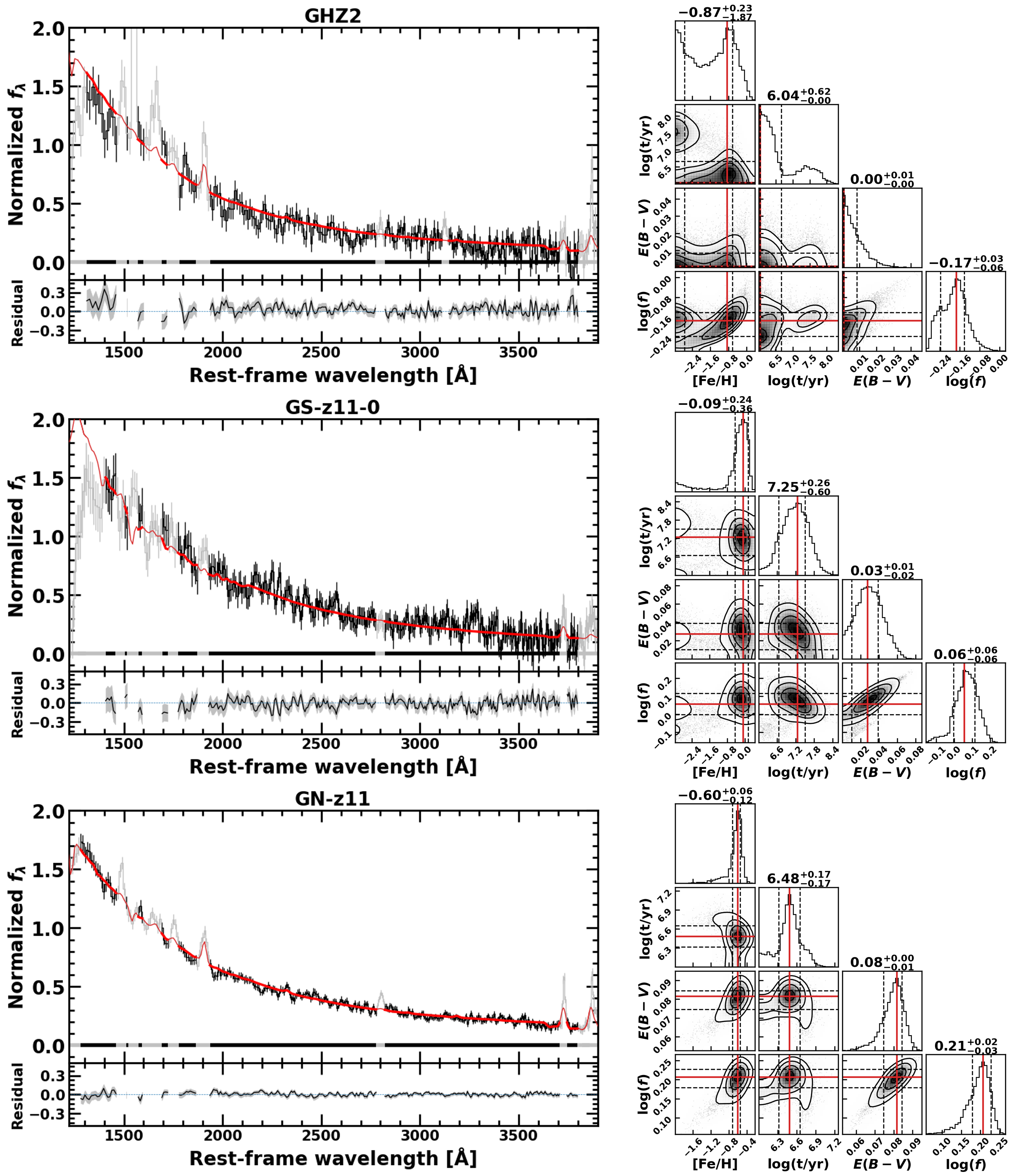}
    \caption{Fitting results of our sample galaxies with the stellar population synthesis models. Left: the upper panel compares the observed and model spectra. The black (gray) line shows the observed spectrum and its $1\sigma$ error, which are used (not used) for the fitting. The red line indicates the best-fit model spectrum. The lower panel presents the residuals from the best-fit model. The black lines and gray shaded regions denote the residuals and their $1\sigma$ errors, respectively. Right: posterior PDFs of the fitting parameters. The red-solid lines and black-dashed lines represent the mode and the boundaries of the $68$\% HPDI, respectively.}
    \label{fig:Fit-PDF_stellar}
\end{figure*}
\setcounter{figure}{3}
\begin{figure*}
    \centering
    \includegraphics[width=0.95\linewidth]{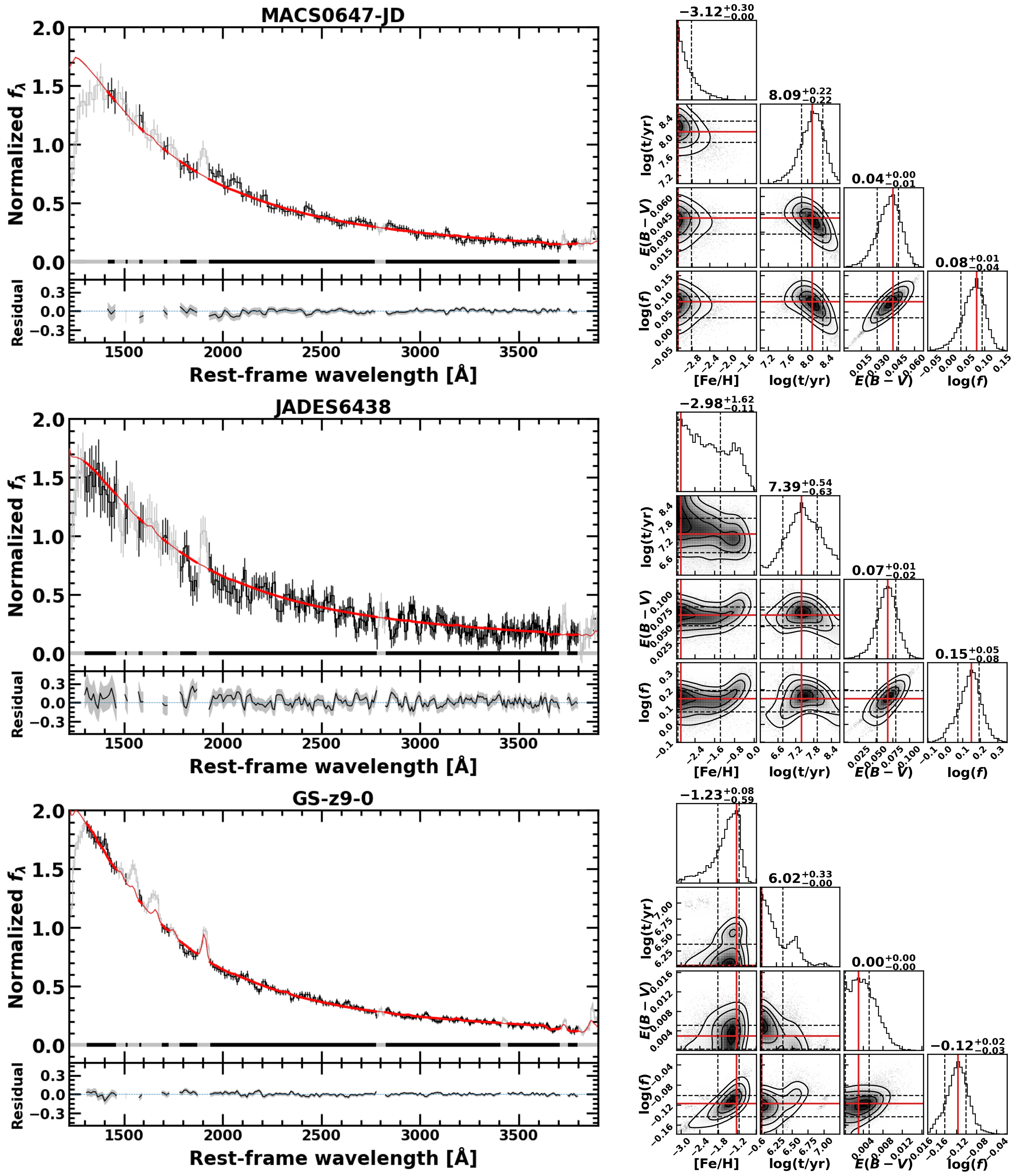}
    \caption{Continued.}
\end{figure*}
\setcounter{figure}{3}
\begin{figure*}
    \centering
    \includegraphics[width=0.95\linewidth]{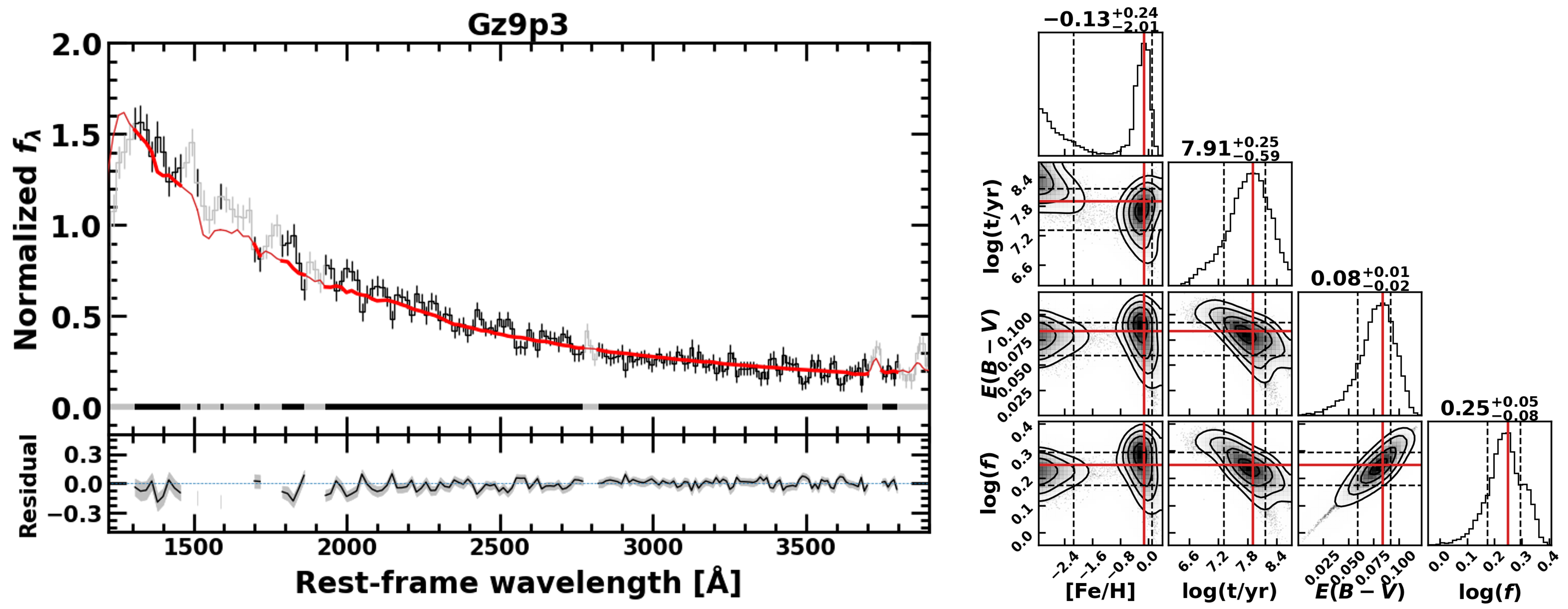}
    \caption{Continued.}
\end{figure*}

\begin{figure}
    \centering
    \includegraphics[width=0.99\linewidth]{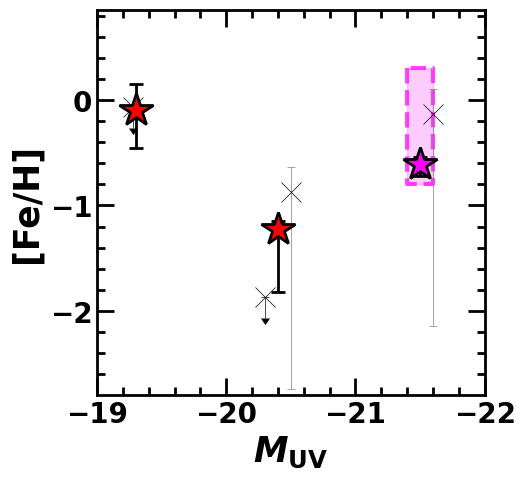}
    \caption{Fe/H ratio as a function of absolute UV magnitude. The magenta (red) star symbols represent well-constrained stellar abundance ratios for GN-z11 (GS-z11-0 and GS-z9-0). The cross symbols indicate the stellar abundance ratios whose posterior distributions are either bimodal or truncated near the lower bound. The magenta-shaded region \red{represents} our measurement of GN-z11 with the AGN models.}
    \label{fig:M_UV-Fe_H}
\end{figure}

We conduct the spectral fitting for our sample of the $7$ galaxies at $z=9-12$. Note that while the iron abundance of GN-z11 in our sample is reported by \citet{Nakane2024}, we reestimate it with our improved fitting and mask ranges. Figure \ref{fig:Fit-PDF_stellar} shows the best-fit model spectra and posterior PDFs for our sample galaxies. The best-fit model spectra are broadly consistent with the observed spectra within the uncertainties, with no significant wavelength-dependent residuals. Based on the posterior PDFs, we estimate the iron abundances, spanning $\mathrm{[Fe/H]}\sim-3\text{--}0$ \red{dex}. The posteriors present unimodal distributions of [Fe/H] for GS-z11-0, GN-z11, and GS-z9-0, yielding good constraints of $\mathrm{[Fe/H]}=-0.09^{+0.24}_{-0.36}$, $-0.60^{+0.06}_{-0.12}$, and $-1.23^{+0.09}_{-0.59}$ \red{dex}, respectively. Our [Fe/H] measurement of GN-z11 is consistent with that of \citet{Nakane2024} and provides tighter constraint by extending the fitting range up to $3800$ \AA.  For GHZ2 and Gz9p3, the posterior PDFs of [Fe/H] are bimodal, which are due to the degeneracy between [Fe/H] and stellar age, resulting in the large uncertainties. We report the best-fit value corresponding to the higher-probability peak for these objects. We discuss the interpretation of these degeneracies in Section \ref{subsubsec:comparison}. The posterior PDFs of [Fe/H] for MACS0647-JD and JADES6438 are accumulated near the lower bounds of the prior range ($\mathrm{[Fe/H]=-3.15}$ \red{dex}) due to the low sensitivity of this low metallicity range to the spectral shape. Therefore, although we report the best-fit values as our fitting results, we conservatively adopt the upper limits of [Fe/H] to discuss chemical abundances. In Figure \ref{fig:M_UV-Fe_H}, we present [Fe/H] of our sample galaxies, among which GS-z11-0 and GN-z11 show relatively high [Fe/H] ($>25\%$ solar metallicity) with well-constrained values. For the other parameters, we obtain the young ages ($t\sim1-30$ Myr) except for MACS0647-JD and Gz9p3 ($t\sim100$ Myr), and low dust attenuation ($E(B-V)\sim0.00-0.08$). We summarize the best-fit parameters of our sample galaxies in Table \ref{tab:iron_fit_result}.


\subsubsection{Comparison with Previous Studies}
\label{subsubsec:comparison}

In this section, we compare the stellar populations derived from our fitting method with those estimated from Spectral Energy Distribution (SED) fitting in the literature (GHZ2: \citealt{Castellano2024,Zavala2024a}, GSz11-0: \citealt{Hainline2024}, GN-z11: \citealt{Bunker2023,Tacchella2023}, MACS0647-JD: \citealt{Hsiao2024a,Hsiao2024b}, GS-z9-0: \citealt{Curti2024}, and Gz9p3: \citealt{Boyett2024}). It is important to note that while our fitting method mainly traces the properties of the young massive stellar components, the standard SED fitting reflects the population-averaged galaxy properties. Overall, the derived low dust attenuation of $E(B-V)\sim0.00-0.08$ is good agreement with the SED fitting results. The measured very young stellar age of GHZ2, GN-z11, and GS-z9-0 ($t\sim1-3$ Myr) are broadly consistent with the SED fitting results and detections of high ionization lines (e.g., N \textsc{iv}] $\lambda\lambda1483,1486$; C \textsc{iv} $\lambda\lambda1548,1550$) \citep{Bunker2023,Tacchella2023,Castellano2024,Curti2024,Zavala2024a}. Although [Fe/H] and stellar age are degenerated for GHZ2 as described in Section \ref{subsubsec:fit_res}, it may be difficult to break the degeneracy since both the stellar ages of $t\sim1$ and $\sim30$ Myr are preferred with the current data. For GS-z11-0, the derived young stellar age of $t=18_{-13}^{+53}$ Myr is lower than those from the SED fitting ($t=158_{-32}^{+42}$ Myr; \citealt{Hainline2024}). This may be because light from younger stellar populations dominates in the UV spectrum compared to that from older populations. The derived stellar ages of MACS0647-JD and Gz9p3 ($\sim100$ Myr) suggest the existence of a certain number of older stellar populations, which is also discussed in the literature \citep{Hsiao2024a,Hsiao2024b,Boyett2024}. While the degeneracy between [Fe/H] and stellar age in Gz9p3 is also difficult to break, as in GHZ2, the solution with higher metallicity is consistent with UV metal absorption features in the high-resolution grating spectrum of Gz9p3 reported by \citet{Boyett2024}.

\begin{table*}
    \caption{Best-fit Parameters for Fitting with the Stellar Population Synthesis Models}
    \begin{center}
    \begin{tabular}{ccccc}
    \hline
    \hline
    Name&[Fe/H] \red{(dex)}&$\log(t/\mathrm{yr})$&E(B-V)&$\log(f)$\\
    (1)&(2)&(3)&(4)&(5)\\
    \hline
    GHZ2&$-0.87_{-1.87}^{+0.23}$&$6.04_{-0.00}^{+0.62}$&$0.00_{-0.00}^{+0.01}$&$-0.17_{-0.06}^{+0.03}$\\
    GS-z11-0&$-0.09_{-0.36}^{+0.24}$&$7.25_{-0.60}^{+0.26}$&$0.03_{-0.02}^{+0.01}$&$0.06_{-0.06}^{+0.06}$\\
    GN-z11&$-0.60_{-0.12}^{+0.06}$&$6.48_{-0.17}^{+0.17}$&$0.08_{-0.01}^{+0.00}$&$0.21_{-0.03}^{+0.02}$\\
    MACS0647-JD&$-3.12_{-0.00}^{+0.30}$&$8.09_{-0.22}^{+0.22}$&$0.04_{-0.01}^{+0.00}$&$0.08_{-0.04}^{+0.01}$\\
    JADES6438&$-2.98_{-0.11}^{+1.62}$&$7.39_{-0.63}^{+0.54}$&$0.07_{-0.02}^{+0.01}$&$0.15_{-0.08}^{+0.05}$\\
    GS-z9-0&$-1.23_{-0.59}^{+0.08}$&$6.02_{-0.00}^{+0.33}$&$0.00_{-0.00}^{+0.00}$&$-0.12_{-0.03}^{+0.02}$\\
    Gz9p3&$-0.13_{-2.01}^{+0.24}$&$7.91_{-0.59}^{+0.25}$&$0.08_{-0.02}^{+0.01}$&$0.25_{-0.08}^{+0.05}$\\
    \hline
    \end{tabular}
    \par
    \vspace{0.05\hsize}
    \footnotesize{(1) Name. (2) Iron abundance. (3) Stellar age. (4) Color excess. (5) Normalization factor.}
    \label{tab:iron_fit_result}
    \end{center}
\end{table*}

\subsection{Iron Abundances from AGN Template Fitting}
\label{subsec:AGN}

\begin{figure*}
    \centering
    \includegraphics[width=0.99\linewidth]{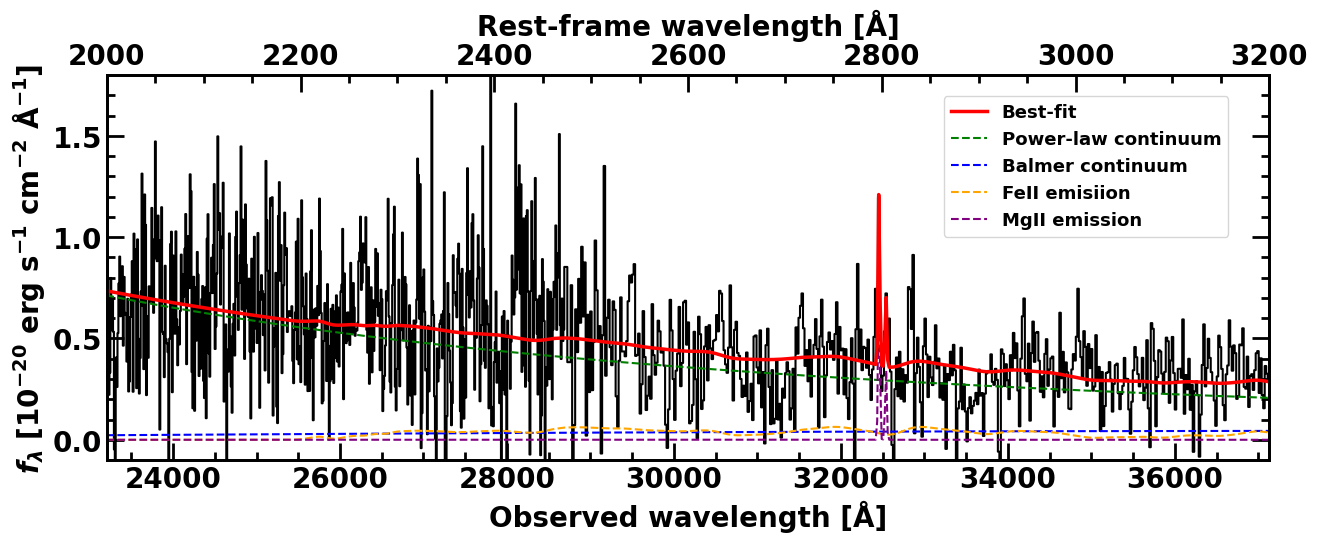}
    \caption{Results of the fitting for GN-z11 with the AGN models. The black and red solid lines present the observed grating spectrum and best-fit model spectrum, respectively. The green, blue, yellow, and purple dashed lines indicate the best-fit power-law continuum, Balmer continuum, Fe \textsc{ii} emission, and Mg \textsc{ii} emission, respectively.}
    \label{fig:fit_AGN}
\end{figure*}

\begin{figure*}
    \centering
    \includegraphics[width=0.99\linewidth]{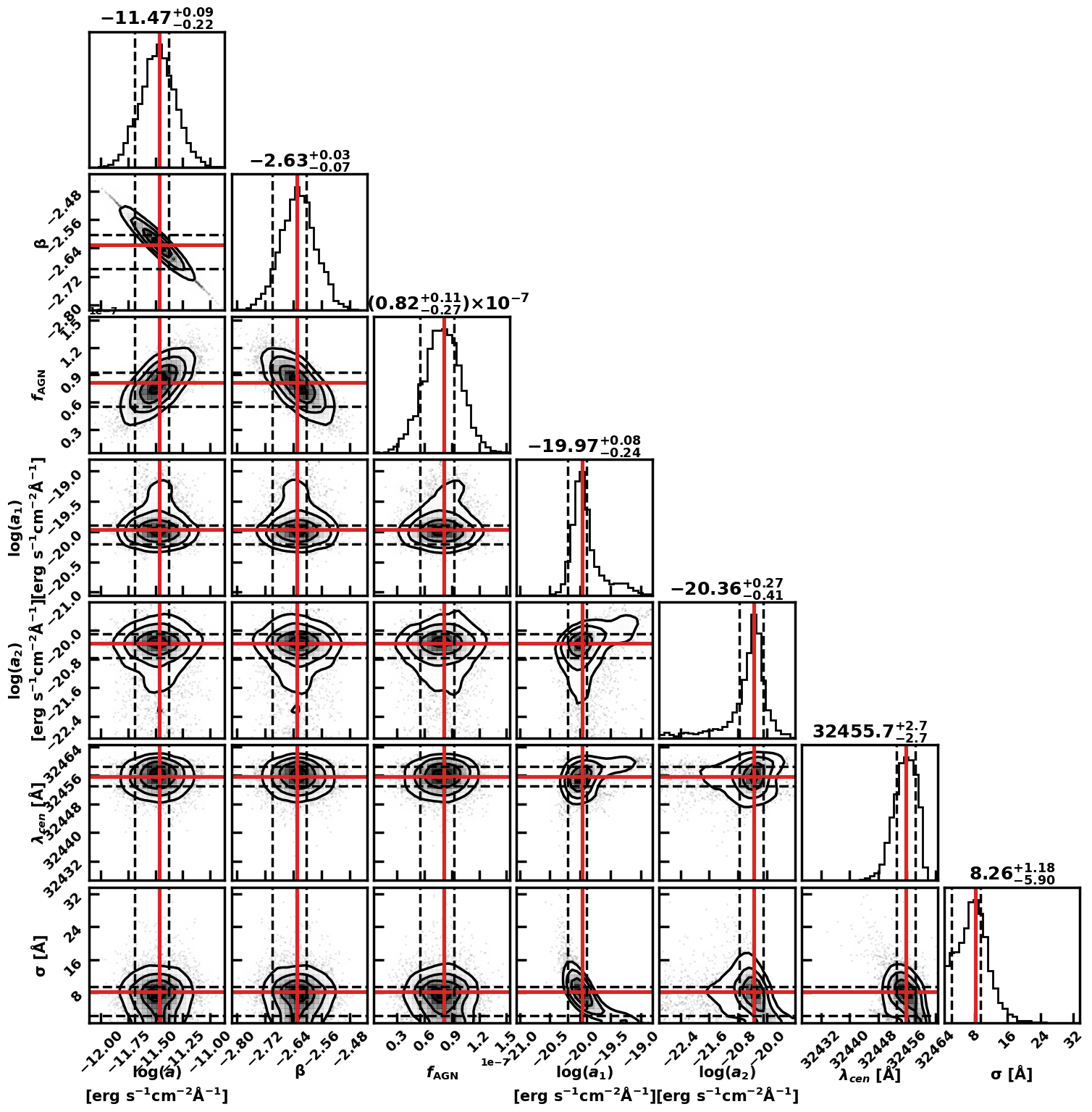}
    \caption{Posterior PDF of the fitting parameters obtained for the grating spectrum of GN-z11 with the AGN models. The symbols are the same as in Figure \ref{fig:Fit-PDF_stellar}.}
    \label{fig:PDF_AGN}
\end{figure*}

\begin{figure}
    \centering
    \includegraphics[width=0.99\linewidth]{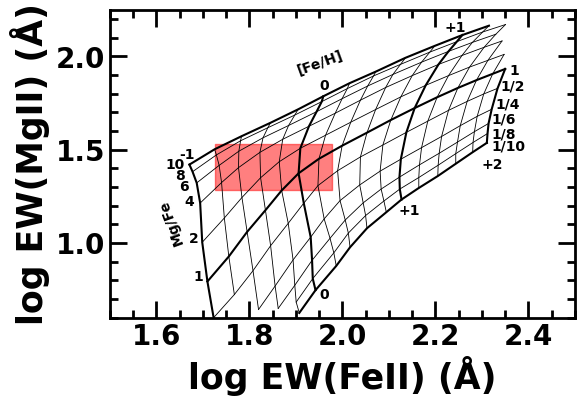}
    \caption{Abundance diagnostic diagram. The grid lines show the calculated EW(Mg \textsc{ii}) and EW(Fe \textsc{ii}) as functions of Mg/Fe ($=10^{\mathrm{[Mg/Fe]}}$) and [Fe/H] (\citealt{Sameshima2017}, see Section \ref{subsec:AGN}). The red-shaded region denotes the measurements of GN-z11 with the AGN models.} 
    \label{fig:FeH_AGN_model}
\end{figure}

In Section \ref{subsec:Stellar_FeH}, we derive the iron abundances, assuming that the UV continuum is dominated by stellar radiation. Out of the $2$ galaxies, which have high [Fe/H] values, GN-z11 \red{has} a possibility of hosting an AGN, as suggested based on the high ionization emission lines and signatures of high electron density (e.g., \citealt{Maiolino2024}). Actually, the Fe abundance of GN-z11 in the case where the UV continuum is dominated by AGN radiation is reported \citep{Ji2024b}. \citet{Ji2024b} investigate the continuum excess between $3000-3550$ \AA\ in the rest-frame wavelengths for the prism spectrum of GN-z11. Comparing the observed spectrum with the AGN models consisting of a power-law continuum, Balmer continuum, and Fe \textsc{ii} complex at $3000-3550$ \AA\ calculated with \textsc{Cloudy}, \citet{Ji2024b} obtain the lower limit of $\mathrm{[Fe/H]}\gtrsim-0.5$ \red{dex}. To further constrain [Fe/H] of GN-z11 in the AGN case, we adopt an independent method, which compares the observed \red{equivalent widths (EWs)} of Fe \textsc{ii} emission around $2200-3090$ \AA\ and Mg \textsc{ii}$\lambda\lambda2796,2803$ lines with photoionization models, following \citet{Sameshima2017,Sameshima2020} and \citet{Onoue2020}. We perform spectral fitting with the following AGN model,
\begin{equation}
    F^\mathrm{C}_{\lambda}=F^\mathrm{PL}_\mathrm{\lambda}+F^\mathrm{BC}_{\lambda}+F^\mathrm{Fe\textsc{ii}+Fe\textsc{iii}}_{\lambda}+F^\mathrm{Mg\textsc{ii}}_{\lambda},
    \label{eq:continuum_model}
\end{equation}
where $F^\mathrm{PL}_\mathrm{\lambda}=a\lambda^{-\beta}$, $F^\mathrm{BC}_{\lambda}$, $F^\mathrm{Fe\textsc{ii}+Fe\textsc{iii}}_{\lambda}$, and $F^\mathrm{Mg\textsc{ii}}_{\lambda}$ represent the power-law continuum flux emitted from an accretion disk, Balmer continuum flux, iron pseudocontinuum flux, and Mg \textsc{ii}$\lambda\lambda2796,2803$ line fluxes, respectively. For the Balmer continuum, we utilize the following formula by \citet{Grandi1982},
\begin{equation}
    F^\mathrm{BC}_{\lambda}=F^\mathrm{BE}_0 B_\lambda(T_e)\qty[1-e^{-\tau_\mathrm{BE}\qty(\frac{\lambda}{\lambda_\mathrm{BE}})^3}],
\end{equation}
where $F^\mathrm{BE}_0$, $B_\lambda(T_e)$, and $\tau_\mathrm{BE}$ are the normalization factor, Planck function at electron temperature $T_e$, and optical depth at the Blamer edge $\lambda_\mathrm{BE}=3646$ \AA. In this study, we adopt the fixed values of $F^\mathrm{BE}_0$, $T_e$, and $\tau_\mathrm{BE}$, which are also used in the literature (\citealt{Dietrich2003,Kurk2007,DeRosa2011,Mazzucchelli2017,Sameshima2017,Shin2019,Sameshima2020,Onoue2020}).
$F^\mathrm{BE}_0$ is fixed so that the Balmer continuum flux at $\lambda=3675$ \AA\ is $30$\% of the power-law continuum flux at $\lambda=3675$ \AA. The other parameters of $T_e$ and $\tau_\mathrm{BE}$ are fixed to be $T_e=\SI{15000}{K}$ and $\tau_\mathrm{BE}=1$. For the iron pseudocontinuum, we use the Fe \textsc{ii} template \citep{Tsuzuki2006} broadened by convolution with a Gaussian function, for which the full width at half maximum (FWHM) is fixed at $\SI{800}{km\ s^{-1}}$. Note that the FWHM of the convolved Gaussian function makes negligible effects on the measured Fe \textsc{ii} flux, as pointed out in \citet{DeRosa2011}. For the Mg \textsc{ii} $\lambda\lambda2796,2803$ lines, we use double Gaussian functions, for which we fix the line widths and wavelength separation in the rest-frame of the two lines. To correct for the instrumental broadening, we convolve the double Gaussian functions with the line spread function derive by \citet{Isobe2023a}, scaling its FWHM by a factor of $0.5$ because the spectral resolution is approximately twice as high for compact sources \citep{de_Graaff2024}. There are \red{$7$} free parameters in our AGN models (Equation \ref{eq:continuum_model}), which are the amplitude $a$ and slope $\beta$ of the power-law continuum, normalization factor $f_\mathrm{AGN}$ of the iron pseudocontinuum, amplitudes $a_1$ and $a_2$ of the Mg \textsc{ii} $\lambda2796$ and Mg \textsc{ii} $\lambda2803$ lines, respectively, center wavelength $\lambda_\mathrm{cen}$ of Mg \textsc{ii} $\lambda2796$, and line width $\sigma$ of Mg \textsc{ii} $\lambda\lambda2796,2803$. We utilize the grating spectrum of GN-z11 to exactly measure the Mg \textsc{ii} emissions. We fit our AGN models to the grating spectrum, using the same fitting range ($1270-3800$ \AA) and masking the same emission lines (except for Mg \textsc{ii}) as in Section \ref{subsubsec:procedure}. To obtain the PDFs of the parameters, we conduct MCMC simulations with \texttt{emcee} \citep{Foreman2013}. We apply flat priors to the \red{parameters} of $\log(a)$, $\beta$, $f_\mathrm{AGN}$, $\log(a_1)$, $\log(a_2)$, $\lambda_\mathrm{cen}$, and $\sigma$. Here, we check whether Fe \textsc{ii} template is needed for the fitting based on the statistical indicator of Widely Applicable Information Criterion (WAIC; \citealt{Watanabe2010}). WAIC is applicable to parameter estimation with posterior distributions unlike Akaike Information Criterion (AIC; \citealt{Akaike1973}), which is used for maximum likelihood estimation. We calculate $\Delta$WAIC, which is the difference between WAICs for the fitting with and without Fe \textsc{ii} template. The obtained value is $\Delta\mathrm{WAIC}=-12<-10$, which means that including Fe \textsc{ii} template enhances the goodness of the fitting. We determine the best-fit parameter and $1\sigma$ uncertainty by the mode and $68$\% HPDI of the posterior distribution, respectively. The best-fit parameters are $\log{(a/\mathrm{erg}\ \mathrm{s}^{-1}}\ \mathrm{cm}^{-2}$ \AA$^{-1})=-11.47_{-0.22}^{+0.09}$,
$\beta=-2.63_{-0.07}^{+0.03}$, $f_\mathrm{AGN}=(0.82_{-0.27}^{+0.11})\times10^{-7}$, $\log{(a_1/\mathrm{erg}\ \mathrm{s}^{-1}}\ \mathrm{cm}^{-2}$ \AA$^{-1})=-19.97_{-0.24}^{+0.08}$, $\log{(a_2/\mathrm{erg}\ \mathrm{s}^{-1}}\ \mathrm{cm}^{-2}$ \AA$^{-1})=-20.36_{-0.41}^{+0.27}$, 
$\lambda_\mathrm{cen}=32455.7_{-2.7}^{+2.7}$ \AA, and $\sigma=8.26_{-5.90}^{+1.18}$ \AA.
In Figures \ref{fig:fit_AGN} and \ref{fig:PDF_AGN}, we present the best-fit AGN models and posterior PDFs of the parameters, respectively. We calculate the Fe \textsc{ii} flux by integrating the best-fit Fe \textsc{ii} template over the wavelength range of $2200-3090$ \AA. We then derive the EW of Fe \textsc{ii}, dividing the Fe \textsc{ii} flux by the continuum flux at $\SI{3000}{\AA}$. The EW of Mg \textsc{ii} is obtained from the Mg \textsc{ii} flux divided by the underneath continuum flux. We estimate the EWs to be $\mathrm{EW(Fe\ \textsc{ii})}=122.5_{-33.7}^{+28.7}$ \AA\ and $\mathrm{EW(Mg\ \textsc{ii})}=7.6_{-2.0}^{+1.7}$ \AA. To obtain Fe abundances, we compare the EWs of Fe \textsc{ii} and Mg \textsc{ii} with the models of \citet{Sameshima2017}. Here, we need to correct the observed EWs for the Eddington ratio and continuum luminosity around $3000$ \AA. This is because the EWs of Fe \textsc{ii} and Mg \textsc{ii} are correlated with Eddington ratio (e.g., \citealt{Dong2011,Sameshima2017}) and anticorrelated with continuum luminosity (Baldwin effect; e.g., \citealt{Baldwin1977,Baldwin1978}), which may not be due to metallicity (e.g., \citealt{Laor1995,Dietrich2002}). We correct EWs of Fe \textsc{ii} and Mg \textsc{ii} with Equation (1) of \citet{Yoshii2022}. In Figure \ref{fig:FeH_AGN_model}, we compare our corrected EWs with the abundance diagnostic diagram introduced by \citet{Sameshima2017}. \citet{Sameshima2017} calculate each grid of the diagram by modeling AGN photoionization with \textsc{Cloudy}, varying abundance ratios of [Fe/H] and [Mg/Fe]. See \citet{Sameshima2017} for further details of the models. We obtain the constraints of $-0.8\lesssim\mathrm{[Fe/H]}\lesssim0.3$ \red{dex}, which is not conclusive, but still consistent with iron-rich abundances \citep{Ji2024b}. In Figure \ref{fig:M_UV-Fe_H}, we present our [Fe/H] measurement of GN-z11 with the AGN models, which is comparable to those with the stellar population synthesis models.

\subsection{Oxygen Abundance Measurement} 
\label{subsec:OH}


To measure oxygen abundances, we measure emission line fluxes by performing spectral fitting with the linear combination of a Gaussian function and a constant value, $A\exp[-(\lambda-\lambda_\mathrm{cen})^2/2\sigma^2]+F_\mathrm{cont}$, where $A$, $\lambda_\mathrm{cen}$, $\sigma$, and $F_\mathrm{cont}$ are the line amplitude, center wavelength, dispersion, and continuum flux, respectively. To obtain the PDFs of the parameters, we perform MCMC simulations with \texttt{emcee} \citep{Foreman2013}. We obtain the line flux ($1\sigma$ uncertainty) from the mode ($68$\% HPDI) of the probability distribution calculated with the posterior distributions of the free parameters.

\begin{figure}
    \centering
    \includegraphics[width=0.99\linewidth]{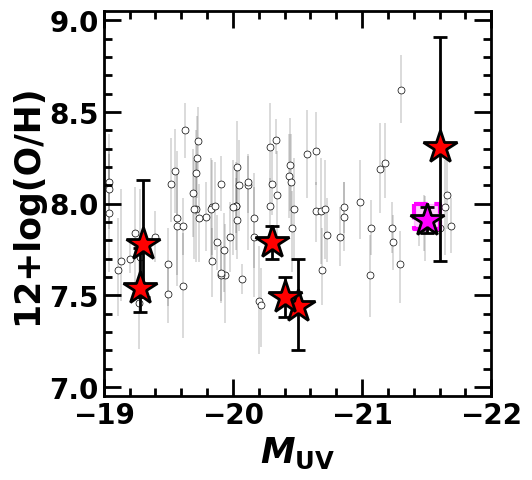}
    \caption{O/H ratio as a function of absolute UV magnitude. The star symbols show the measurements of our sample galaxies (this work, \citealt{Calabro2024,Curti2024,Hainline2024,Hsiao2024b,Alvarez-Marquez2025}). The open circles indicate the measurements of galaxies at $z\sim4-9$ \citep{Nakajima2023}.}
    \label{fig:M_UV-O_H}
\end{figure}

For the galaxies with the [O \textsc{iii}] $\lambda4363$ or O \textsc{iii}] $\lambda\lambda1661,1666$ lines detected, the O/H values are measured with the direct-$T_e$ method: GHZ2 \citep{Calabro2024}, GN-z11 (stellar and AGN cases; \citealt{Alvarez-Marquez2025}), MACS0647-JD \citep{Hsiao2024b}, and GS-z9-0 \citep{Curti2024}. For the other $3$ galaxies with no detection of [O \textsc{iii}] $\lambda4363$ and O \textsc{iii}] $\lambda\lambda1661,1666$ lines (GS-z11-0, JADES6438, and Gz9p3), we measure the O/H values with the strong line method. The strong line methods are based on the empirical relations between line ratios of strong emissions and oxygen abundances derived with the direct-$T_e$ method (e.g., \citealt{Nakajima2022,Curti2023,Sanders2024}). In this work, we use the metallicity relationships with the R2 ($=$[O \textsc{ii}] $\lambda\lambda3726,3729$/H$\beta$) and Ne3O2 ($=$[Ne \textsc{iii}]$\lambda3869$/[O \textsc{ii}] $\lambda\lambda3726,3729$) indices of \citet{Curti2017,Curti2020,Curti2023}. We estimate the oxygen abundances and their errors by conducting Monte Carlo simulations. We derive $1000$ values of $12+\log(\mathrm{O/H})$, fluctuating the observed fluxes and the metallicity relationship with the R2 (Ne3O2) index by their $1\sigma$ uncertainties based on the normal distribution. We then determine the $12+\log(\mathrm{O/H})$ values and $1\sigma$ errors from the median and $16$th/$84$th percentiles of the distribution for the $1000$ values of $12+\log(\mathrm{O/H})$, respectively. We obtain $12+\log\mathrm{(O/H)}=7.78_{-0.29}^{+0.35}$ for GS-z11-0 from $\mathrm{Ne3O2}=0.72_{-0.41}^{+0.41}$ \citep{Hainline2024}. Although the metallicity relationship with the Ne3O2 index is scattered compared to the other indices (e.g., Figure 4 in \citealt{Nakajima2022}), the effects of different calibrations on the O/H measurements are within the uncertainties (see also Figure 5 in \citealt{Curti2023}), which does not affect our conclusion. For JADES6438, we derive $12+\log\mathrm{(O/H)}=7.54_{-0.13}^{+0.22}$ from $\mathrm{R2}=0.39_{-0.14}^{+0.16}$, which is consistent with the O/H value of $7.74_{-0.32}^{+0.21}$ derived from the Ne3O2 index within the uncertainty. We measure the O/H value of Gz9p3 based on $\mathrm{R2}=2.45_{-0.94}^{+2.42}$ and obtain $12+\log\mathrm{(O/H)}=8.31_{-0.62}^{+0.60}$. Given the large uncertainty, our O/H measurement is consistent with that reported in the literature ($12+\log\mathrm{(O/H)}=7.6\pm0.5$; \citealt{Boyett2024}), which is derived from the Ne3O2 index \citep{Shi2007,Maiolino2008,Jones2015,Bian2018}. While our estimate is slightly higher than that of \citet{Boyett2024}, \citet{Pollock2025} also report the high value of $12+\log\mathrm{(O/H)}>8.34$ based on the upper limit of $T_e$ constrained by no detection of the [O \textsc{iii}] $\lambda4363$ line. We also measure the R23 ($=$([O \textsc] $\lambda\lambda4959,5007$+[O \textsc{ii}] $\lambda3726,3729$)/H$\beta$) and R3 ($=$[O \textsc{iii}] $\lambda5007$/H$\beta$) indices for Gz9p3. The measured values are beyond the metallcity relationships, preventing us from deriving the O/H value. We thus adopt the result with the R2 index as our O/H measurement for Gz9p3. In Table \ref{tab:abundance_ratio} and Figure \ref{fig:M_UV-O_H}, we show the oxygen abundance measurements for our sample galaxies, which are comparable to those of $z\sim4-9$ galaxies \citep{Nakajima2023}.

\subsection{Abundance Ratios of [O/Fe]} 
\label{subsec:abundance}
\begin{figure*}[t]
    \centering
    \includegraphics[width=0.99\hsize]{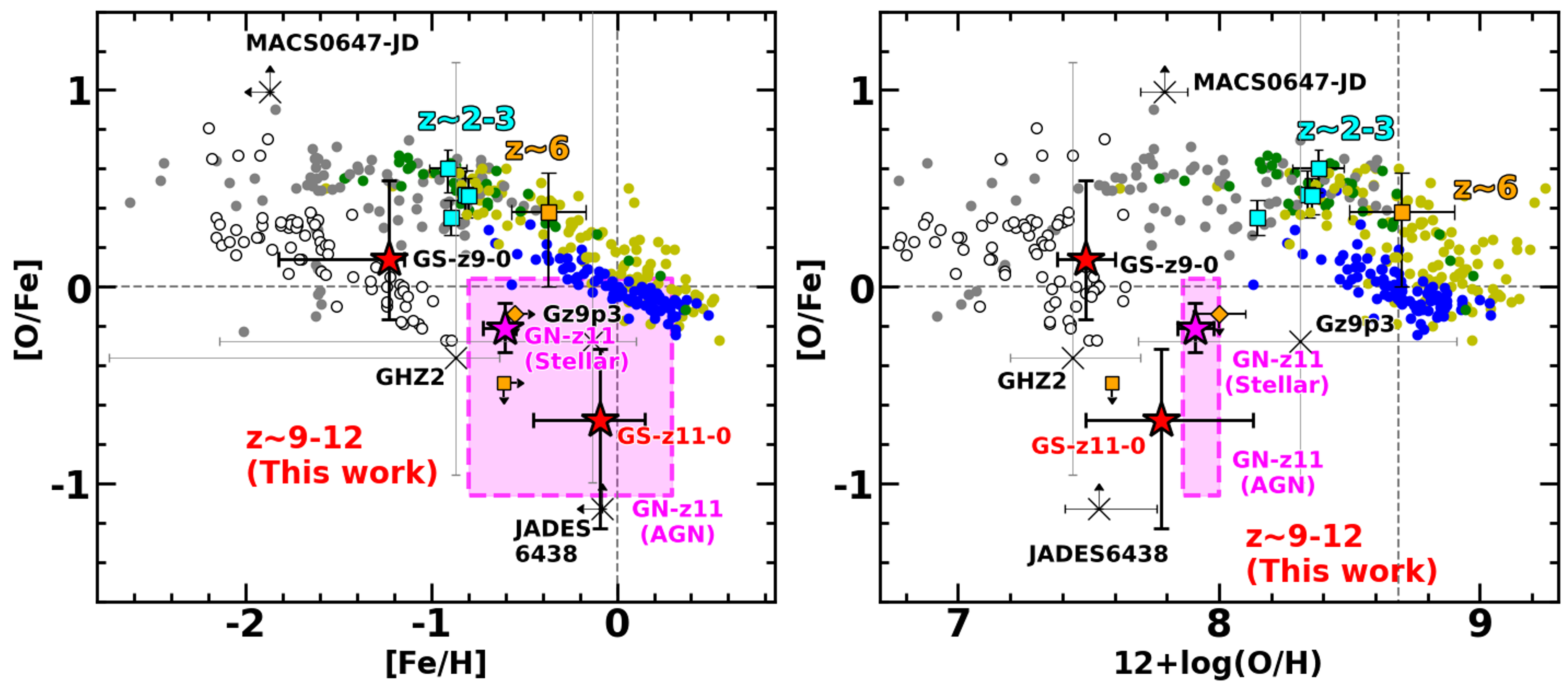}
    \caption{[O/Fe] as a function of [Fe/H] (left) and $12+\log(\mathrm{O/H})$ (right). The star symbols, cross symbols, and magenta-shaded region show the same as in Figure \ref{fig:M_UV-Fe_H}. The cyan and orange squares represent the measurements for the composite spectra of star-forming galaxies at $z\sim2-3$ \citep{Steidel2016,Cullen2021,Kashino2022} and $6$ \citep{Harikane2020}, respectively. The orange circle and diamond indicate the measurements for $z\sim6$ galaxies of GS\_3073 \citep{Ji2024a} and GS9422 \citep{Tacchella2024}, respectively. The yellow, green, blue, and gray circles denote the measurements for the MW stars \citep{Melendez2003,Carretta2005,Yong2005,Lecureur2007,Pasquini2008,Yong2008,Carretta2010,Valenti2011,Bensby2013,Zhao2016,Amarsi2019} in the bulge, thick disk, thin disk, and halo, respectively. The white circles present the measurements of the metal-poor stars in the Sculptor galaxy \citep{Hill2019,Tang2023}. The gray-dashed lines denote the solar abundance ratios.}
    \label{fig:OFe}
\end{figure*}

To derive [O/Fe] ratios, we combine [Fe/H] and O/H measurements. Note that chemical abundances of massive stars are expected to be similar to those of the surrounding ionized gas, from which the massive stars formed, because massive stars have short lifetimes \citep{Steidel2016,Cullen2019,Harikane2020,Cullen2021,Kashino2022,Nakane2024}. We thus adopt the ratio of the gas-phase O/H to the stellar [Fe/H] as a proxy of the instantaneous [O/Fe] ratio in the galaxy. We obtain the distributions of [O/Fe] by combining the posterior distributions of [Fe/H] and distributions of O/H, which are approximated by normal distributions, following the relation of $\mathrm{[O/Fe]}=\mathrm{[O/H]}-\mathrm{[Fe/H]}$. The [O/Fe] values and associated $1\sigma$ errors are determined by the mode and $68$\% HPDI of the resulting [O/Fe] distributions, respectively. We note that for MACS0647-JD and JADES6438, we adopt the $3\sigma$ upper limits rather than the best-fit values as iron abundances (see Section \ref{subsubsec:fit_res}). We thus derive $3\sigma$ lower limits of [O/Fe] from the distributions for these objects. We summarize the chemical abundance measurements of our sample in Table \ref{tab:abundance_ratio}. In Figure \ref{fig:OFe}, we compare our [O/Fe] measurements with those of MW stars \citep{Melendez2003,Carretta2005,Yong2005,Lecureur2007,Pasquini2008,Yong2008,Carretta2010,Valenti2011,Bensby2013,Zhao2016,Amarsi2019}, metal-poor stars in Sculptor galaxy \citep{Hill2019,Tang2023}, composite spectra of star-forming galaxies at $z\sim2-3$ \citep{Steidel2016,Cullen2021,Kashino2022} and $6$ \citep{Harikane2020}, and individual $z\sim6$ galaxies of GS\_3073 \citep{Ji2024a} and GS9422 \citep{Tacchella2024}. We find that $2$ out of $7$ galaxies, GS-z11-0 and GN-z11 (stellar case), show Fe enhancements ($\mathrm{[O/Fe]}<0$ \red{dex}) like GS\_3073 and GS9422 compared to the MW stars and $z\sim2-6$ galaxies at the same [Fe/H] and 12+log(O/H) values at $\sim2\sigma$ level. The [O/Fe] ratio of GN-z11 in the AGN case also indicates the Fe enhancement. Although GHZ2, Gz9p3, and JADES6438 appear to be in the $\mathrm{[O/Fe]}<0$ \red{dex} region, the [O/Fe] ratios are still consistent with $\mathrm{[O/Fe]}>0$ \red{dex}, given the uncertainties, as well as MACS0647-JD and GS-z9-0.

\begin{table*}
    \caption{Abundance Ratio Measurements}
    \begin{center}
    \resizebox{\textwidth}{!}{
    \begin{tabular}{ccccc}
    \hline
    \hline
    Name&[Fe/H] \red{(dex)}&$12+\log(\mathrm{O/H})$&[O/Fe] \red{(dex)}&Ref.\\
    (1)&(2)&(3)&(4)&(5)\\
    \hline
    GHZ2$^a$&$-0.87_{-1.87}^{+0.23}$&$7.44_{-0.24}^{+0.26}$&$-0.36_{-0.60}^{+1.50}$&\citet{Calabro2024}\\
    GS-z11-0&$-0.09_{-0.36}^{+0.24}$&$7.78_{-0.29}^{+0.35}$&$-0.68_{-0.55}^{+0.37}$&This work, \citet{Hainline2024}\\
    GN-z11 (Stellar)&$-0.60_{-0.12}^{+0.06}$&$7.91\pm0.07$&$-0.21_{-0.13}^{+0.13}$&\citet{Alvarez-Marquez2025}\\
    GN-z11 (AGN)&$-0.8\text{-}0.3$&$7.93\pm0.07^c$&$-1.1\text{-}0.0$&\citet{Alvarez-Marquez2025}\\
    MACS0647-JD$^b$&$<-1.87$&$7.79\pm0.09$&$>0.99$&\citet{Hsiao2024b}\\
    JADES6438$^b$&$<-0.08$&$7.54_{-0.13}^{+0.22}$&$>-1.13$&This work\\
    GS-z9-0&$-1.23_{-0.59}^{+0.09}$&$7.49\pm0.11$&$0.14_{-0.30}^{+0.40}$&\citet{Curti2024}\\
    Gz9p3$^a$&$-0.13_{-2.01}^{+0.24}$&$8.31_{-0.62}^{+0.60}$&$-0.28_{-0.72}^{+1.92}$&This work\\    
    \hline
    \end{tabular}}
    \par
    \vspace{0.05\hsize}
    \footnotesize{(1) Name. (2) Iron abundance and $3\sigma$ upper limit. (3) Oxygen abundance. (4) [O/Fe] abundance ratio and $3\sigma$ lower limit. (5) References for oxygen abundances.\\
    $^a$ The posterior distributions of [Fe/H] exhibit bimodality (see Figure \ref{fig:Fit-PDF_stellar}). \\
    $^b$ The posterior distributions of [Fe/H] are truncated at the lower prior bounds (see Figure \ref{fig:Fit-PDF_stellar}).\\
    $^c$ The uncertainty of 12+log(O/H) measurement based on the stellar radiation \citep{Alvarez-Marquez2025} is adopted.}
    \label{tab:abundance_ratio}
    \end{center}
\end{table*}

\section{Discussion} \label{sec:discussion}
\subsection{Fe Enhancement by Supernovae}
\label{subsec:yields}

As described in Section \ref{subsec:abundance}, we find the low [O/Fe] ratios of GS-z11-0 and GN-z11 compared to the MW stars and $z\sim2-6$ galaxies (Figure \ref{fig:OFe}). To explore the origins of such low [O/Fe], we compare [O/Fe] measurements with the supernova yield models of CCSNe, HNe, BrHNe, PISNe, and SNeIa. We use the CCSN and HN yield models of \citet{Nomoto2013} with the progenitor mass of $13-40\ M_\odot$ and metallicity of $Z=0.004$, which is comparable to that of GN-z11 in the stellar case. Although the yield models depend on the metallicity, the range of [O/Fe] ratios for the progenitor mass of $13-40\ M_\odot$ does not vary significantly for different metallicities. The CCSN and HN yield models have different explosion energies $E$. While the explosion energies are $E_{51}=E/10^{51}\ \mathrm{erg}=1$ for all CCSN yield models, those of the HN yield models are higher values of $E_{51}=10$, $10$, $20$, and $30$ for the progenitor mass of $20$, $25$, $30$, and $40$ $M_\odot$, respectively. The amount of Fe ejected by the supernovae also depends on the mass cut that divides the ejecta and compact remnant as well as the explosion energy. The mass cuts of the HN yield models are set to explain the yield of the observed HNe (e.g., \citealt{Nomoto2004}). We utilize the BrHN yields of \citet{Umeda2008} with the highest explosion energy of $E_{51}=50$, $150$, $100$, $110$, and $210$ for the progenitor mass of $30$, $50$, $80$, $90$, and $100$ $M_\odot$, respectively, and the lowest mass cut just above the Fe core for a certain progenitor mass within $30-100$ $M_\odot$ (see Table 3 in \citealt{Umeda2008}). The [O/Fe] ratios of BrHNe are thus the lowest among the HNe at the same progenitor mass. We take the non-rotating PISNe yields of \citet{Takahashi2018} with the progenitor mass of $220-280$ $M_\odot$. For SN Ia yields, we use the models with delayed detonation of Chandrasekhar mass C+O white dwarf in a single degenerate system (W7 model; \citealt{Nomoto1984}), whose data are taken from \citet{Iwamoto1999}.

\begin{figure}[t]
    \centering
    \includegraphics[width=0.99\hsize]{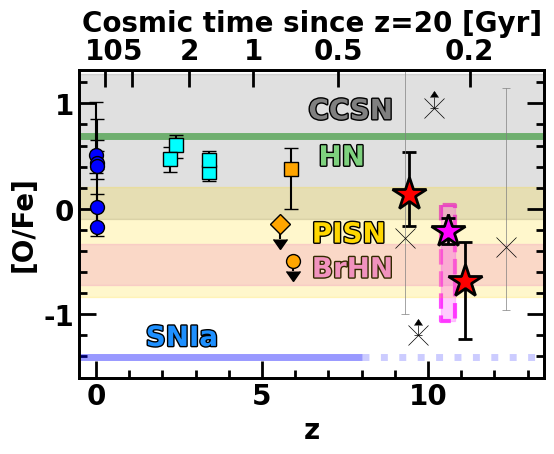}
    \caption{Redshift evolution of [O/Fe]. The cyan, orange, red, magenta, and black symbols are the same as in Figure \ref{fig:OFe}. The blue circles denote the measurements for local EMPGs \citep{Izotov2018,Kojima2020,Isobe2022}. The gray, green, pink, yellow and blue shaded regions represent the yield models of CCSNe, HNe, BrHNe, PISNe, and SNe Ia, respectively.}
    \label{fig:z-O_Fe}
\end{figure} 

In Figure \ref{fig:z-O_Fe}, we present a redshift evolution of [O/Fe] for galaxies with the yield models and cosmic age since $z=20$ ($\sim200$ Myr after the Big Bang), when the first star formation is expected to occur based on the standard $\Lambda$-CDM scenario (e.g., \citealt{Tegmark1997}). While the Fe enhancements of GS-z11-0 and GN-z11 require BrHNe, PISNe, or SNe Ia, the other $5$ galaxies are still consistent with chemical enrichment by CCSNe as well as the $z\sim2-6$ galaxies \citep{Steidel2016,Cullen2019,Harikane2020,Cullen2021}. Here, the timescale of these supernovae is important. While BrHNe and PISNe instantaneously occur after the formation of massive stars ($\sim$ a few Myr; e.g., \citealt{Nomoto2013}), SNe Ia need the delay time (typically $\sim$ a few hundreds of Myr; e.g., \citealt{Chen2021}) for white dwarf formation and gas accretion/white dwarf merger. The short timescale of BrHNe/PISNe makes it difficult to observe low [O/Fe] ratios for at least $2$ out of $7$ galaxies during $z=9.3-12.3$ ($\sim\SI{150}{Myr}$). Conversely, based on the long timescale due to the delay time, SNe Ia may have difficulty in causing Fe enhancement at as early as $z\sim10$, when the age of the universe is only $\sim450$ Myr. In the following sections, we examine which SNe mainly contribute to the observed low [O/Fe] ratios with chemical evolution models.

\subsection{Chemical Evolution Models}
\label{subsec:chem_model}

\begin{figure*}
    \centering
    \includegraphics[width=0.99\linewidth]{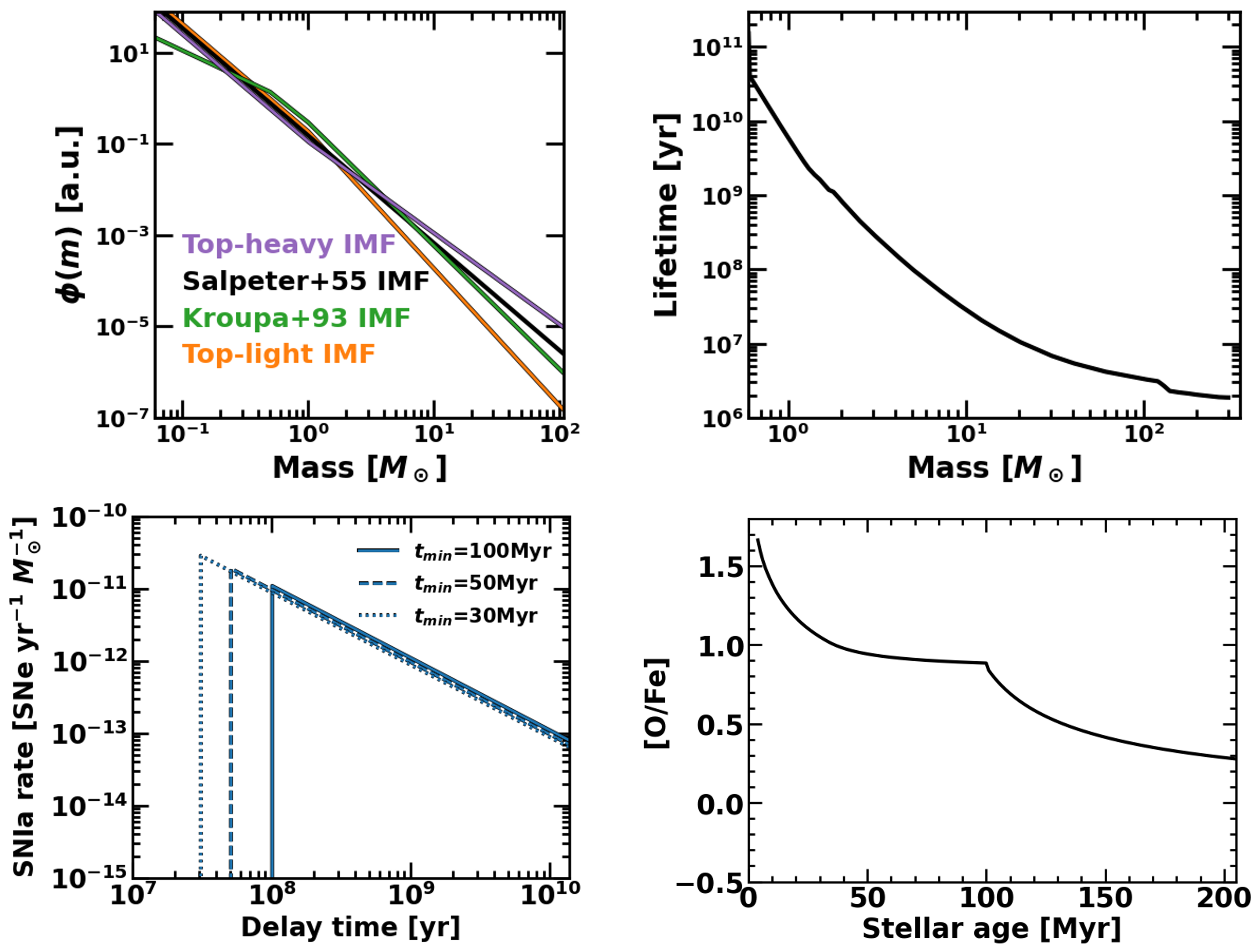}
    \caption{An overview of the construction processes for the PISN and SN Ia models. Top left: Initial mass function. The purple, black, green, and orange lines show the top-heavy, \citet{Salpeter1955}, \citet{Kroupa1993}, and top-light IMF, respectively. Top right: Stellar lifetime as a function of stellar mass. The black line is derived with stellar evolution models \citep{Padovani1993,Takahashi2018}. Bottom left: Delay time distribution for SNe Ia. \red{The distributions are approximated by power-law functions of $\propto t^{-1}$ (e.g., \citealt{Rodney2014})}. The solid, dashed, and dotted line correspond to the minimum delay time of $t_\mathrm{min}=100$, $50$, and $\SI{30}{Myr}$, respectively. Bottom right: [O/Fe] ratio as a function of stellar age. The black line represents an example of the SN Ia model with the delay time of $t_\mathrm{min}=\SI{100}{Myr}$, assuming a constant star formation history.}
    \label{fig:model_construction}
\end{figure*}

To evaluate the Fe enhancement by PISNe or SNe Ia, we construct PISN and SN Ia models with reference to \citet{Suzuki2018}. Although we do not construct BrHN models, we discuss the possibility of Fe enhancements by BrHNe in Section \ref{subsec:origin}. We define the [O/Fe] ratio at each age in the PISN (SN Ia) model as the ratio of the total numbers of oxygen and iron ejected by CCSNe and PISNe (SNe Ia) which have already occurred before the age. We present an overview for the processes of the model construction in Figure \ref{fig:model_construction}. In the PISN and SN Ia models, stars instantaneously formed based on the IMF (top left panel of Figure \ref{fig:model_construction}). We use the IMF given by \citet{Salpeter1955}, which is expressed by the power-law function of $\xi(M)=dN/dM\propto M^{-\alpha}$ with $\alpha=2.35$ for entire mass ranges. We normalize the IMF to be the numbers of stars with mass $[M,M+dM]$ per $1M_\odot$ of star formation. The distribution $\Phi(M)$ is given by
\begin{align}
    \Phi(M)&=C\xi(M)\notag\\
    C^{-1}&=\int_{M_\mathrm{min}}^{M_\mathrm{max}}M\xi(M)dM,
\end{align}
where $M_\mathrm{min}$ and $M_\mathrm{max}$ are the lower and upper limits for the masses of stars, respectively. We use the fixed values of $M_\mathrm{min}=0.08M_\odot$ and $M_\mathrm{max}=300M_\odot$ ($100M_\odot$) for the PISN (SN Ia) model. For comparison, we also use the IMF given by \citet{Kroupa1993},  
\begin{equation}
    \alpha=\begin{cases}
        2.35 &\mathrm{for}\ M<2M_\odot,\\
        2.7&\mathrm{for}\ M\geq 2M_\odot,
    \end{cases}
\end{equation}
top-heavy IMF,
\begin{equation}
    \alpha=\begin{cases}
        2.35 &\mathrm{for}\ M<2M_\odot,\\
        2.0&\mathrm{for}\ M\geq 2M_\odot,
    \end{cases}
\end{equation}
and top-light IMF,
\begin{equation}
    \alpha=\begin{cases}
        2.35 &\mathrm{for}\ M<2M_\odot,\\
        3.0&\mathrm{for}\ M\geq 2M_\odot.
    \end{cases}
\end{equation}
In both the PISN and SN Ia models, the stars with masses of $9M_\odot<M<100M_\odot$ cause CCSNe at the end of their lifetimes (top right panel of Figure \ref{fig:model_construction}) expressed by the following formula from \citet{Padovani1993},
\begin{equation}
    \tau(M)=\begin{cases}
        160 &\mathrm{Gyr}\ (M<0.6M_\odot),\\
        10^{(0.334-q(M)^{0.5})/0.116}&\mathrm{Gyr}\ (0.6\leq M\leq 6.0M_\odot),\\
        1.2M^{-1.85}+0.003&\mathrm{Gyr}\ (M\geq 6.0M_\odot),
        \label{eq:lifetime}
    \end{cases}
\end{equation}
with 
\begin{equation}
    q(M)=1.790-0.2232[7.764-\log_{10}(M)].
\end{equation}
We derive the IMF-averaged mass ejecta for an element $i$ produced by CCSN with
\begin{equation}
    M^\mathrm{CC}_i=\int_0^{t}dt'\int_{9M_\odot}^{100M_\odot}dM Y^\mathrm{CC}_i(M)\Phi(M)\mathrm{SFR}(t'-\tau(M)),
    \label{eq:CCSN}
\end{equation}
where $Y^\mathrm{CC}_i$, $t$, and $\mathrm{SFR}(t)$ are the yield mass of the element $i$ produced by CCSN, stellar age, and star formation rate at $t$, respectively. We use the models with the metallicity $Z=0.004$ provided by \citet{Nomoto2013} for $Y^\mathrm{CC}_i$. We note that the resulting variation in [O/Fe] of our chemical evolution models is only $\sim0.1$ dex for different metallicities (see also Figure 9 in \citealt{Nomoto2013}), which does not change our conclusion. We assume a constant star formation history to calculate $\mathrm{SFR}(t)$ for our fiducial models, which is aligned with that our fiducial measurements of [O/Fe] are based on the constant star formation history. To investigate the effects of the star formation history, we also adopt the decreasing and increasing star formation histories, obtained with the SED fitting for GS-11-0 \citep{Hainline2024} and GN-z11 \citep{Tacchella2023}, respectively. In the PISN models, we also calculate the IMF-averaged yields of PISNe, which is caused by the stars with the masses of $140M_\odot<M<300M_\odot$ at the end of their lifetimes (top right panel of Figure \ref{fig:model_construction}; \citealt{Takahashi2018}), in the same way as CCSNe. We consider two types of yields for the PISN models, CCSNe + PISNe with and without failed supernovae, which directly collapse into black holes and do not eject any gas (e.g., \citealt{Ebinger2020}). We assume no ejecta for the stars with masses of $25M_\odot<M<140M_\odot$ if we add failed supernovae. For SNe Ia, we consider delay time distribution (DTD), which is the probability distribution of the number of SNe Ia occurring with a delay time (bottom left panel of Figure \ref{fig:model_construction}). We use the DTD with a power-law function form of $\propto t^{-1}$ (e.g., \citealt{Rodney2014}),
\begin{equation}
    \mathrm{DTD(t)}=\frac{N_\mathrm{Ia}t^{-1}}{\ln(t_\mathrm{max}/t_\mathrm{min})} \Theta(t-t_\mathrm{min}),
    \label{eq:DTD}
\end{equation}
where $N_\mathrm{Ia}$, $t_\mathrm{min}$, $t_\mathrm{max}$, and $\Theta(x)$ indicate the SN Ia rate, minimum delay time, cosmic age at present ($=\SI{13.8}{Gyr}$), and Heaviside step function, respectively. The SN Ia rate, the time-integrated number of SNe Ia per $1M_\odot$, is measured to be $\sim(1-7)\times10^{-3}M_\odot^{-1}$ (e.g., \citealt{Rodney2014,Maoz2017}) and fixed to be $N_\mathrm{Ia}=5.4\times10^{-3}M_\odot^{-1}$ \citep{Maoz2017} in our SN Ia models. We set three types of the minimum delay time of $t_\mathrm{min}=\SI{30}{Myr}$, $\SI{50}{Myr}$, and $\SI{100}{Myr}$ (e.g., \citealt{Totani2008,Maoz2014,Rodney2014,Maiolino2019}). We calculate the mass for an element $i$ ejected by SNe Ia with the delay time distribution, described as
\begin{equation}
    M_i^\mathrm{Ia}=\int_{t_\mathrm{min}}^{t}d\tau \int_{t_\mathrm{min}}^tdt' Y_i^\mathrm{Ia} \mathrm{DTD}(t') \mathrm{SFR}(\tau-t'),
    \label{eq:SNIa}
\end{equation}
where $Y_i^\mathrm{Ia}$ is the yield mass of the element $i$ produced by SN Ia. We use the W7 model (\citealt{Nomoto1984,Iwamoto1999}; see Section \ref{subsec:yields}) for $Y_i^\mathrm{Ia}$. We derive [O/Fe] ratios as \red{functions} of $t$ for the PISN (SN Ia) models by converting the ratios of the mass ejected by CCSNe and PISNe (SNe Ia) to the number ratios. In the bottom right panel of Figure \ref{fig:model_construction}, we present an example of the SNe Ia model with the \citet{Salpeter1955} IMF and delay time of $t_\mathrm{min}=\SI{100}{Myr}$, assuming a constant star formation history. At $t\sim0-30$ Myr, the [O/Fe] ratio decreases as relatively lower-mass massive stars (with longer lifetimes) eject gas with lower [O/Fe] through CCSNe. Between $t\sim30-100$ Myr, the ejecta from all massive stars in the allowed mass range are mixed, resulting in a flat [O/Fe] ratio. At $t>\SI{100}{Myr}$, SNe Ia begin to contribute significantly, leading a sharp decrease in the [O/Fe] ratio.

\begin{figure*}
    \centering
    \includegraphics[width=0.99\linewidth]{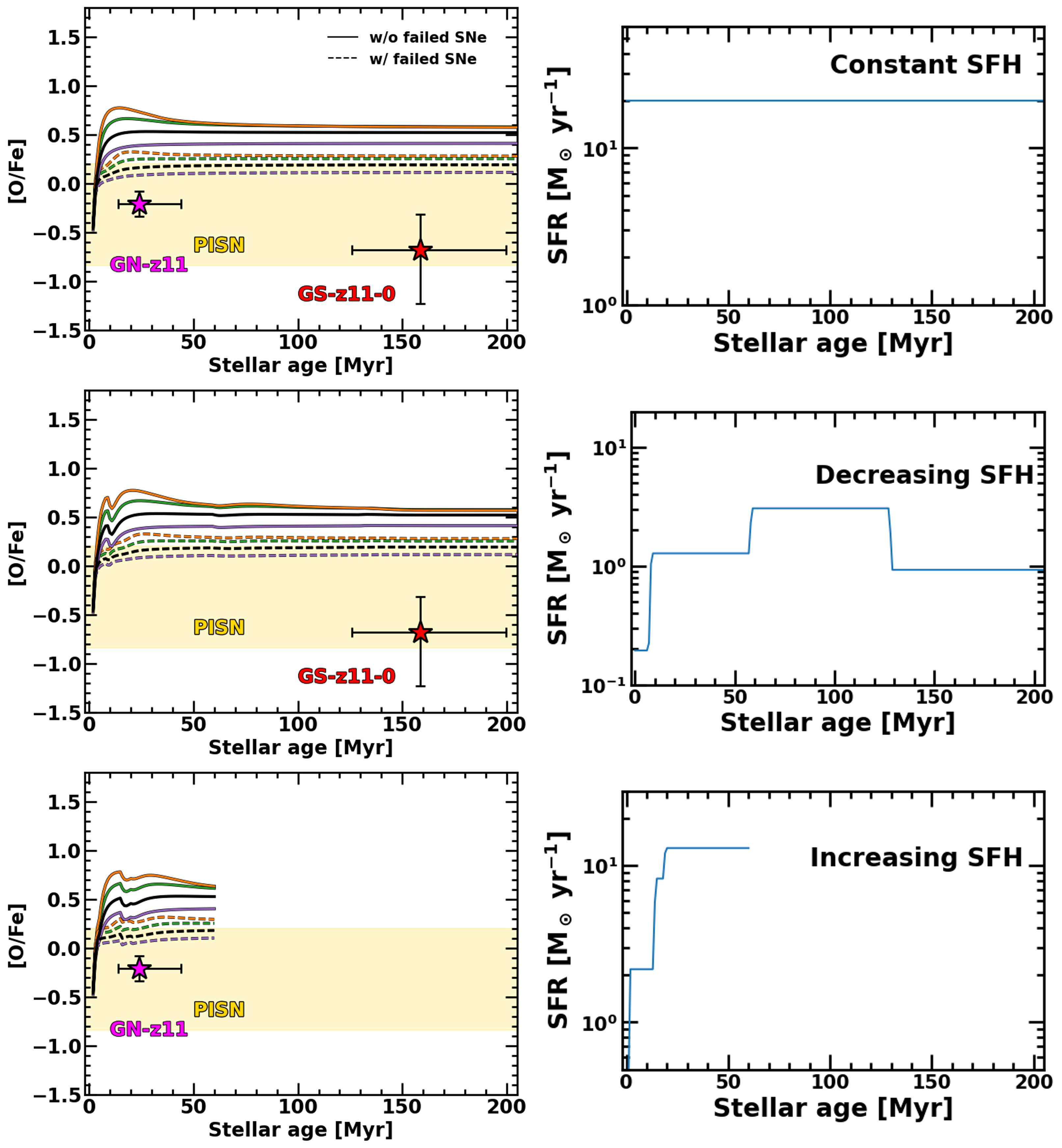}
    \caption{Comparison of the observed abundance ratios with the PISN models. The top, middle, and bottom panels show [O/Fe] as a function of stellar age (left) with our PISN models constructed under the assumption of the constant star formation history (SFH), decreasing SFH \citep{Hainline2024}, and increasing SFH \citep{Tacchella2023} (right), respectively. The red and magenta star symbols represent the measurements of GS-z11-0 and GN-z11, respectively. The purple, black, green, and orange lines indicate the PISN models with the top-heavy, \citet{Salpeter1955}, \citet{Kroupa1993}, and top-light IMFs, respectively. The solid and dashed lines denote the PISN models without and with failed SNe, respectively. The yellow-shaded regions present the yield models of PISNe \citep{Takahashi2018}.}
    \label{fig:PISN_model}
\end{figure*}
\begin{figure*}
    \centering
    \includegraphics[width=0.99\linewidth]{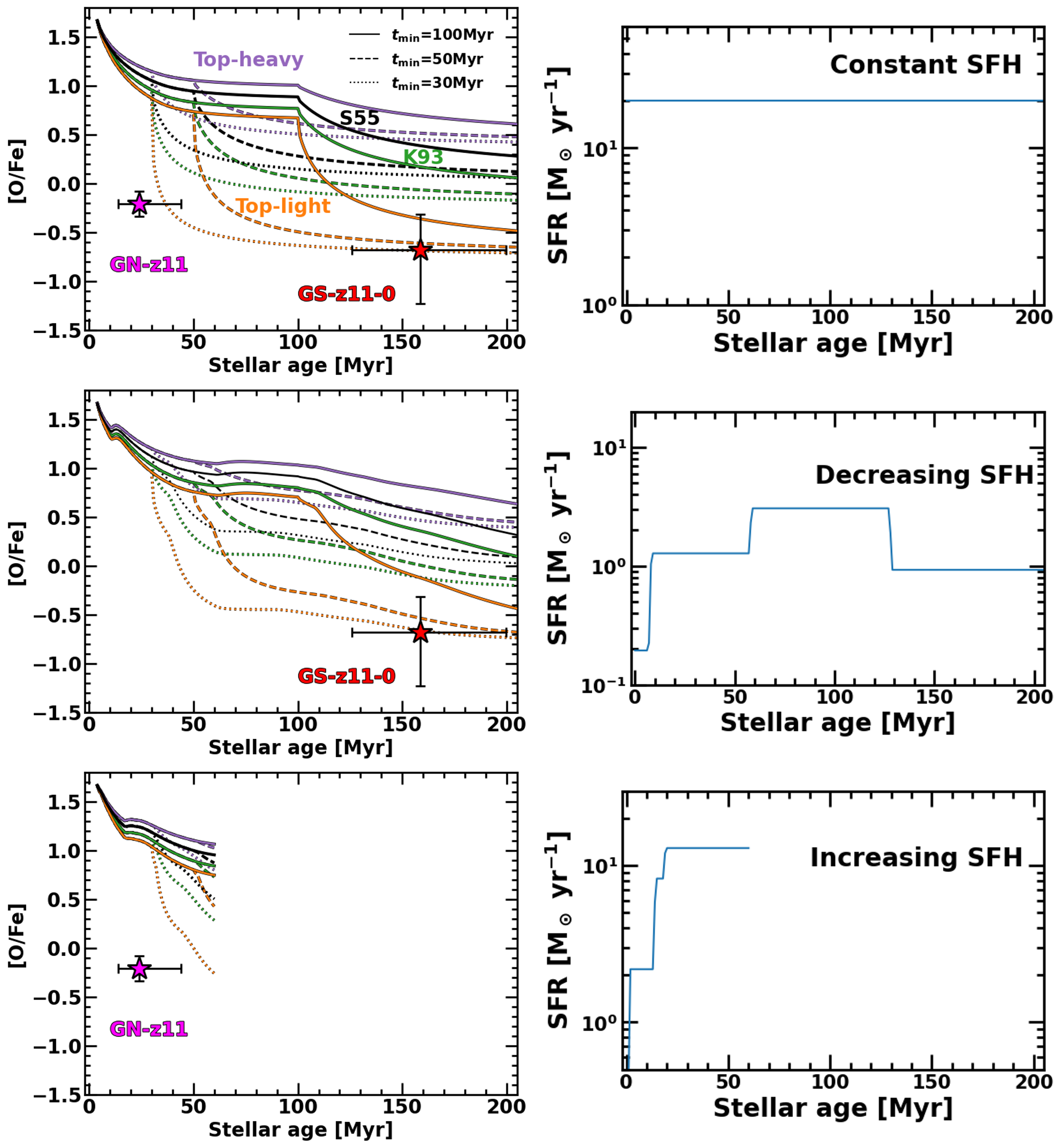}
    \caption{Same as Figure \ref{fig:PISN_model}, but for the SN Ia models. The solid, dashed, and dotted lines present our SN Ia models with the delay times of $t_\mathrm{min}=\SI{100}{Myr}$, $\SI{50}{Myr}$, and $\SI{30}{Myr}$, respectively. }
    \label{fig:SNIa_model}
\end{figure*}

\subsection{Origins of Low [O/Fe] Ratios}
\label{subsec:origin}
In Figures \ref{fig:PISN_model} and \ref{fig:SNIa_model}, we compare the [O/Fe] ratios of GS-z11-0/GN-z11 with the PISN and SN Ia models, respectively. We take the values of stellar age measured from the SED fitting in the literature ($t=158_{-32}^{+42}\ \SI{}{Myr}$ for GS-z11-0; \citealt{Hainline2024} and $t=24_{-10}^{+20}\ \SI{}{Myr}$ for GN-z11; \citealt{Tacchella2023}) rather than our fitting results. This is because stellar age from the SED fitting reflects more galaxy-averaged properties including older populations compared to our measurements, mainly tracing massive young populations (see \red{Section} \ref{subsubsec:comparison}). Note that for GN-z11 in the stellar case, we adopt the value measured from the combined photometry of the point source and extended component. The point source and extended component have stellar ages of $t=11_{-7}^{+58}$ and $35_{-19}^{+15}$ Myr, respectively. On the other hand, even under the assumption of dominant AGN radiation for GN-z11, the relative contribution from AGN and stellar components to the observed fluxes remains highly uncertain, preventing a reliable estimate of the stellar age. We thus only compare with the stellar case for GN-z11. The top panels present our fiducial models with the assumption of the constant star formation history. While the PISN models cannot explain the low [O/Fe] values regardless of the IMF slope and existence of failed SNe, the yields of only PISN (yellow shaded region in Figure \ref{fig:PISN_model}) can explain the low [O/Fe] values. In this case, BrHNe may also be the cause of Fe enhancement (see Figure \ref{fig:z-O_Fe}). In the SN Ia case, it is difficult to explain the low [O/Fe] ratio except with the models incorporating the short delay time ($t_\mathrm{min}\sim30-50\ \SI{}{Myr}$) and top-light IMF. We also develop the PISN and SN Ia models under the assumption of the decreasing and increasing star formation histories, as shown in the middle and bottom panels, respectively. The decreasing star formation history has negligible effects on both the PISN and SNIa models. While the increasing star formation history does not largely change the PISN models, it elevates the [O/Fe] ratios of the SN Ia models, which makes it somewhat difficult to explain the low [O/Fe] ratio of GN-z11. We note that the increasing star formation history is obtained for the point source in GN-z11. If the Fe enhancement of GN-z11 occurs in the point source region, PISNe or BrHNe may be likely causes. In summary, the star formation history has minor effects on our results.

We discuss the star formation in the early galaxy with the Fe enhancement based on the comparison with the PISN and SN Ia models. To explain the observed low [O/Fe] ratios, it is necessary both to produce iron by PISNe/BrHNe/SNe Ia and to suppress the oxygen enhancement by reducing the contribution from CCSNe. Comparison with the PISNe models suggests that the observed Fe enhancement may be caused by only PISNe or BrHNe. This unusual condition can be interpreted as follows. Under the metal-poor environments in the early epoch of the Universe, massive stars are expected to be predominant based on the hydrodynamical simulations (e.g., \citealt{Hirano2014,Hirano2015,Chon2021}). If only massive stars with $M>140\ M_\odot$ form as the first generation stars and cause PISNe, Fe-enhanced gas is produced. Alternatively, massive stars ($M\gtrsim30\ M_\odot$) in the first generation cause not CCSNe but BrHNe, which also results in the Fe enhancements of the gas. The second generation stars then form from the gas, for which we can observe the low [O/Fe] ratios. Actually, this kind of dual bursty star formation scenario is suggested to explain the abundance ratios of high-$z$ galaxies \citep{Kobayashi&Ferrara2024}. For GN-z11, the stellar age of the point source is smaller than that of the extended component, as described above. This suggests a younger starburst in a central, compact region, which may support the dual burst scenario.

The comparison with the SN Ia models suggests that the short delay time and top-light IMF are required to explain the observed low [O/Fe] ratios. The short delay time may be accomplished by the fast gas accretion for single degenerate systems or fast white dwarf merger for double degenerate systems in the dense star formation region, which is suggested by the compact morphologies of GS-z11-0 and GN-z11 with the half-light radii of $\SI{108}{pc}$ and $\SI{55}{pc}$, respectively \citep{Ono2025}. The top-light IMF contrasts with the predominant formation of massive stars expected in the early epoch of the Universe as described above. One possibility is that hydrogen deuteride (HD) radiation, which is a more efficient coolant for star-forming gas, plays a greater role than the standard coolant of molecular hydrogen (\ce{H_2}), resulting in fewer massive stars (e.g., \citealt{Hirano2014,Nishijima2024}). In the SNe Ia case, we note that the metallicity dependence of the delay time distribution for the single degenerate systems is suggested (e.g., \citealt{Kobayashi2009}). For the accreting white dwarf, the optically thick winds, which regulate the mass accretion rate, are mainly due to the opacity of iron and can be critical to cause SNe Ia (e.g., \citealt{Hachisu1996,Hachisu1999,Kobayashi1998}). Based on the results of \citet{Kobayashi2009}, short delay time for the single degenerate systems requires the high iron abundances ($\mathrm{[Fe/H]}\gtrsim-1$ \red{dex}) for the progenitor of SNe Ia. However, it may not be possible to realize both to form the iron-rich ($\mathrm{[Fe/H]}\gtrsim-1$ \red{dex}) progenitors and to cause SNe Ia in the epoch as early as $z\sim10$. This \red{implies} that if the Fe enhancement is caused by SNe Ia with short delay time, the progenitors of SNe Ia may be the double-degenerate systems. For iron enrichment by SNe Ia, it is worth to compare with the stars in the MW and Sculptor galaxies. In the [Fe/H]-[O/Fe] plot (Figure \ref{fig:OFe}), our high-$z$ galaxies are on a different sequence from the MW stars, but appear to be on the same sequence as the metal-poor stars in the Sculptor galaxy. While [O/Fe] ratios of the MW stars begin to decrease at $\mathrm{[Fe/H]}\sim-1$ \red{dex}, those of the stars in the Sculptor galaxy start to decrease at $\mathrm{[Fe/H]}\sim-2$ \red{dex} (e.g., \citealt{Tang2023}). This may imply that the high-$z$ galaxies experience a similar Fe enrichment by SNe Ia to the Sculptor galaxy, which is earlier than the MW.

For further exploration of the low [O/Fe] ratios in the early galaxies, it is important to examine other elemental abundance ratios. While GS-z11-0 show only a few emission lines \citep{Hainline2024}, many emission lines are detected for GN-z11 \citep{Bunker2023} (see Figure \ref{fig:individual_spectra}). In particular, a nitrogen-enhanced ratio of $\mathrm{[N/O]}>0.61$ \red{dex} for GN-z11 \citep{Cameron2023} is also puzzling. To accomplish the high [N/O] ratio, it is necessary to suppress the oxygen enrichment as well as the low [O/Fe] ratios. One possible scenario to explain both the high [N/O] and low [O/Fe] ratios is the dual starbursts (e.g., \citealt{Kobayashi&Ferrara2024,Nakane2024}). One of the other possible scenarios is the chemically differential wind (e.g., \citealt{Rizzuti2024}). In this scenario, elements such as nitrogen produced by low/intermediate mass stars or iron originating in SNe Ia, all formed in isolation, are less ejected from the galaxy relative to $\alpha$-elements (e.g., oxygen) selectively ejected by CCSNe, which are clustered together. The high [N/O] ratios of GN-z11 and other high-$z$ galaxies can be explained by the differential winds in \citet{Rizzuti2024}. For the low [O/Fe] ratios, we may need to reconcile the weak winds from isolated SNe Ia with short delay time, which requires fast gas accretion/white dwarf merger, possibly triggered in the dense star formation. In any cases, all of PISNe, BrHNe, and SNe Ia have possibility to reproduce the observed low [O/Fe] and high [N/O] ratios. To distinguish the origins of the low [O/Fe] ratios, it is necessary to use intermediate-mass elements, such as S and Ar, demonstrated in the recent studies (e.g., \citealt{Watanabe2024,Leung2024}).

\section{Summary} 
\label{sec:summary}
In this paper, we present our measurements of Fe abundances for $7$ galaxies at $9.3<z<12.3$ with $-22<M_\mathrm{UV}<-19$ selected based on the high S/N spectra obtained from multiple JWST/NIRSpec programs. To measure the iron abundances of [Fe/H], we fit the observed spectra with the stellar population synthesis model spectra in the rest-frame UV wavelength ranges, masking out the nebular emission lines. For GN-z11, we also conduct spectral fitting with the AGN models in the case where the UV continuum is dominated by the AGN radiation. Combined with oxygen abundance measurements from emission lines, we obtain [O/Fe] ratios. Our major findings are summarized below:
\begin{itemize}
    \item[1.] We obtain [Fe/H] values for $5$ galaxies and the $3\sigma$ upper limits for $2$ galaxies. From [O/Fe] measurements, we find that majority of ($5$ out of $7$) galaxies are consistent with $\mathrm{[O/Fe]}\gtrsim0$ \red{dex} while that $2$ out of $7$ galaxies show Fe enhancements
    (i.e., low [O/Fe] ratios) of $\mathrm{[O/Fe]}=-0.68_{-0.55}^{+0.37}$ \red{dex} and $-0.21_{-0.13}^{+0.13}$ \red{dex} ($-1.1\text{--}0.0$ \red{dex}) for GS-z11-0 and GN-z11 in the stellar (AGN) case, respectively.
    
    \item[2.] The [O/Fe] ratios of  the $5$ galaxies are comparable to those of CCSNe yields as well as $z\sim2-6$ galaxies. In contrast, the Fe enhancements of GS-z11-0 and GN-z11 require the contribution of theoretical PISNe/BrHNe or SNeIa. To explore the origins of low [O/Fe] ratios, we develop chemical evolution models, incorporating the yields of CCSNe, PISNe, and SNe Ia. By comparing our [O/Fe] measurements with the models, we suggest that the Fe enhancements are accomplished by 1) PISNe/BrHNe in the first star formation with little contribution from CCSNe, or 2) SNe Ia with short delay time ($\sim30-50$ Myr) and a top-light IMF.
\end{itemize}


\section*{Acknowledgements} We thank the anonymous referee for through reading and valuable comments that improved our manuscript. We thank Andrea Ferrara, Souradeep Bhattcharya, Andrew Bunker, Hajime Fukushima, Koki Kakiichi, Chiaki Kobayashi, Roberto Maiolino, Charlotte Mason, and Hidenobu Yajima for the valuable discussions on this work. This work is based on observations made with the NASA/ESA/CSA James Webb Space Telescope. The data were obtained from the Mikulski Archive for Space Telescopes at the Space Telescope Science Institute, which is operated by the Association of Universities for Research in Astronomy, Inc., under NASA contract NAS 5-03127 for JWST. These observations are associated with programs ERS-1345 and DDT-2750 (CEERS), GTOs-1180, 1181, 1210, 1286, and GO-3215 (JADES), GO-2561 (UNCOVER), DDT-2767, GO-1433, GO-2198, and GO-3073. We acknowledge the CEERS, JADES, UNCOVER, DDT-2767, GO-1433, GO-2198, and GO-3073 teams led by Steven L. Finkelstein, Daniel Eisenstein \& Nora L\"uetzgendorf, Ivo Labb{\'e} \& Rachel Bezanson, Patrick L. Kelly, Dan Coe, Laia Barrufet \& Pascal Oesch, and Marco Castellano, respectively, for developing their observation programs. The JWST data presented in this paper were obtained from Mikulski Archive for Space Telescope at the Space Telescope Institute. The calibration level $3$ data in MAST is available at doi: \dataset[10.17909/fxtr-3677]{https://doi.org/10.17909/fxtr-3677}. In
our analysis, we begin with the versions of level $2$. Some of the data products presented herein were retrieved from the Dawn JWST Archive (DJA). DJA is an initiative of the Cosmic Dawn Center (DAWN), which is funded by the Danish National Research Foundation under grant DNRF140. This publication is based upon work supported by the World Premier International Research Center Initiative (WPI Initiative), MEXT, Japan, and KAKENHI (20H00180, 21H04467) through Japan Society for the Promotion of Science (JSPS). M.N. and Y.I. are supported by JSPS KAKENHI Grant Nos. 25KJ0828 and 24KJ0202, respectively. K.N. is supported by JSPS KAKENHI Grant Nos. 20K04024, 21H044pp, 23K03452, and 25K01046. This work was supported by the joint research program of the Institute for Cosmic Ray Research (ICRR), the University of Tokyo.

%


\software{NumPy \citep{Harris2020}, matplotlib \citep{Hunter2007}, SciPy \citep{Virtanen2020}, Astropy \citep{Astropy2013,Astropy2018,Astropy2022}, emcee \citep{Foreman2013}, BPASS v2.2.1 \citep{Eldridge2017,Stanway2018}, and \textsc{Cloudy} v23.01 \citep{Ferland1998,Gunasekera2023}}.





\bibliographystyle{aasjournal}
\bibliography{library.bib}

\begin{thebibliography}{}
\expandafter\ifx\csname natexlab\endcsname\relax\def\natexlab#1{#1}\fi
\providecommand{\url}[1]{\href{#1}{#1}}
\providecommand{\dodoi}[1]{doi:~\href{http://doi.org/#1}{\nolinkurl{#1}}}
\providecommand{\doeprint}[1]{\href{http://ascl.net/#1}{\nolinkurl{http://ascl.net/#1}}}
\providecommand{\doarXiv}[1]{\href{https://arxiv.org/abs/#1}{\nolinkurl{https://arxiv.org/abs/#1}}}

\bibitem[{{Abdurro'uf} {et~al.}(2024){Abdurro'uf}, {Larson}, {Coe}, {Hsiao}, {{\'A}lvarez-M{\'a}rquez}, {Crespo G{\'o}mez}, {Adamo}, {Bhatawdekar}, {Bik}, {Bradley}, {Conselice}, {Dayal}, {Diego}, {Fujimoto}, {Furtak}, {Hutchison}, {Jung}, {Killi}, {Kokorev}, {Mingozzi}, {Norman}, {Resseguier}, {Ricotti}, {Rigby}, {Vanzella}, {Welch}, {Windhorst}, {Xu}, \& {Zitrin}}]{Abdurro'uf2024}
{Abdurro'uf}, {Larson}, R.~L., {Coe}, D., {et~al.} 2024, \apj, 973, 47, \dodoi{10.3847/1538-4357/ad6001}

\bibitem[{Akaike(1973)}]{Akaike1973}
Akaike, H. 1973, Information Theory and an Extension of the Maximum Likelihood Principle (New York, NY: Springer New York), 199--213

\bibitem[{{{\'A}lvarez-M{\'a}rquez} {et~al.}(2025){{\'A}lvarez-M{\'a}rquez}, {Crespo G{\'o}mez}, {Colina}, {Langeroodi}, {Marques-Chaves}, {Prieto-Jim{\'e}nez}, {Bik}, {Alonso-Herrero}, {Boogaard}, {Costantin}, {Garc{\'\i}a-Mar{\'\i}n}, {Gillman}, {Hjorth}, {Iani}, {Jermann}, {Labiano}, {Melinder}, {Meyer}, {{\"O}stlin}, {P{\'e}rez-Gonz{\'a}lez}, {Rinaldi}, {Walter}, {van der Werf}, \& {Wright}}]{Alvarez-Marquez2025}
{{\'A}lvarez-M{\'a}rquez}, J., {Crespo G{\'o}mez}, A., {Colina}, L., {et~al.} 2025, \aap, 695, A250, \dodoi{10.1051/0004-6361/202451731}

\bibitem[{{Amarsi} {et~al.}(2019){Amarsi}, {Nissen}, \& {Sk{\'u}lad{\'o}ttir}}]{Amarsi2019}
{Amarsi}, A.~M., {Nissen}, P.~E., \& {Sk{\'u}lad{\'o}ttir}, {\'A}. 2019, \aap, 630, A104, \dodoi{10.1051/0004-6361/201936265}

\bibitem[{{Arellano-C{\'o}rdova} {et~al.}(2022){Arellano-C{\'o}rdova}, {Berg}, {Chisholm}, {Arrabal Haro}, {Dickinson}, {Finkelstein}, {Leclercq}, {Rogers}, {Simons}, {Skillman}, {Trump}, \& {Kartaltepe}}]{Arellano-Cordova2022}
{Arellano-C{\'o}rdova}, K.~Z., {Berg}, D.~A., {Chisholm}, J., {et~al.} 2022, \apjl, 940, L23, \dodoi{10.3847/2041-8213/ac9ab2}

\bibitem[{{Arrabal Haro} {et~al.}(2023{\natexlab{a}}){Arrabal Haro}, {Dickinson}, {Finkelstein}, {Fujimoto}, {Fern{\'a}ndez}, {Kartaltepe}, {Jung}, {Cole}, {Burgarella}, {Chworowsky}, {Hutchison}, {Morales}, {Papovich}, {Simons}, {Amor{\'\i}n}, {Backhaus}, {Bagley}, {Bisigello}, {Calabr{\`o}}, {Castellano}, {Cleri}, {Dav{\'e}}, {Dekel}, {Ferguson}, {Fontana}, {Gawiser}, {Giavalisco}, {Harish}, {Hathi}, {Hirschmann}, {Holwerda}, {Huertas-Company}, {Koekemoer}, {Larson}, {Lucas}, {Mobasher}, {P{\'e}rez-Gonz{\'a}lez}, {Pirzkal}, {Rose}, {Santini}, {Trump}, {de la Vega}, {Wang}, {Weiner}, {Wilkins}, {Yang}, {Yung}, \& {Zavala}}]{Haro2023a}
{Arrabal Haro}, P., {Dickinson}, M., {Finkelstein}, S.~L., {et~al.} 2023{\natexlab{a}}, \apjl, 951, L22, \dodoi{10.3847/2041-8213/acdd54}

\bibitem[{{Arrabal Haro} {et~al.}(2023{\natexlab{b}}){Arrabal Haro}, {Dickinson}, {Finkelstein}, {Kartaltepe}, {Donnan}, {Burgarella}, {Carnall}, {Cullen}, {Dunlop}, {Fern{\'a}ndez}, {Fujimoto}, {Jung}, {Krips}, {Larson}, {Papovich}, {P{\'e}rez-Gonz{\'a}lez}, {Amor{\'\i}n}, {Bagley}, {Buat}, {Casey}, {Chworowsky}, {Cohen}, {Ferguson}, {Giavalisco}, {Huertas-Company}, {Hutchison}, {Kocevski}, {Koekemoer}, {Lucas}, {McLeod}, {McLure}, {Pirzkal}, {Seill{\'e}}, {Trump}, {Weiner}, {Wilkins}, \& {Zavala}}]{Haro2023b}
---. 2023{\natexlab{b}}, \nat, 622, 707, \dodoi{10.1038/s41586-023-06521-7}

\bibitem[{{Asplund} {et~al.}(2009){Asplund}, {Grevesse}, {Sauval}, \& {Scott}}]{Asplund2009}
{Asplund}, M., {Grevesse}, N., {Sauval}, A.~J., \& {Scott}, P. 2009, \araa, 47, 481, \dodoi{10.1146/annurev.astro.46.060407.145222}

\bibitem[{{Astropy Collaboration} {et~al.}(2013){Astropy Collaboration}, {Robitaille}, {Tollerud}, {Greenfield}, {Droettboom}, {Bray}, {Aldcroft}, {Davis}, {Ginsburg}, {Price-Whelan}, {Kerzendorf}, {Conley}, {Crighton}, {Barbary}, {Muna}, {Ferguson}, {Grollier}, {Parikh}, {Nair}, {Unther}, {Deil}, {Woillez}, {Conseil}, {Kramer}, {Turner}, {Singer}, {Fox}, {Weaver}, {Zabalza}, {Edwards}, {Azalee Bostroem}, {Burke}, {Casey}, {Crawford}, {Dencheva}, {Ely}, {Jenness}, {Labrie}, {Lim}, {Pierfederici}, {Pontzen}, {Ptak}, {Refsdal}, {Servillat}, \& {Streicher}}]{Astropy2013}
{Astropy Collaboration}, {Robitaille}, T.~P., {Tollerud}, E.~J., {et~al.} 2013, \aap, 558, A33, \dodoi{10.1051/0004-6361/201322068}

\bibitem[{{Astropy Collaboration} {et~al.}(2018){Astropy Collaboration}, {Price-Whelan}, {Sip{\H{o}}cz}, {G{\"u}nther}, {Lim}, {Crawford}, {Conseil}, {Shupe}, {Craig}, {Dencheva}, {Ginsburg}, {VanderPlas}, {Bradley}, {P{\'e}rez-Su{\'a}rez}, {de Val-Borro}, {Aldcroft}, {Cruz}, {Robitaille}, {Tollerud}, {Ardelean}, {Babej}, {Bach}, {Bachetti}, {Bakanov}, {Bamford}, {Barentsen}, {Barmby}, {Baumbach}, {Berry}, {Biscani}, {Boquien}, {Bostroem}, {Bouma}, {Brammer}, {Bray}, {Breytenbach}, {Buddelmeijer}, {Burke}, {Calderone}, {Cano Rodr{\'\i}guez}, {Cara}, {Cardoso}, {Cheedella}, {Copin}, {Corrales}, {Crichton}, {D'Avella}, {Deil}, {Depagne}, {Dietrich}, {Donath}, {Droettboom}, {Earl}, {Erben}, {Fabbro}, {Ferreira}, {Finethy}, {Fox}, {Garrison}, {Gibbons}, {Goldstein}, {Gommers}, {Greco}, {Greenfield}, {Groener}, {Grollier}, {Hagen}, {Hirst}, {Homeier}, {Horton}, {Hosseinzadeh}, {Hu}, {Hunkeler}, {Ivezi{\'c}}, {Jain}, {Jenness}, {Kanarek}, {Kendrew}, {Kern}, {Kerzendorf}, {Khvalko}, {King}, {Kirkby}, {Kulkarni},
  {Kumar}, {Lee}, {Lenz}, {Littlefair}, {Ma}, {Macleod}, {Mastropietro}, {McCully}, {Montagnac}, {Morris}, {Mueller}, {Mumford}, {Muna}, {Murphy}, {Nelson}, {Nguyen}, {Ninan}, {N{\"o}the}, {Ogaz}, {Oh}, {Parejko}, {Parley}, {Pascual}, {Patil}, {Patil}, {Plunkett}, {Prochaska}, {Rastogi}, {Reddy Janga}, {Sabater}, {Sakurikar}, {Seifert}, {Sherbert}, {Sherwood-Taylor}, {Shih}, {Sick}, {Silbiger}, {Singanamalla}, {Singer}, {Sladen}, {Sooley}, {Sornarajah}, {Streicher}, {Teuben}, {Thomas}, {Tremblay}, {Turner}, {Terr{\'o}n}, {van Kerkwijk}, {de la Vega}, {Watkins}, {Weaver}, {Whitmore}, {Woillez}, {Zabalza}, \& {Astropy Contributors}}]{Astropy2018}
{Astropy Collaboration}, {Price-Whelan}, A.~M., {Sip{\H{o}}cz}, B.~M., {et~al.} 2018, \aj, 156, 123, \dodoi{10.3847/1538-3881/aabc4f}

\bibitem[{{Astropy Collaboration} {et~al.}(2022){Astropy Collaboration}, {Price-Whelan}, {Lim}, {Earl}, {Starkman}, {Bradley}, {Shupe}, {Patil}, {Corrales}, {Brasseur}, {N{\"o}the}, {Donath}, {Tollerud}, {Morris}, {Ginsburg}, {Vaher}, {Weaver}, {Tocknell}, {Jamieson}, {van Kerkwijk}, {Robitaille}, {Merry}, {Bachetti}, {G{\"u}nther}, {Aldcroft}, {Alvarado-Montes}, {Archibald}, {B{\'o}di}, {Bapat}, {Barentsen}, {Baz{\'a}n}, {Biswas}, {Boquien}, {Burke}, {Cara}, {Cara}, {Conroy}, {Conseil}, {Craig}, {Cross}, {Cruz}, {D'Eugenio}, {Dencheva}, {Devillepoix}, {Dietrich}, {Eigenbrot}, {Erben}, {Ferreira}, {Foreman-Mackey}, {Fox}, {Freij}, {Garg}, {Geda}, {Glattly}, {Gondhalekar}, {Gordon}, {Grant}, {Greenfield}, {Groener}, {Guest}, {Gurovich}, {Handberg}, {Hart}, {Hatfield-Dodds}, {Homeier}, {Hosseinzadeh}, {Jenness}, {Jones}, {Joseph}, {Kalmbach}, {Karamehmetoglu}, {Ka{\l}uszy{\'n}ski}, {Kelley}, {Kern}, {Kerzendorf}, {Koch}, {Kulumani}, {Lee}, {Ly}, {Ma}, {MacBride}, {Maljaars}, {Muna}, {Murphy}, {Norman},
  {O'Steen}, {Oman}, {Pacifici}, {Pascual}, {Pascual-Granado}, {Patil}, {Perren}, {Pickering}, {Rastogi}, {Roulston}, {Ryan}, {Rykoff}, {Sabater}, {Sakurikar}, {Salgado}, {Sanghi}, {Saunders}, {Savchenko}, {Schwardt}, {Seifert-Eckert}, {Shih}, {Jain}, {Shukla}, {Sick}, {Simpson}, {Singanamalla}, {Singer}, {Singhal}, {Sinha}, {Sip{\H{o}}cz}, {Spitler}, {Stansby}, {Streicher}, {{\v{S}}umak}, {Swinbank}, {Taranu}, {Tewary}, {Tremblay}, {de Val-Borro}, {Van Kooten}, {Vasovi{\'c}}, {Verma}, {de Miranda Cardoso}, {Williams}, {Wilson}, {Winkel}, {Wood-Vasey}, {Xue}, {Yoachim}, {Zhang}, {Zonca}, \& {Astropy Project Contributors}}]{Astropy2022}
{Astropy Collaboration}, {Price-Whelan}, A.~M., {Lim}, P.~L., {et~al.} 2022, \apj, 935, 167, \dodoi{10.3847/1538-4357/ac7c74}

\bibitem[{{Baldwin}(1977)}]{Baldwin1977}
{Baldwin}, J.~A. 1977, \apj, 214, 679, \dodoi{10.1086/155294}

\bibitem[{{Baldwin} {et~al.}(1978){Baldwin}, {Burke}, {Gaskell}, \& {Wampler}}]{Baldwin1978}
{Baldwin}, J.~A., {Burke}, W.~L., {Gaskell}, C.~M., \& {Wampler}, E.~J. 1978, \nat, 273, 431, \dodoi{10.1038/273431a0}

\bibitem[{{Barrufet} {et~al.}(2025){Barrufet}, {Oesch}, {Marques-Chaves}, {Arellano-Cordova}, {Baggen}, {Carnall}, {Cullen}, {Dunlop}, {Gottumukkala}, {Fudamoto}, {Illingworth}, {Magee}, {McLure}, {McLeod}, {Micha{\l}owski}, {Stefanon}, {van Dokkum}, \& {Weibel}}]{Barrufet2025}
{Barrufet}, L., {Oesch}, P.~A., {Marques-Chaves}, R., {et~al.} 2025, \mnras, 537, 3453, \dodoi{10.1093/mnras/staf013}

\bibitem[{{Bensby} {et~al.}(2013){Bensby}, {Yee}, {Feltzing}, {Johnson}, {Gould}, {Cohen}, {Asplund}, {Mel{\'e}ndez}, {Lucatello}, {Han}, {Thompson}, {Gal-Yam}, {Udalski}, {Bennett}, {Bond}, {Kohei}, {Sumi}, {Suzuki}, {Suzuki}, {Takino}, {Tristram}, {Yamai}, \& {Yonehara}}]{Bensby2013}
{Bensby}, T., {Yee}, J.~C., {Feltzing}, S., {et~al.} 2013, \aap, 549, A147, \dodoi{10.1051/0004-6361/201220678}

\bibitem[{{Bezanson} {et~al.}(2024){Bezanson}, {Labbe}, {Whitaker}, {Leja}, {Price}, {Franx}, {Brammer}, {Marchesini}, {Zitrin}, {Wang}, {Weaver}, {Furtak}, {Atek}, {Coe}, {Cutler}, {Dayal}, {van Dokkum}, {Feldmann}, {F{\"o}rster Schreiber}, {Fujimoto}, {Geha}, {Glazebrook}, {de Graaff}, {Greene}, {Juneau}, {Kassin}, {Kriek}, {Khullar}, {Maseda}, {Mowla}, {Muzzin}, {Nanayakkara}, {Nelson}, {Oesch}, {Pacifici}, {Pan}, {Papovich}, {Setton}, {Shapley}, {Smit}, {Stefanon}, {Taylor}, \& {Williams}}]{Bezanson2024}
{Bezanson}, R., {Labbe}, I., {Whitaker}, K.~E., {et~al.} 2024, \apj, 974, 92, \dodoi{10.3847/1538-4357/ad66cf}

\bibitem[{{Bhattacharya} {et~al.}(2025){Bhattacharya}, {Arnaboldi}, {Gerhard}, {Kobayashi}, \& {Saha}}]{Bhattacharya2025}
{Bhattacharya}, S., {Arnaboldi}, M., {Gerhard}, O., {Kobayashi}, C., \& {Saha}, K. 2025, \apjl, 983, L30, \dodoi{10.3847/2041-8213/adc735}

\bibitem[{{Bian} {et~al.}(2018){Bian}, {Kewley}, \& {Dopita}}]{Bian2018}
{Bian}, F., {Kewley}, L.~J., \& {Dopita}, M.~A. 2018, \apj, 859, 175, \dodoi{10.3847/1538-4357/aabd74}

\bibitem[{{Bogd{\'a}n} {et~al.}(2024){Bogd{\'a}n}, {Goulding}, {Natarajan}, {Kov{\'a}cs}, {Tremblay}, {Chadayammuri}, {Volonteri}, {Kraft}, {Forman}, {Jones}, {Churazov}, \& {Zhuravleva}}]{Bogdan2024}
{Bogd{\'a}n}, {\'A}., {Goulding}, A.~D., {Natarajan}, P., {et~al.} 2024, Nature Astronomy, 8, 126, \dodoi{10.1038/s41550-023-02111-9}

\bibitem[{{Boyett} {et~al.}(2024){Boyett}, {Trenti}, {Leethochawalit}, {Calabr{\'o}}, {Metha}, {Roberts-Borsani}, {Dalmasso}, {Yang}, {Santini}, {Treu}, {Jones}, {Henry}, {Mason}, {Morishita}, {Nanayakkara}, {Roy}, {Wang}, {Fontana}, {Merlin}, {Castellano}, {Paris}, {Brada{\v{c}}}, {Malkan}, {Marchesini}, {Mascia}, {Glazebrook}, {Pentericci}, {Vanzella}, \& {Vulcani}}]{Boyett2024}
{Boyett}, K., {Trenti}, M., {Leethochawalit}, N., {et~al.} 2024, Nature Astronomy, 8, 657, \dodoi{10.1038/s41550-024-02218-7}

\bibitem[{{Brammer}(2023)}]{Brammer2023}
{Brammer}, G. 2023, {msaexp: NIRSpec analyis tools}, 0.6.17,  Zenodo, \dodoi{10.5281/zenodo.7299500}

\bibitem[{{Brandt} {et~al.}(1998){Brandt}, {Heap}, {Beaver}, {Boggess}, {Carpenter}, {Ebbets}, {Hutchings}, {Jura}, {Leckrone}, {Linsky}, {Maran}, {Savage}, {Smith}, {Trafton}, {Walter}, {Weymann}, {Snow}, {Ake}, \& {Hogen}}]{Brandt1998}
{Brandt}, J.~C., {Heap}, S.~R., {Beaver}, E.~A., {et~al.} 1998, \aj, 116, 941, \dodoi{10.1086/300446}

\bibitem[{{Bunker} {et~al.}(2023){Bunker}, {Saxena}, {Cameron}, {Willott}, {Curtis-Lake}, {Jakobsen}, {Carniani}, {Smit}, {Maiolino}, {Witstok}, {Curti}, {D'Eugenio}, {Jones}, {Ferruit}, {Arribas}, {Charlot}, {Chevallard}, {Giardino}, {de Graaff}, {Looser}, {L{\"u}tzgendorf}, {Maseda}, {Rawle}, {Rix}, {Del Pino}, {Alberts}, {Egami}, {Eisenstein}, {Endsley}, {Hainline}, {Hausen}, {Johnson}, {Rieke}, {Rieke}, {Robertson}, {Shivaei}, {Stark}, {Sun}, {Tacchella}, {Tang}, {Williams}, {Willmer}, {Baker}, {Baum}, {Bhatawdekar}, {Bowler}, {Boyett}, {Chen}, {Circosta}, {Helton}, {Ji}, {Kumari}, {Lyu}, {Nelson}, {Parlanti}, {Perna}, {Sandles}, {Scholtz}, {Suess}, {Topping}, {{\"U}bler}, {Wallace}, \& {Whitler}}]{Bunker2023}
{Bunker}, A.~J., {Saxena}, A., {Cameron}, A.~J., {et~al.} 2023, \aap, 677, A88, \dodoi{10.1051/0004-6361/202346159}

\bibitem[{{Bunker} {et~al.}(2024){Bunker}, {Cameron}, {Curtis-Lake}, {Jakobsen}, {Carniani}, {Curti}, {Witstok}, {Maiolino}, {D'Eugenio}, {Looser}, {Willott}, {Bonaventura}, {Hainline}, {{\"U}bler}, {Willmer}, {Saxena}, {Smit}, {Alberts}, {Arribas}, {Baker}, {Baum}, {Bhatawdekar}, {Bowler}, {Boyett}, {Charlot}, {Chen}, {Chevallard}, {Circosta}, {DeCoursey}, {de Graaff}, {Egami}, {Eisenstein}, {Endsley}, {Ferruit}, {Giardino}, {Hausen}, {Helton}, {Hviding}, {Ji}, {Johnson}, {Jones}, {Kumari}, {Laseter}, {L{\"u}tzgendorf}, {Maseda}, {Nelson}, {Parlanti}, {Perna}, {Rauscher}, {Rawle}, {Rix}, {Rieke}, {Robertson}, {Rodr{\'\i}guez Del Pino}, {Sandles}, {Scholtz}, {Sharpe}, {Skarbinski}, {Stark}, {Sun}, {Tacchella}, {Topping}, {Villanueva}, {Wallace}, {Williams}, \& {Woodrum}}]{Bunker2024}
{Bunker}, A.~J., {Cameron}, A.~J., {Curtis-Lake}, E., {et~al.} 2024, \aap, 690, A288, \dodoi{10.1051/0004-6361/202347094}

\bibitem[{{Calabr{\`o}} {et~al.}(2024){Calabr{\`o}}, {Castellano}, {Zavala}, {Pentericci}, {Arrabal Haro}, {Bakx}, {Burgarella}, {Casey}, {Dickinson}, {Finkelstein}, {Fontana}, {Llerena}, {Mascia}, {Merlin}, {Mitsuhashi}, {Napolitano}, {Paris}, {P{\'e}rez-Gonz{\'a}lez}, {Roberts-Borsani}, {Santini}, {Treu}, \& {Vanzella}}]{Calabro2024}
{Calabr{\`o}}, A., {Castellano}, M., {Zavala}, J.~A., {et~al.} 2024, \apj, 975, 245, \dodoi{10.3847/1538-4357/ad7602}

\bibitem[{{Calzetti} {et~al.}(2000){Calzetti}, {Armus}, {Bohlin}, {Kinney}, {Koornneef}, \& {Storchi-Bergmann}}]{Calzetti2000}
{Calzetti}, D., {Armus}, L., {Bohlin}, R.~C., {et~al.} 2000, \apj, 533, 682, \dodoi{10.1086/308692}

\bibitem[{{Cameron} {et~al.}(2023){Cameron}, {Katz}, {Rey}, \& {Saxena}}]{Cameron2023}
{Cameron}, A.~J., {Katz}, H., {Rey}, M.~P., \& {Saxena}, A. 2023, \mnras, 523, 3516, \dodoi{10.1093/mnras/stad1579}

\bibitem[{{Carretta} {et~al.}(2010){Carretta}, {Bragaglia}, {Gratton}, {Recio-Blanco}, {Lucatello}, {D'Orazi}, \& {Cassisi}}]{Carretta2010}
{Carretta}, E., {Bragaglia}, A., {Gratton}, R.~G., {et~al.} 2010, \aap, 516, A55, \dodoi{10.1051/0004-6361/200913451}

\bibitem[{{Carretta} {et~al.}(2005){Carretta}, {Gratton}, {Lucatello}, {Bragaglia}, \& {Bonifacio}}]{Carretta2005}
{Carretta}, E., {Gratton}, R.~G., {Lucatello}, S., {Bragaglia}, A., \& {Bonifacio}, P. 2005, \aap, 433, 597, \dodoi{10.1051/0004-6361:20041892}

\bibitem[{{Castellano} {et~al.}(2024){Castellano}, {Napolitano}, {Fontana}, {Roberts-Borsani}, {Treu}, {Vanzella}, {Zavala}, {Arrabal Haro}, {Calabr{\`o}}, {Llerena}, {Mascia}, {Merlin}, {Paris}, {Pentericci}, {Santini}, {Bakx}, {Bergamini}, {Cupani}, {Dickinson}, {Filippenko}, {Glazebrook}, {Grillo}, {Kelly}, {Malkan}, {Mason}, {Morishita}, {Nanayakkara}, {Rosati}, {Sani}, {Wang}, \& {Yoon}}]{Castellano2024}
{Castellano}, M., {Napolitano}, L., {Fontana}, A., {et~al.} 2024, \apj, 972, 143, \dodoi{10.3847/1538-4357/ad5f88}

\bibitem[{{Chartab} {et~al.}(2024){Chartab}, {Newman}, {Rudie}, {Blanc}, \& {Kelson}}]{Chartab2024}
{Chartab}, N., {Newman}, A.~B., {Rudie}, G.~C., {Blanc}, G.~A., \& {Kelson}, D.~D. 2024, \apj, 960, 73, \dodoi{10.3847/1538-4357/ad0554}

\bibitem[{{Chen} {et~al.}(2021){Chen}, {Hu}, \& {Wang}}]{Chen2021}
{Chen}, X., {Hu}, L., \& {Wang}, L. 2021, \apj, 922, 15, \dodoi{10.3847/1538-4357/ac178d}

\bibitem[{{Chisholm} {et~al.}(2019){Chisholm}, {Rigby}, {Bayliss}, {Berg}, {Dahle}, {Gladders}, \& {Sharon}}]{Chisholm2019}
{Chisholm}, J., {Rigby}, J.~R., {Bayliss}, M., {et~al.} 2019, \apj, 882, 182, \dodoi{10.3847/1538-4357/ab3104}

\bibitem[{{Chon} {et~al.}(2021){Chon}, {Omukai}, \& {Schneider}}]{Chon2021}
{Chon}, S., {Omukai}, K., \& {Schneider}, R. 2021, \mnras, 508, 4175, \dodoi{10.1093/mnras/stab2497}

\bibitem[{{Cullen} {et~al.}(2019){Cullen}, {McLure}, {Dunlop}, {Khochfar}, {Dav{\'e}}, {Amor{\'\i}n}, {Bolzonella}, {Carnall}, {Castellano}, {Cimatti}, {Cirasuolo}, {Cresci}, {Fynbo}, {Fontanot}, {Gargiulo}, {Garilli}, {Guaita}, {Hathi}, {Hibon}, {Mannucci}, {Marchi}, {McLeod}, {Pentericci}, {Pozzetti}, {Shapley}, {Talia}, \& {Zamorani}}]{Cullen2019}
{Cullen}, F., {McLure}, R.~J., {Dunlop}, J.~S., {et~al.} 2019, \mnras, 487, 2038, \dodoi{10.1093/mnras/stz1402}

\bibitem[{{Cullen} {et~al.}(2021){Cullen}, {Shapley}, {McLure}, {Dunlop}, {Sanders}, {Topping}, {Reddy}, {Amor{\'\i}n}, {Begley}, {Bolzonella}, {Calabr{\`o}}, {Carnall}, {Castellano}, {Cimatti}, {Cirasuolo}, {Cresci}, {Fontana}, {Fontanot}, {Garilli}, {Guaita}, {Hamadouche}, {Hathi}, {Mannucci}, {McLeod}, {Pentericci}, {Saxena}, {Talia}, \& {Zamorani}}]{Cullen2021}
{Cullen}, F., {Shapley}, A.~E., {McLure}, R.~J., {et~al.} 2021, \mnras, 505, 903, \dodoi{10.1093/mnras/stab1340}

\bibitem[{{Curti} {et~al.}(2017){Curti}, {Cresci}, {Mannucci}, {Marconi}, {Maiolino}, \& {Esposito}}]{Curti2017}
{Curti}, M., {Cresci}, G., {Mannucci}, F., {et~al.} 2017, \mnras, 465, 1384, \dodoi{10.1093/mnras/stw2766}

\bibitem[{{Curti} {et~al.}(2020){Curti}, {Mannucci}, {Cresci}, \& {Maiolino}}]{Curti2020}
{Curti}, M., {Mannucci}, F., {Cresci}, G., \& {Maiolino}, R. 2020, \mnras, 491, 944, \dodoi{10.1093/mnras/stz2910}

\bibitem[{{Curti} {et~al.}(2023){Curti}, {D'Eugenio}, {Carniani}, {Maiolino}, {Sandles}, {Witstok}, {Baker}, {Bennett}, {Piotrowska}, {Tacchella}, {Charlot}, {Nakajima}, {Maheson}, {Mannucci}, {Amiri}, {Arribas}, {Belfiore}, {Bonaventura}, {Bunker}, {Chevallard}, {Cresci}, {Curtis-Lake}, {Hayden-Pawson}, {Jones}, {Kumari}, {Laseter}, {Looser}, {Marconi}, {Maseda}, {Scholtz}, {Smit}, {{\"U}bler}, \& {Wallace}}]{Curti2023}
{Curti}, M., {D'Eugenio}, F., {Carniani}, S., {et~al.} 2023, \mnras, 518, 425, \dodoi{10.1093/mnras/stac2737}

\bibitem[{{Curti} {et~al.}(2024){Curti}, {Witstok}, {Jakobsen}, {Kobayashi}, {Curtis-Lake}, {Hainline}, {Ji}, {D'Eugenio}, {Chevallard}, {Maiolino}, {Scholtz}, {Carniani}, {Arribas}, {Baker}, {Bhatawdekar}, {Boyett}, {Bunker}, {Cameron}, {Cargile}, {Charlot}, {Eisenstein}, {Ji}, {Johnson}, {Kumari}, {Maseda}, {Robertson}, {Silcock}, {Tacchella}, {Ubler}, {Venturi}, {Williams}, {Willmer}, \& {Willott}}]{Curti2024}
{Curti}, M., {Witstok}, J., {Jakobsen}, P., {et~al.} 2024, arXiv e-prints, arXiv:2407.02575, \dodoi{10.48550/arXiv.2407.02575}

\bibitem[{{de Graaff} {et~al.}(2024){de Graaff}, {Rix}, {Carniani}, {Suess}, {Charlot}, {Curtis-Lake}, {Arribas}, {Baker}, {Boyett}, {Bunker}, {Cameron}, {Chevallard}, {Curti}, {Eisenstein}, {Franx}, {Hainline}, {Hausen}, {Ji}, {Johnson}, {Jones}, {Maiolino}, {Maseda}, {Nelson}, {Parlanti}, {Rawle}, {Robertson}, {Tacchella}, {{\"U}bler}, {Williams}, {Willmer}, \& {Willott}}]{de_Graaff2024}
{de Graaff}, A., {Rix}, H.-W., {Carniani}, S., {et~al.} 2024, \aap, 684, A87, \dodoi{10.1051/0004-6361/202347755}

\bibitem[{{de Graaff} {et~al.}(2025){de Graaff}, {Brammer}, {Weibel}, {Lewis}, {Maseda}, {Oesch}, {Bezanson}, {Boogaard}, {Cleri}, {Cooper}, {Gottumukkala}, {Greene}, {Hirschmann}, {Hviding}, {Katz}, {Labb{\'e}}, {Leja}, {Matthee}, {McConachie}, {Miller}, {Naidu}, {Price}, {Rix}, {Setton}, {Suess}, {Wang}, {Whitaker}, \& {Williams}}]{de_Graaff2025}
{de Graaff}, A., {Brammer}, G., {Weibel}, A., {et~al.} 2025, \aap, 697, A189, \dodoi{10.1051/0004-6361/202452186}

\bibitem[{{De Rosa} {et~al.}(2011){De Rosa}, {Decarli}, {Walter}, {Fan}, {Jiang}, {Kurk}, {Pasquali}, \& {Rix}}]{DeRosa2011}
{De Rosa}, G., {Decarli}, R., {Walter}, F., {et~al.} 2011, \apj, 739, 56, \dodoi{10.1088/0004-637X/739/2/56}

\bibitem[{{Dean} \& {Bruhweiler}(1985)}]{Dean&Bruhweiler1985}
{Dean}, C.~A., \& {Bruhweiler}, F.~C. 1985, \apjs, 57, 133, \dodoi{10.1086/190998}

\bibitem[{{D'Eugenio} {et~al.}(2024{\natexlab{a}}){D'Eugenio}, {Maiolino}, {Carniani}, {Chevallard}, {Curtis-Lake}, {Witstok}, {Charlot}, {Baker}, {Arribas}, {Boyett}, {Bunker}, {Curti}, {Eisenstein}, {Hainline}, {Ji}, {Johnson}, {Kumari}, {Looser}, {Nakajima}, {Nelson}, {Rieke}, {Robertson}, {Scholtz}, {Smit}, {Sun}, {Venturi}, {Tacchella}, {{\"U}bler}, {Willmer}, \& {Willott}}]{D'Eugenio2024a}
{D'Eugenio}, F., {Maiolino}, R., {Carniani}, S., {et~al.} 2024{\natexlab{a}}, \aap, 689, A152, \dodoi{10.1051/0004-6361/202348636}

\bibitem[{{D'Eugenio} {et~al.}(2024{\natexlab{b}}){D'Eugenio}, {Cameron}, {Scholtz}, {Carniani}, {Willott}, {Curtis-Lake}, {Bunker}, {Parlanti}, {Maiolino}, {Willmer}, {Jakobsen}, {Robertson}, {Johnson}, {Tacchella}, {Cargile}, {Rawle}, {Arribas}, {Chevallard}, {Curti}, {Egami}, {Eisenstein}, {Kumari}, {Looser}, {Rieke}, {Rodr{\'\i}guez Del Pino}, {Saxena}, {{\"U}bler}, {Venturi}, {Witstok}, {Baker}, {Bhatawdekar}, {Bonaventura}, {Boyett}, {Charlot}, {Danhaive}, {Hainline}, {Hausen}, {Helton}, {Ji}, {Ji}, {Jones}, {Joud{\v{z}}balis}, {Maseda}, {P{\'e}rez-Gonz{\'a}lez}, {Perna}, {Pusk{\'a}s}, {Shivaei}, {Silcock}, {Simmonds}, {Smit}, {Sun}, {Villanueva}, {Williams}, \& {Zhu}}]{D'Eugenio2024}
{D'Eugenio}, F., {Cameron}, A.~J., {Scholtz}, J., {et~al.} 2024{\natexlab{b}}, arXiv e-prints, arXiv:2404.06531, \dodoi{10.48550/arXiv.2404.06531}

\bibitem[{{Dietrich} {et~al.}(2003){Dietrich}, {Hamann}, {Appenzeller}, \& {Vestergaard}}]{Dietrich2003}
{Dietrich}, M., {Hamann}, F., {Appenzeller}, I., \& {Vestergaard}, M. 2003, \apj, 596, 817, \dodoi{10.1086/378045}

\bibitem[{{Dietrich} {et~al.}(2002){Dietrich}, {Hamann}, {Shields}, {Constantin}, {Vestergaard}, {Chaffee}, {Foltz}, \& {Junkkarinen}}]{Dietrich2002}
{Dietrich}, M., {Hamann}, F., {Shields}, J.~C., {et~al.} 2002, \apj, 581, 912, \dodoi{10.1086/344410}

\bibitem[{{Dong} {et~al.}(2011){Dong}, {Wang}, {Ho}, {Wang}, {Fan}, {Wang}, {Zhou}, \& {Yuan}}]{Dong2011}
{Dong}, X.-B., {Wang}, J.-G., {Ho}, L.~C., {et~al.} 2011, \apj, 736, 86, \dodoi{10.1088/0004-637X/736/2/86}

\bibitem[{{Ebinger} {et~al.}(2020){Ebinger}, {Curtis}, {Ghosh}, {Fr{\"o}hlich}, {Hempel}, {Perego}, {Liebend{\"o}rfer}, \& {Thielemann}}]{Ebinger2020}
{Ebinger}, K., {Curtis}, S., {Ghosh}, S., {et~al.} 2020, \apj, 888, 91, \dodoi{10.3847/1538-4357/ab5dcb}

\bibitem[{{Eisenstein} {et~al.}(2023{\natexlab{a}}){Eisenstein}, {Willott}, {Alberts}, {Arribas}, {Bonaventura}, {Bunker}, {Cameron}, {Carniani}, {Charlot}, {Curtis-Lake}, {D'Eugenio}, {Endsley}, {Ferruit}, {Giardino}, {Hainline}, {Hausen}, {Jakobsen}, {Johnson}, {Maiolino}, {Rieke}, {Rieke}, {Rix}, {Robertson}, {Stark}, {Tacchella}, {Williams}, {Willmer}, {Baker}, {Baum}, {Bhatawdekar}, {Boyett}, {Chen}, {Chevallard}, {Circosta}, {Curti}, {Danhaive}, {DeCoursey}, {de Graaff}, {Dressler}, {Egami}, {Helton}, {Hviding}, {Ji}, {Jones}, {Kumari}, {L{\"u}tzgendorf}, {Laseter}, {Looser}, {Lyu}, {Maseda}, {Nelson}, {Parlanti}, {Perna}, {Pusk{\'a}s}, {Rawle}, {Rodr{\'\i}guez Del Pino}, {Sandles}, {Saxena}, {Scholtz}, {Sharpe}, {Shivaei}, {Silcock}, {Simmonds}, {Skarbinski}, {Smit}, {Stone}, {Suess}, {Sun}, {Tang}, {Topping}, {{\"U}bler}, {Villanueva}, {Wallace}, {Whitler}, {Witstok}, \& {Woodrum}}]{Eisenstein2023a}
{Eisenstein}, D.~J., {Willott}, C., {Alberts}, S., {et~al.} 2023{\natexlab{a}}, arXiv e-prints, arXiv:2306.02465, \dodoi{10.48550/arXiv.2306.02465}

\bibitem[{{Eisenstein} {et~al.}(2023{\natexlab{b}}){Eisenstein}, {Johnson}, {Robertson}, {Tacchella}, {Hainline}, {Jakobsen}, {Maiolino}, {Bonaventura}, {Bunker}, {Cameron}, {Cargile}, {Curtis-Lake}, {Hausen}, {Pusk{\'a}s}, {Rieke}, {Sun}, {Willmer}, {Willott}, {Alberts}, {Arribas}, {Baker}, {Baum}, {Bhatawdekar}, {Carniani}, {Charlot}, {Chen}, {Chevallard}, {Curti}, {DeCoursey}, {D'Eugenio}, {de Graaff}, {Egami}, {Helton}, {Ji}, {Jones}, {Kumari}, {L{\"u}tzgendorf}, {Laseter}, {Looser}, {Lyu}, {Maseda}, {Nelson}, {Parlanti}, {Rauscher}, {Rawle}, {Rieke}, {Rix}, {Rujopakarn}, {Sandles}, {Saxena}, {Scholtz}, {Sharpe}, {Shivaei}, {Simmonds}, {Smit}, {Topping}, {{\"U}bler}, {Venturi}, {Williams}, {Witstok}, \& {Woodrum}}]{Eisenstein2023b}
{Eisenstein}, D.~J., {Johnson}, B.~D., {Robertson}, B., {et~al.} 2023{\natexlab{b}}, arXiv e-prints, arXiv:2310.12340, \dodoi{10.48550/arXiv.2310.12340}

\bibitem[{{Eldridge} {et~al.}(2017){Eldridge}, {Stanway}, {Xiao}, {McClelland}, {Taylor}, {Ng}, {Greis}, \& {Bray}}]{Eldridge2017}
{Eldridge}, J.~J., {Stanway}, E.~R., {Xiao}, L., {et~al.} 2017, \pasa, 34, e058, \dodoi{10.1017/pasa.2017.51}

\bibitem[{{Ferland} {et~al.}(1998){Ferland}, {Korista}, {Verner}, {Ferguson}, {Kingdon}, \& {Verner}}]{Ferland1998}
{Ferland}, G.~J., {Korista}, K.~T., {Verner}, D.~A., {et~al.} 1998, \pasp, 110, 761, \dodoi{10.1086/316190}

\bibitem[{{Finkelstein} {et~al.}(2023){Finkelstein}, {Bagley}, {Ferguson}, {Wilkins}, {Kartaltepe}, {Papovich}, {Yung}, {Haro}, {Behroozi}, {Dickinson}, {Kocevski}, {Koekemoer}, {Larson}, {Le Bail}, {Morales}, {P{\'e}rez-Gonz{\'a}lez}, {Burgarella}, {Dav{\'e}}, {Hirschmann}, {Somerville}, {Wuyts}, {Bromm}, {Casey}, {Fontana}, {Fujimoto}, {Gardner}, {Giavalisco}, {Grazian}, {Grogin}, {Hathi}, {Hutchison}, {Jha}, {Jogee}, {Kewley}, {Kirkpatrick}, {Long}, {Lotz}, {Pentericci}, {Pierel}, {Pirzkal}, {Ravindranath}, {Ryan}, {Trump}, {Yang}, {Bhatawdekar}, {Bisigello}, {Buat}, {Calabr{\`o}}, {Castellano}, {Cleri}, {Cooper}, {Croton}, {Daddi}, {Dekel}, {Elbaz}, {Franco}, {Gawiser}, {Holwerda}, {Huertas-Company}, {Jaskot}, {Leung}, {Lucas}, {Mobasher}, {Pandya}, {Tacchella}, {Weiner}, \& {Zavala}}]{Finkelstein2023}
{Finkelstein}, S.~L., {Bagley}, M.~B., {Ferguson}, H.~C., {et~al.} 2023, \apjl, 946, L13, \dodoi{10.3847/2041-8213/acade4}

\bibitem[{{Foreman-Mackey} {et~al.}(2013){Foreman-Mackey}, {Hogg}, {Lang}, \& {Goodman}}]{Foreman2013}
{Foreman-Mackey}, D., {Hogg}, D.~W., {Lang}, D., \& {Goodman}, J. 2013, \pasp, 125, 306, \dodoi{10.1086/670067}

\bibitem[{{Fujimoto} {et~al.}(2023){Fujimoto}, {Arrabal Haro}, {Dickinson}, {Finkelstein}, {Kartaltepe}, {Larson}, {Burgarella}, {Bagley}, {Behroozi}, {Chworowsky}, {Hirschmann}, {Trump}, {Wilkins}, {Yung}, {Koekemoer}, {Papovich}, {Pirzkal}, {Ferguson}, {Fontana}, {Grogin}, {Grazian}, {Kewley}, {Kocevski}, {Lotz}, {Pentericci}, {Ravindranath}, {Somerville}, {Wilkins}, {Amor{\'\i}n}, {Backhaus}, {Calabr{\`o}}, {Casey}, {Cooper}, {Fern{\'a}ndez}, {Franco}, {Giavalisco}, {Hathi}, {Harish}, {Hutchison}, {Iyer}, {Jung}, {Lucas}, \& {Zavala}}]{Fujimoto2023}
{Fujimoto}, S., {Arrabal Haro}, P., {Dickinson}, M., {et~al.} 2023, \apjl, 949, L25, \dodoi{10.3847/2041-8213/acd2d9}

\bibitem[{{Gardner} {et~al.}(2023){Gardner}, {Mather}, {Abbott}, {Abell}, {Abernathy}, {Abney}, {Abraham}, {Abraham}, {Abul-Huda}, {Acton}, {Adams}, {Adams}, {Adler}, {Adriaensen}, {Aguilar}, {Ahmed}, {Ahmed}, {Ahmed}, {Albat}, {Albert}, {Alberts}, {Aldridge}, {Allen}, {Allen}, {Altenburg}, {Altunc}, {Alvarez}, {{\'A}lvarez-M{\'a}rquez}, {Alves de Oliveira}, {Ambrose}, {Anandakrishnan}, {Andersen}, {Anderson}, {Anderson}, {Anderson}, {Anderson}, {Aprea}, {Archer}, {Arenberg}, {Argyriou}, {Arribas}, {Artigau}, {Arvai}, {Atcheson}, {Atkinson}, {Averbukh}, {Aymergen}, {Bacinski}, {Baggett}, {Bagnasco}, {Baker}, {Balzano}, {Banks}, {Baran}, {Barker}, {Barrett}, {Barringer}, {Barto}, {Bast}, {Baudoz}, {Baum}, {Beatty}, {Beaulieu}, {Bechtold}, {Beck}, {Beddard}, {Beichman}, {Bellagama}, {Bely}, {Berger}, {Bergeron}, {Bernier}, {Bertch}, {Beskow}, {Betz}, {Biagetti}, {Birkmann}, {Bjorklund}, {Blackwood}, {Blazek}, {Blossfeld}, {Bluth}, {Boccaletti}, {Boegner}, {Bohlin}, {Boia}, {B{\"o}ker}, {Bonaventura}, {Bond},
  {Bosley}, {Boucarut}, {Bouchet}, {Bouwman}, {Bower}, {Bowers}, {Bowers}, {Boyce}, {Boyer}, {Boyer}, {Boyer}, {Boyer}, {Bradley}, {Brady}, {Brandl}, {Brannen}, {Breda}, {Bremmer}, {Brennan}, {Bresnahan}, {Bright}, {Broiles}, {Bromenschenkel}, {Brooks}, {Brooks}, {Brown}, {Brown}, {Brown}, {Bruce}, {Bryson}, {Bujanda}, {Bullock}, {Bunker}, {Bureo}, {Burt}, {Bush}, {Bushouse}, {Bussman}, {Cabaud}, {Cale}, {Calhoon}, {Calvani}, {Canipe}, {Caputo}, {Cara}, {Carey}, {Case}, {Cesari}, {Cetorelli}, {Chance}, {Chandler}, {Chaney}, {Chapman}, {Charlot}, {Chayer}, {Cheezum}, {Chen}, {Chen}, {Cherinka}, {Chichester}, {Chilton}, {Chittiraibalan}, {Clampin}, {Clark}, {Clark}, {Clark}, {Claybrooks}, {Cleveland}, {Cohen}, {Cohen}, {Col{\'o}n}, {Coleman}, {Colina}, {Comber}, {Comeau}, {Comer}, {Conde Reis}, {Connolly}, {Conroy}, {Contos}, {Contreras}, {Cook}, {Cooper}, {Cooper}, {Correia}, {Correnti}, {Cossou}, {Costanza}, {Coulais}, {Cox}, {Coyle}, {Cracraft}, {Crew}, {Curtis}, {Cusveller}, {Da Costa Maciel}, {Dailey},
  {Daugeron}, {Davidson}, {Davies}, {Davis}, {Davis}, {Day}, {de Chambure}, {de Jong}, {De Marchi}, {Dean}, {Decker}, {Delisa}, {Dell}, {Dellagatta}, {Dembinska}, {Demosthenes}, {Dencheva}, {Deneu}, {DePriest}, {Deschenes}, {Dethienne}, {Detre}, {Diaz}, {Dicken}, {DiFelice}, {Dillman}, {Disharoon}, {Dixon}, {Doggett}, {Dominguez}, {Donaldson}, {Doria-Warner}, {Santos}, {Doty}, {Douglas}, {Doyon}, {Dressler}, {Driggers}, {Driggers}, {Dunn}, {DuPrie}, {Dupuis}, {Durning}, {Dutta}, {Earl}, {Eccleston}, {Ecobichon}, {Egami}, {Ehrenwinkler}, {Eisenhamer}, {Eisenhower}, {Eisenstein}, {El Hamel}, {Elie}, {Elliott}, {Elliott}, {Engesser}, {Espinoza}, {Etienne}, {Etxaluze}, {Evans}, {Fabreguettes}, {Falcolini}, {Falini}, {Fatig}, {Feeney}, {Feinberg}, {Fels}, {Ferdous}, {Ferguson}, {Ferrarese}, {Ferreira}, {Ferruit}, {Ferry}, {Filippazzo}, {Firre}, {Fix}, {Flagey}, {Flanagan}, {Fleming}, {Florian}, {Flynn}, {Foiadelli}, {Fontaine}, {Fontanella}, {Forshay}, {Fortner}, {Fox}, {Framarini}, {Francisco}, {Franck}, {Franx},
  {Franz}, {Friedman}, {Friend}, {Frost}, {Fu}, {Fullerton}, {Gaillard}, {Galkin}, {Gallagher}, {Galyer}, {Garc{\'\i}a Mar{\'\i}n}, {Gardner}, {Garland}, {Garrett}, {Gasman}, {G{\'a}sp{\'a}r}, {Gastaud}, {Gaudreau}, {Gauthier}, {Geers}, {Geithner}, {Gennaro}, {Gerber}, {Gereau}, {Giampaoli}, {Giardino}, {Gibbons}, {Gilbert}, {Gilman}, {Girard}, {Giuliano}, {Gkountis}, {Glasse}, {Glassmire}, {Glauser}, {Glazer}, {Goldberg}, {Golimowski}, {Gonzaga}, {Gordon}, {Gordon}, {Goudfrooij}, {Gough}, {Graham}, {Grau}, {Green}, {Greene}, {Greene}, {Greenfield}, {Greenhouse}, {Greve}, {Greville}, {Grimaldi}, {Groe}, {Groebner}, {Grumm}, {Grundy}, {G{\"u}del}, {Guillard}, {Guldalian}, {Gunn}, {Gurule}, {Gutman}, {Guy}, {Guyot}, {Hack}, {Haderlein}, {Hagan}, {Hagedorn}, {Hainline}, {Haley}, {Hami}, {Hamilton}, {Hammann}, {Hammel}, {Hanley}, {Hansen}, {Hardy}, {Harnisch}, {Harr}, {Harris}, {Hart}, {Hartig}, {Hasan}, {Hashim}, {Hashimoto}, {Haskins}, {Hawkins}, {Hayden}, {Hayden}, {Healy}, {Hecht}, {Heeg}, {Hejal}, {Helm},
  {Hengemihle}, {Henning}, {Henry}, {Henry}, {Henshaw}, {Hernandez}, {Herrington}, {Heske}, {Hesman}, {Hickey}, {Hilbert}, {Hines}, {Hinz}, {Hirsch}, {Hitcho}, {Hodapp}, {Hodge}, {Hoffman}, {Holfeltz}, {Holler}, {Hoppa}, {Horner}, {Howard}, {Howard}, {Huber}, {Hunkeler}, {Hunter}, {Hunter}, {Hurd}, {Hurst}, {Hutchings}, {Hylan}, {Ignat}, {Illingworth}, {Irish}, {Isaacs}, {Jackson}, {Jaffe}, {Jahic}, {Jahromi}, {Jakobsen}, {James}, {James}, {James}, {Jamieson}, {Jandra}, {Jayawardhana}, {Jedrzejewski}, {Jeffers}, {Jensen}, {Joanne}, {Johns}, {Johnson}, {Johnson}, {Johnson}, {Johnson}, {Johnson}, {Johnson}, {Johnstone}, {Jollet}, {Jones}, {Jones}, {Jones}, {Jones}, {Jones}, {Jordan}, {Jordan}, {Jue}, {Jurkowski}, {Justis}, {Justtanont}, {Kaleida}, {Kalirai}, {Kalmanson}, {Kaltenegger}, {Kammerer}, {Kan}, {Kanarek}, {Kao}, {Karakla}, {Karl}, {Kassin}, {Kauffman}, {Kavanagh}, {Kelley}, {Kelly}, {Kendrew}, {Kennedy}, {Kenny}, {Keski-Kuha}, {Keyes}, {Khan}, {Kidwell}, {Kimble}, {King}, {King}, {Kinzel}, {Kirk},
  {Kirkpatrick}, {Klaassen}, {Klingemann}, {Klintworth}, {Knapp}, {Knight}, {Knollenberg}, {Knutsen}, {Koehler}, {Koekemoer}, {Kofler}, {Kontson}, {Kovacs}, {Kozhurina-Platais}, {Krause}, {Kriss}, {Krist}, {Kristoffersen}, {Krogel}, {Krueger}, {Kulp}, {Kumari}, {Kwan}, {Kyprianou}, {Labador}, {Labiano}, {Lafreni{\`e}re}, {Lagage}, {Laidler}, {Laine}, {Laird}, {Lajoie}, {Lallo}, {Lam}, {LaMassa}, {Lambros}, {Lampenfield}, {Lander}, {Langston}, {Larson}, {Larson}, {LaVerghetta}, {Law}, {Lawrence}, {Lee}, {Lee}, {Lee}, {Leisenring}, {Leveille}, {Levenson}, {Levi}, {Levine}, {Lewis}, {Lewis}, {Lewis}, {Libralato}, {Lidon}, {Liebrecht}, {Lightsey}, {Lilly}, {Lim}, {Lim}, {Ling}, {Link}, {Link}, {Lipinski}, {Liu}, {Lo}, {Lobmeyer}, {Logue}, {Long}, {Long}, {Long}, {Long}, {L{\'o}pez-Caniego}, {Lotz}, {Love-Pruitt}, {Lubskiy}, {Luers}, {Luetgens}, {Luevano}, {Lui}, {Lund}, {Lundquist}, {Lunine}, {L{\"u}tzgendorf}, {Lynch}, {MacDonald}, {MacDonald}, {Macias}, {Macklis}, {Maghami}, {Maharaja}, {Maiolino},
  {Makrygiannis}, {Malla}, {Malumuth}, {Manjavacas}, {Marini}, {Marrione}, {Marston}, {Martel}, {Martin}, {Martin}, {Martinez}, {Maschmann}, {Masci}, {Masetti}, {Maszkiewicz}, {Matthews}, {Matuskey}, {McBrayer}, {McCarthy}, {McCaughrean}, {McClare}, {McClare}, {McCloskey}, {McClurg}, {McCoy}, {McElwain}, {McGregor}, {McGuffey}, {McKay}, {McKenzie}, {McLean}, {McMaster}, {McNeil}, {De Meester}, {Mehalick}, {Meixner}, {Mel{\'e}ndez}, {Menzel}, {Menzel}, {Merz}, {Mesterharm}, {Meyer}, {Meyett}, {Meza}, {Midwinter}, {Milam}, {Miller}, {Miller}, {Miskey}, {Misselt}, {Mitchell}, {Mohan}, {Montoya}, {Moran}, {Morishita}, {Moro-Mart{\'\i}n}, {Morrison}, {Morrison}, {Morse}, {Moschos}, {Moseley}, {Mosier}, {Mosner}, {Mountain}, {Muckenthaler}, {Mueller}, {Mueller}, {Muhiem}, {M{\"u}hlmann}, {Mullally}, {Mullen}, {Munger}, {Murphy}, {Murray}, {Muzerolle}, {Mycroft}, {Myers}, {Myers}, {Myers}, {Myers}, {Myrick}, {Nagle}, {Nayak}, {Naylor}, {Neff}, {Nelan}, {Nella}, {Nguyen}, {Nguyen}, {Nickson}, {Nidhiry}, {Niedner},
  {Nieto-Santisteban}, {Nikolov}, {Nishisaka}, {Noriega-Crespo}, {Nota}, {O'Mara}, {Oboryshko}, {O'Brien}, {Ochs}, {Offenberg}, {Ogle}, {Ohl}, {Olmsted}, {Osborne}, {O'Shaughnessy}, {{\"O}stlin}, {O'Sullivan}, {Otor}, {Ottens}, {Ouellette}, {Outlaw}, {Owens}, {Pacifici}, {Page}, {Paranilam}, {Park}, {Parrish}, {Paschal}, {Patapis}, {Patel}, {Patrick}, {Pattishall}, {Paul}, {Paul}, {Pauly}, {Pavlovsky}, {Pe{\~n}a-Guerrero}, {Pedder}, {Peek}, {Pelham}, {Penanen}, {Perriello}, {Perrin}, {Perrine}, {Perrygo}, {Peslier}, {Petach}, {Peterson}, {Pfarr}, {Pierson}, {Pietraszkiewicz}, {Pilchen}, {Pipher}, {Pirzkal}, {Pitman}, {Player}, {Plesha}, {Plitzke}, {Pohner}, {Poletis}, {Pollizzi}, {Polster}, {Pontius}, {Pontoppidan}, {Porges}, {Potter}, {Prescott}, {Proffitt}, {Pueyo}, {Quispe Neira}, {Radich}, {Rager}, {Rameau}, {Ramey}, {Ramos Alarcon}, {Rampini}, {Rapp}, {Rashford}, {Rauscher}, {Ravindranath}, {Rawle}, {Rawlings}, {Ray}, {Regan}, {Rehm}, {Rehm}, {Reid}, {Reis}, {Renk}, {Reoch}, {Ressler}, {Rest},
  {Reynolds}, {Richon}, {Richon}, {Ridgaway}, {Riedel}, {Rieke}, {Rieke}, {Rifelli}, {Rigby}, {Riggs}, {Ringel}, {Ritchie}, {Rix}, {Robberto}, {Robinson}, {Robinson}, {Robinson}, {Rock}, {Rodriguez}, {Rodr{\'\i}guez del Pino}, {Roellig}, {Rohrbach}, {Roman}, {Romelfanger}, {Romo}, {Rosales}, {Rose}, {Roteliuk}, {Roth}, {Rothwell}, {Rouzaud}, {Rowe}, {Rowlands}, {Roy}, {Royer}, {Rui}, {Rumler}, {Rumpl}, {Russ}, {Ryan}, {Ryan}, {Saad}, {Sabata}, {Sabatino}, {Sabbi}, {Sabelhaus}, {Sabia}, {Sahu}, {Saif}, {Salvignol}, {Samara-Ratna}, {Samuelson}, {Sanders}, {Sappington}, {Sargent}, {Sauer}, {Savadkin}, {Sawicki}, {Schappell}, {Scheffer}, {Scheithauer}, {Scherer}, {Schiff}, {Schlawin}, {Schmeitzky}, {Schmitz}, {Schmude}, {Schneider}, {Schreiber}, {Schroeven-Deceuninck}, {Schultz}, {Schwab}, {Schwartz}, {Scoccimarro}, {Scott}, {Scott}, {Seaton}, {Seely}, {Seery}, {Seidleck}, {Sembach}, {Shanahan}, {Shaughnessy}, {Shaw}, {Shay}, {Sheehan}, {Sheth}, {Shih}, {Shivaei}, {Siegel}, {Sienkiewicz}, {Simmons}, {Simon},
  {Sirianni}, {Sivaramakrishnan}, {Slade}, {Sloan}, {Slocum}, {Slowinski}, {Smith}, {Smith}, {Smith}, {Smith}, {Smith}, {Smith}, {Smolik}, {Soderblom}, {Sohn}, {Sokol}, {Sonneborn}, {Sontag}, {Sooy}, {Soummer}, {Southwood}, {Spain}, {Sparmo}, {Speer}, {Spencer}, {Sprofera}, {Stallcup}, {Stanley}, {Stansberry}, {Stark}, {Starr}, {Stassi}, {Steck}, {Steeley}, {Stephens}, {Stephenson}, {Stewart}, {Stiavelli}, {}, {Strada}, {Straughn}, {Streetman}, {Strickland}, {Strobele}, {Stuhlinger}, {Stys}, {Such}, {Sukhatme}, {Sullivan}, {Sullivan}, {Sumner}, {Sun}, {Sunnquist}, {Swade}, {Swam}, {Swenton}, {Swoish}, {Tam Litten}, {Tamas}, {Tao}, {Taylor}, {Taylor}, {te Plate}, {Van Tea}, {Teague}, {Telfer}, {Temim}, {Texter}, {Thatte}, {Thompson}, {Thompson}, {Thomson}, {Thronson}, {Tierney}, {Tikkanen}, {Tinnin}, {Tippet}, {Todd}, {Tran}, {Trauger}, {Trejo}, {Vinh Truong}, {Tsukamoto}, {Tufail}, {Tumlinson}, {Tustain}, {Tyra}, {Ubeda}, {Underwood}, {Uzzo}, {Vaclavik}, {Valenduc}, {Valenti}, {Van Campen}, {van de Wetering},
  {Van Der Marel}, {van Haarlem}, {Vandenbussche}, {van Dishoeck}, {Vanterpool}, {Vernoy}, {Vila Costas}, {Volk}, {Voorzaat}, {Voyton}, {Vydra}, {Waddy}, {Waelkens}, {Wahlgren}, {Walker}, {Wander}, {Warfield}, {Warner}, {Wasiak}, {Wasiak}, {Wehner}, {Weiler}, {Weilert}, {Weiss}, {Wells}, {Welty}, {Wheate}, {Wheeler}, {White}, {Whitehouse}, {Whiteleather}, {Whitman}, {Williams}, {Willmer}, {Willott}, {Willoughby}, {Wilson}, {Wilson}, {Wilson}, {Windhorst}, {Wislowski}, {Wolfe}, {Wolfe}, {Wolff}, {Wondel}, {Woo}, {Woods}, {Worden}, {Workman}, {Wright}, {Wu}, {Wu}, {Wun}, {Wymer}, {Yadetie}, {Yan}, {Yang}, {Yates}, {Yeager}, {Yerger}, {Young}, {Young}, {Yu}, {Yu}, {Zak}, {Zeidler}, {Zepp}, {Zhou}, {Zincke}, {Zonak}, \& {Zondag}}]{Gardner2023}
{Gardner}, J.~P., {Mather}, J.~C., {Abbott}, R., {et~al.} 2023, \pasp, 135, 068001, \dodoi{10.1088/1538-3873/acd1b5}

\bibitem[{{Goulding} {et~al.}(2023){Goulding}, {Greene}, {Setton}, {Labbe}, {Bezanson}, {Miller}, {Atek}, {Bogd{\'a}n}, {Brammer}, {Chemerynska}, {Cutler}, {Dayal}, {Fudamoto}, {Fujimoto}, {Furtak}, {Kokorev}, {Khullar}, {Leja}, {Marchesini}, {Natarajan}, {Nelson}, {Oesch}, {Pan}, {Papovich}, {Price}, {van Dokkum}, {Wang}, {Weaver}, {Whitaker}, \& {Zitrin}}]{Goulding2023}
{Goulding}, A.~D., {Greene}, J.~E., {Setton}, D.~J., {et~al.} 2023, \apjl, 955, L24, \dodoi{10.3847/2041-8213/acf7c5}

\bibitem[{{Grandi}(1982)}]{Grandi1982}
{Grandi}, S.~A. 1982, \apj, 255, 25, \dodoi{10.1086/159799}

\bibitem[{{Gunasekera} {et~al.}(2023){Gunasekera}, {van Hoof}, {Chatzikos}, \& {Ferland}}]{Gunasekera2023}
{Gunasekera}, C.~M., {van Hoof}, P. A.~M., {Chatzikos}, M., \& {Ferland}, G.~J. 2023, Research Notes of the American Astronomical Society, 7, 246, \dodoi{10.3847/2515-5172/ad0e75}

\bibitem[{{Hachisu} {et~al.}(1996){Hachisu}, {Kato}, \& {Nomoto}}]{Hachisu1996}
{Hachisu}, I., {Kato}, M., \& {Nomoto}, K. 1996, \apjl, 470, L97, \dodoi{10.1086/310303}

\bibitem[{{Hachisu} {et~al.}(1999){Hachisu}, {Kato}, \& {Nomoto}}]{Hachisu1999}
---. 1999, \apj, 522, 487, \dodoi{10.1086/307608}

\bibitem[{{Hainline} {et~al.}(2024){Hainline}, {D'Eugenio}, {Jakobsen}, {Chevallard}, {Carniani}, {Witstok}, {Ji}, {Curtis-Lake}, {Johnson}, {Robertson}, {Tacchella}, {Curti}, {Charlot}, {Helton}, {Arribas}, {Bhatawdekar}, {Bunker}, {Cameron}, {Egami}, {Eisenstein}, {Hausen}, {Kumari}, {Maiolino}, {Perez-Gonzalez}, {Rieke}, {Saxena}, {Scholtz}, {Smit}, {Sun}, {Williams}, {Willmer}, \& {Willott}}]{Hainline2024}
{Hainline}, K.~N., {D'Eugenio}, F., {Jakobsen}, P., {et~al.} 2024, arXiv e-prints, arXiv:2404.04325, \dodoi{10.48550/arXiv.2404.04325}

\bibitem[{{Harikane} {et~al.}(2020){Harikane}, {Laporte}, {Ellis}, \& {Matsuoka}}]{Harikane2020}
{Harikane}, Y., {Laporte}, N., {Ellis}, R.~S., \& {Matsuoka}, Y. 2020, \apj, 902, 117, \dodoi{10.3847/1538-4357/abb597}

\bibitem[{{Harikane} {et~al.}(2024){Harikane}, {Nakajima}, {Ouchi}, {Umeda}, {Isobe}, {Ono}, {Xu}, \& {Zhang}}]{Harikane2024}
{Harikane}, Y., {Nakajima}, K., {Ouchi}, M., {et~al.} 2024, \apj, 960, 56, \dodoi{10.3847/1538-4357/ad0b7e}

\bibitem[{Harris {et~al.}(2020)Harris, Millman, van~der Walt, Gommers, Virtanen, Cournapeau, Wieser, Taylor, Berg, Smith, Kern, Picus, Hoyer, van Kerkwijk, Brett, Haldane, del R{\'{i}}o, Wiebe, Peterson, G{\'{e}}rard-Marchant, Sheppard, Reddy, Weckesser, Abbasi, Gohlke, \& Oliphant}]{Harris2020}
Harris, C.~R., Millman, K.~J., van~der Walt, S.~J., {et~al.} 2020, Nature, 585, 357, \dodoi{10.1038/s41586-020-2649-2}

\bibitem[{{Heintz} {et~al.}(2025){Heintz}, {Brammer}, {Watson}, {Oesch}, {Keating}, {Hayes}, {Abdurro'uf}, {Arellano-C{\'o}rdova}, {Carnall}, {Christiansen}, {Cullen}, {Dav{\'e}}, {Dayal}, {Ferrara}, {Finlator}, {Fynbo}, {Flury}, {Gelli}, {Gillman}, {Gottumukkala}, {Gould}, {Greve}, {Hardin}, {Hsiao}, {Hutter}, {Jakobsson}, {Killi}, {Khosravaninezhad}, {Laursen}, {Lee}, {Magdis}, {Matthee}, {Naidu}, {Narayanan}, {Pollock}, {Prescott}, {Rusakov}, {Shuntov}, {Sneppen}, {Smit}, {Tanvir}, {Terp}, {Toft}, {Valentino}, {Vijayan}, {Weaver}, {Wise}, \& {Witstok}}]{Heintz2025}
{Heintz}, K.~E., {Brammer}, G.~B., {Watson}, D., {et~al.} 2025, \aap, 693, A60, \dodoi{10.1051/0004-6361/202450243}

\bibitem[{{Hill} {et~al.}(2019){Hill}, {Sk{\'u}lad{\'o}ttir}, {Tolstoy}, {Venn}, {Shetrone}, {Jablonka}, {Primas}, {Battaglia}, {de Boer}, {Fran{\c{c}}ois}, {Helmi}, {Kaufer}, {Letarte}, {Starkenburg}, \& {Spite}}]{Hill2019}
{Hill}, V., {Sk{\'u}lad{\'o}ttir}, {\'A}., {Tolstoy}, E., {et~al.} 2019, \aap, 626, A15, \dodoi{10.1051/0004-6361/201833950}

\bibitem[{{Hirano} {et~al.}(2015){Hirano}, {Hosokawa}, {Yoshida}, {Omukai}, \& {Yorke}}]{Hirano2015}
{Hirano}, S., {Hosokawa}, T., {Yoshida}, N., {Omukai}, K., \& {Yorke}, H.~W. 2015, \mnras, 448, 568, \dodoi{10.1093/mnras/stv044}

\bibitem[{{Hirano} {et~al.}(2014){Hirano}, {Hosokawa}, {Yoshida}, {Umeda}, {Omukai}, {Chiaki}, \& {Yorke}}]{Hirano2014}
{Hirano}, S., {Hosokawa}, T., {Yoshida}, N., {et~al.} 2014, \apj, 781, 60, \dodoi{10.1088/0004-637X/781/2/60}

\bibitem[{{Hsiao} {et~al.}(2024{\natexlab{a}}){Hsiao}, {Abdurro'uf}, {Coe}, {Larson}, {Jung}, {Mingozzi}, {Dayal}, {Kumari}, {Kokorev}, {Vikaeus}, {Brammer}, {Furtak}, {Adamo}, {Andrade-Santos}, {Antwi-Danso}, {Brada{\v{c}}}, {Bradley}, {Broadhurst}, {Carnall}, {Conselice}, {Diego}, {Donahue}, {Eldridge}, {Fujimoto}, {Henry}, {Hernandez}, {Hutchison}, {James}, {Norman}, {Park}, {Pirzkal}, {Postman}, {Ricotti}, {Rigby}, {Vanzella}, {Welch}, {Wilkins}, {Windhorst}, {Xu}, {Zackrisson}, \& {Zitrin}}]{Hsiao2024a}
{Hsiao}, T. Y.-Y., {Abdurro'uf}, {Coe}, D., {et~al.} 2024{\natexlab{a}}, \apj, 973, 8, \dodoi{10.3847/1538-4357/ad5da8}

\bibitem[{{Hsiao} {et~al.}(2024{\natexlab{b}}){Hsiao}, {{\'A}lvarez-M{\'a}rquez}, {Coe}, {Crespo G{\'o}mez}, {Abdurro'uf}, {Dayal}, {Larson}, {Bik}, {Blanco-Prieto}, {Colina}, {P{\'e}rez-Gonz{\'a}lez}, {Costantin}, {Prieto-Jim{\'e}nez}, {Adamo}, {Bradley}, {Conselice}, {Fujimoto}, {Furtak}, {Hutchison}, {James}, {Jim{\'e}nez-Teja}, {Jung}, {Kokorev}, {Mingozzi}, {Norman}, {Ricotti}, {Rigby}, {Sharon}, {Vanzella}, {Welch}, {Xu}, {Zackrisson}, \& {Zitrin}}]{Hsiao2024b}
{Hsiao}, T. Y.-Y., {{\'A}lvarez-M{\'a}rquez}, J., {Coe}, D., {et~al.} 2024{\natexlab{b}}, \apj, 973, 81, \dodoi{10.3847/1538-4357/ad6562}

\bibitem[{Hunter(2007)}]{Hunter2007}
Hunter, J.~D. 2007, Computing in Science \& Engineering, 9, 90, \dodoi{10.1109/MCSE.2007.55}

\bibitem[{{Inoue} {et~al.}(2014){Inoue}, {Shimizu}, {Iwata}, \& {Tanaka}}]{Inoue2014}
{Inoue}, A.~K., {Shimizu}, I., {Iwata}, I., \& {Tanaka}, M. 2014, \mnras, 442, 1805, \dodoi{10.1093/mnras/stu936}

\bibitem[{{Isobe} {et~al.}(2023{\natexlab{a}}){Isobe}, {Ouchi}, {Nakajima}, {Harikane}, {Ono}, {Xu}, {Zhang}, \& {Umeda}}]{Isobe2023a}
{Isobe}, Y., {Ouchi}, M., {Nakajima}, K., {et~al.} 2023{\natexlab{a}}, \apj, 956, 139, \dodoi{10.3847/1538-4357/acf376}

\bibitem[{{Isobe} {et~al.}(2022){Isobe}, {Ouchi}, {Suzuki}, {Moriya}, {Nakajima}, {Nomoto}, {Rauch}, {Harikane}, {Kojima}, {Ono}, {Fujimoto}, {Inoue}, {Kim}, {Komiyama}, {Kusakabe}, {Lee}, {Maseda}, {Matthee}, {Michel-Dansac}, {Nagao}, {Nanayakkara}, {Nishigaki}, {Onodera}, {Sugahara}, \& {Xu}}]{Isobe2022}
{Isobe}, Y., {Ouchi}, M., {Suzuki}, A., {et~al.} 2022, \apj, 925, 111, \dodoi{10.3847/1538-4357/ac3509}

\bibitem[{{Isobe} {et~al.}(2023{\natexlab{b}}){Isobe}, {Ouchi}, {Tominaga}, {Watanabe}, {Nakajima}, {Umeda}, {Yajima}, {Harikane}, {Fukushima}, {Xu}, {Ono}, \& {Zhang}}]{Isobe2023b}
{Isobe}, Y., {Ouchi}, M., {Tominaga}, N., {et~al.} 2023{\natexlab{b}}, \apj, 959, 100, \dodoi{10.3847/1538-4357/ad09be}

\bibitem[{{Iwamoto} {et~al.}(1999){Iwamoto}, {Brachwitz}, {Nomoto}, {Kishimoto}, {Umeda}, {Hix}, \& {Thielemann}}]{Iwamoto1999}
{Iwamoto}, K., {Brachwitz}, F., {Nomoto}, K., {et~al.} 1999, \apjs, 125, 439, \dodoi{10.1086/313278}

\bibitem[{{Izotov} {et~al.}(2018){Izotov}, {Worseck}, {Schaerer}, {Guseva}, {Thuan}, {Fricke}, \& {Orlitov{\'a}}}]{Izotov2018}
{Izotov}, Y.~I., {Worseck}, G., {Schaerer}, D., {et~al.} 2018, \mnras, 478, 4851, \dodoi{10.1093/mnras/sty1378}

\bibitem[{{Jakobsen} {et~al.}(2022){Jakobsen}, {Ferruit}, {Alves de Oliveira}, {Arribas}, {Bagnasco}, {Barho}, {Beck}, {Birkmann}, {B{\"o}ker}, {Bunker}, {Charlot}, {de Jong}, {de Marchi}, {Ehrenwinkler}, {Falcolini}, {Fels}, {Franx}, {Franz}, {Funke}, {Giardino}, {Gnata}, {Holota}, {Honnen}, {Jensen}, {Jentsch}, {Johnson}, {Jollet}, {Karl}, {Kling}, {K{\"o}hler}, {Kolm}, {Kumari}, {Lander}, {Lemke}, {L{\'o}pez-Caniego}, {L{\"u}tzgendorf}, {Maiolino}, {Manjavacas}, {Marston}, {Maschmann}, {Maurer}, {Messerschmidt}, {Moseley}, {Mosner}, {Mott}, {Muzerolle}, {Pirzkal}, {Pittet}, {Plitzke}, {Posselt}, {Rapp}, {Rauscher}, {Rawle}, {Rix}, {R{\"o}del}, {Rumler}, {Sabbi}, {Salvignol}, {Schmid}, {Sirianni}, {Smith}, {Strada}, {te Plate}, {Valenti}, {Wettemann}, {Wiehe}, {Wiesmayer}, {Willott}, {Wright}, {Zeidler}, \& {Zincke}}]{Jakobsen2022}
{Jakobsen}, P., {Ferruit}, P., {Alves de Oliveira}, C., {et~al.} 2022, \aap, 661, A80, \dodoi{10.1051/0004-6361/202142663}

\bibitem[{{Ji} {et~al.}(2024{\natexlab{a}}){Ji}, {{\"U}bler}, {Maiolino}, {D'Eugenio}, {Arribas}, {Bunker}, {Charlot}, {Perna}, {Rodr{\'\i}guez Del Pino}, {B{\"o}ker}, {Cresci}, {Curti}, {Kumari}, \& {Lamperti}}]{Ji2024a}
{Ji}, X., {{\"U}bler}, H., {Maiolino}, R., {et~al.} 2024{\natexlab{a}}, \mnras, \dodoi{10.1093/mnras/stae2375}

\bibitem[{{Ji} {et~al.}(2024{\natexlab{b}}){Ji}, {Maiolino}, {Ferland}, {D'Eugenio}, {Bhatawdekar}, {Charlot}, {Chevallard}, {Curti}, {Curtis-Lake}, {Hainline}, {Ji}, {Robertson}, {Rodr{\'\i}guez Del Pino}, {Scholtz}, {Tacchella}, {Williams}, \& {Witstok}}]{Ji2024b}
{Ji}, X., {Maiolino}, R., {Ferland}, G., {et~al.} 2024{\natexlab{b}}, arXiv e-prints, arXiv:2405.05772, \dodoi{10.48550/arXiv.2405.05772}

\bibitem[{{Jones} {et~al.}(2024){Jones}, {Bunker}, {Saxena}, {Witstok}, {Stark}, {Arribas}, {Baker}, {Bhatawdekar}, {Bowler}, {Boyett}, {Cameron}, {Carniani}, {Charlot}, {Chevallard}, {Curti}, {Curtis-Lake}, {Eisenstein}, {Hainline}, {Hausen}, {Ji}, {Johnson}, {Kumari}, {Looser}, {Maiolino}, {Maseda}, {Parlanti}, {Rix}, {Robertson}, {Sandles}, {Scholtz}, {Smit}, {Tacchella}, {{\"U}bler}, {Williams}, \& {Willott}}]{Jones2024}
{Jones}, G.~C., {Bunker}, A.~J., {Saxena}, A., {et~al.} 2024, \aap, 683, A238, \dodoi{10.1051/0004-6361/202347099}

\bibitem[{{Jones} {et~al.}(2015){Jones}, {Martin}, \& {Cooper}}]{Jones2015}
{Jones}, T., {Martin}, C., \& {Cooper}, M.~C. 2015, \apj, 813, 126, \dodoi{10.1088/0004-637X/813/2/126}

\bibitem[{{Kageura} {et~al.}(2025){Kageura}, {Ouchi}, {Nakane}, {Umeda}, {Harikane}, {Yoshiura}, {Nakajima}, {Yajima}, \& {Thai}}]{Kageura2025}
{Kageura}, Y., {Ouchi}, M., {Nakane}, M., {et~al.} 2025, \apjs, 278, 33, \dodoi{10.3847/1538-4365/adc690}

\bibitem[{{Kashino} {et~al.}(2022){Kashino}, {Lilly}, {Renzini}, {Daddi}, {Zamorani}, {Silverman}, {Ilbert}, {Peng}, {Mainieri}, {Bardelli}, {Zucca}, {Kartaltepe}, \& {Sanders}}]{Kashino2022}
{Kashino}, D., {Lilly}, S.~J., {Renzini}, A., {et~al.} 2022, \apj, 925, 82, \dodoi{10.3847/1538-4357/ac399e}

\bibitem[{{Kobayashi} \& {Ferrara}(2024)}]{Kobayashi&Ferrara2024}
{Kobayashi}, C., \& {Ferrara}, A. 2024, \apjl, 962, L6, \dodoi{10.3847/2041-8213/ad1de1}

\bibitem[{{Kobayashi} \& {Nomoto}(2009)}]{Kobayashi2009}
{Kobayashi}, C., \& {Nomoto}, K. 2009, \apj, 707, 1466, \dodoi{10.1088/0004-637X/707/2/1466}

\bibitem[{{Kobayashi} {et~al.}(1998){Kobayashi}, {Tsujimoto}, {Nomoto}, {Hachisu}, \& {Kato}}]{Kobayashi1998}
{Kobayashi}, C., {Tsujimoto}, T., {Nomoto}, K., {Hachisu}, I., \& {Kato}, M. 1998, \apjl, 503, L155, \dodoi{10.1086/311556}

\bibitem[{{Kojima} {et~al.}(2020){Kojima}, {Ouchi}, {Rauch}, {Ono}, {Nakajima}, {Isobe}, {Fujimoto}, {Harikane}, {Hashimoto}, {Hayashi}, {Komiyama}, {Kusakabe}, {Kim}, {Lee}, {Mukae}, {Nagao}, {Onodera}, {Shibuya}, {Sugahara}, {Umemura}, \& {Yabe}}]{Kojima2020}
{Kojima}, T., {Ouchi}, M., {Rauch}, M., {et~al.} 2020, \apj, 898, 142, \dodoi{10.3847/1538-4357/aba047}

\bibitem[{{Kojima} {et~al.}(2021){Kojima}, {Ouchi}, {Rauch}, {Ono}, {Nakajima}, {Isobe}, {Fujimoto}, {Harikane}, {Hashimoto}, {Hayashi}, {Komiyama}, {Kusakabe}, {Kim}, {Lee}, {Mukae}, {Nagao}, {Onodera}, {Shibuya}, {Sugahara}, {Umemura}, \& {Yabe}}]{Kojima2021}
---. 2021, \apj, 913, 22, \dodoi{10.3847/1538-4357/abec3d}

\bibitem[{{Kroupa} {et~al.}(1993){Kroupa}, {Tout}, \& {Gilmore}}]{Kroupa1993}
{Kroupa}, P., {Tout}, C.~A., \& {Gilmore}, G. 1993, \mnras, 262, 545, \dodoi{10.1093/mnras/262.3.545}

\bibitem[{{Kurk} {et~al.}(2007){Kurk}, {Walter}, {Fan}, {Jiang}, {Riechers}, {Rix}, {Pentericci}, {Strauss}, {Carilli}, \& {Wagner}}]{Kurk2007}
{Kurk}, J.~D., {Walter}, F., {Fan}, X., {et~al.} 2007, \apj, 669, 32, \dodoi{10.1086/521596}

\bibitem[{{Laor} {et~al.}(1995){Laor}, {Bahcall}, {Jannuzi}, {Schneider}, \& {Green}}]{Laor1995}
{Laor}, A., {Bahcall}, J.~N., {Jannuzi}, B.~T., {Schneider}, D.~P., \& {Green}, R.~F. 1995, \apjs, 99, 1, \dodoi{10.1086/192177}

\bibitem[{{Lecureur} {et~al.}(2007){Lecureur}, {Hill}, {Zoccali}, {Barbuy}, {G{\'o}mez}, {Minniti}, {Ortolani}, \& {Renzini}}]{Lecureur2007}
{Lecureur}, A., {Hill}, V., {Zoccali}, M., {et~al.} 2007, \aap, 465, 799, \dodoi{10.1051/0004-6361:20066036}

\bibitem[{{Leung} \& {Nomoto}(2024)}]{Leung2024}
{Leung}, S.-C., \& {Nomoto}, K. 2024, \apj, 974, 310, \dodoi{10.3847/1538-4357/ad6ddb}

\bibitem[{{Maiolino} \& {Mannucci}(2019)}]{Maiolino2019}
{Maiolino}, R., \& {Mannucci}, F. 2019, \aapr, 27, 3, \dodoi{10.1007/s00159-018-0112-2}

\bibitem[{{Maiolino} {et~al.}(2008){Maiolino}, {Nagao}, {Grazian}, {Cocchia}, {Marconi}, {Mannucci}, {Cimatti}, {Pipino}, {Ballero}, {Calura}, {Chiappini}, {Fontana}, {Granato}, {Matteucci}, {Pastorini}, {Pentericci}, {Risaliti}, {Salvati}, \& {Silva}}]{Maiolino2008}
{Maiolino}, R., {Nagao}, T., {Grazian}, A., {et~al.} 2008, \aap, 488, 463, \dodoi{10.1051/0004-6361:200809678}

\bibitem[{{Maiolino} {et~al.}(2024){Maiolino}, {Scholtz}, {Witstok}, {Carniani}, {D'Eugenio}, {de Graaff}, {{\"U}bler}, {Tacchella}, {Curtis-Lake}, {Arribas}, {Bunker}, {Charlot}, {Chevallard}, {Curti}, {Looser}, {Maseda}, {Rawle}, {Rodr{\'\i}guez del Pino}, {Willott}, {Egami}, {Eisenstein}, {Hainline}, {Robertson}, {Williams}, {Willmer}, {Baker}, {Boyett}, {DeCoursey}, {Fabian}, {Helton}, {Ji}, {Jones}, {Kumari}, {Laporte}, {Nelson}, {Perna}, {Sandles}, {Shivaei}, \& {Sun}}]{Maiolino2024}
{Maiolino}, R., {Scholtz}, J., {Witstok}, J., {et~al.} 2024, \nat, 627, 59, \dodoi{10.1038/s41586-024-07052-5}

\bibitem[{{Maoz} \& {Graur}(2017)}]{Maoz2017}
{Maoz}, D., \& {Graur}, O. 2017, \apj, 848, 25, \dodoi{10.3847/1538-4357/aa8b6e}

\bibitem[{{Maoz} {et~al.}(2014){Maoz}, {Mannucci}, \& {Nelemans}}]{Maoz2014}
{Maoz}, D., {Mannucci}, F., \& {Nelemans}, G. 2014, \araa, 52, 107, \dodoi{10.1146/annurev-astro-082812-141031}

\bibitem[{{Mazzucchelli} {et~al.}(2017){Mazzucchelli}, {Ba{\~n}ados}, {Venemans}, {Decarli}, {Farina}, {Walter}, {Eilers}, {Rix}, {Simcoe}, {Stern}, {Fan}, {Schlafly}, {De Rosa}, {Hennawi}, {Chambers}, {Greiner}, {Burgett}, {Draper}, {Kaiser}, {Kudritzki}, {Magnier}, {Metcalfe}, {Waters}, \& {Wainscoat}}]{Mazzucchelli2017}
{Mazzucchelli}, C., {Ba{\~n}ados}, E., {Venemans}, B.~P., {et~al.} 2017, \apj, 849, 91, \dodoi{10.3847/1538-4357/aa9185}

\bibitem[{{Mel{\'e}ndez} {et~al.}(2003){Mel{\'e}ndez}, {Barbuy}, {Bica}, {Zoccali}, {Ortolani}, {Renzini}, \& {Hill}}]{Melendez2003}
{Mel{\'e}ndez}, J., {Barbuy}, B., {Bica}, E., {et~al.} 2003, \aap, 411, 417, \dodoi{10.1051/0004-6361:20031357}

\bibitem[{{Naidu} {et~al.}(2025){Naidu}, {Oesch}, {Brammer}, {Weibel}, {Li}, {Matthee}, {Chisholm}, {Pollock}, {Heintz}, {Johnson}, {Shen}, {Hviding}, {Leja}, {Tacchella}, {Ganguly}, {Witten}, {Atek}, {Belli}, {Bose}, {Bouwens}, {Dayal}, {Decarli}, {de Graaff}, {Fudamoto}, {Giovinazzo}, {Greene}, {Illingworth}, {Inoue}, {Kane}, {Labbe}, {Leonova}, {Marques-Chaves}, {Meyer}, {Nelson}, {Roberts-Borsani}, {Schaerer}, {Simcoe}, {Stefanon}, {Sugahara}, {Toft}, {van der Wel}, {van Dokkum}, {Walter}, {Watson}, {Weaver}, \& {Whitaker}}]{Naidu2025}
{Naidu}, R.~P., {Oesch}, P.~A., {Brammer}, G., {et~al.} 2025, arXiv e-prints, arXiv:2505.11263, \dodoi{10.48550/arXiv.2505.11263}

\bibitem[{{Nakajima} {et~al.}(2023){Nakajima}, {Ouchi}, {Isobe}, {Harikane}, {Zhang}, {Ono}, {Umeda}, \& {Oguri}}]{Nakajima2023}
{Nakajima}, K., {Ouchi}, M., {Isobe}, Y., {et~al.} 2023, \apjs, 269, 33, \dodoi{10.3847/1538-4365/acd556}

\bibitem[{{Nakajima} {et~al.}(2022){Nakajima}, {Ouchi}, {Xu}, {Rauch}, {Harikane}, {Nishigaki}, {Isobe}, {Kusakabe}, {Nagao}, {Ono}, {Onodera}, {Sugahara}, {Kim}, {Komiyama}, {Lee}, \& {Zahedy}}]{Nakajima2022}
{Nakajima}, K., {Ouchi}, M., {Xu}, Y., {et~al.} 2022, \apjs, 262, 3, \dodoi{10.3847/1538-4365/ac7710}

\bibitem[{{Nakane} {et~al.}(2024){Nakane}, {Ouchi}, {Nakajima}, {Harikane}, {Tominaga}, {Takahashi}, {Kashino}, {Yanagisawa}, {Watanabe}, {Nomoto}, {Isobe}, {Nishigaki}, {Ishigaki}, {Ono}, \& {Takeda}}]{Nakane2024}
{Nakane}, M., {Ouchi}, M., {Nakajima}, K., {et~al.} 2024, \apj, 976, 122, \dodoi{10.3847/1538-4357/ad84e810.1134/S1063773708080045}

\bibitem[{{Napolitano} {et~al.}(2025){Napolitano}, {Castellano}, {Pentericci}, {Arrabal Haro}, {Fontana}, {Treu}, {Bergamini}, {Calabr{\`o}}, {Mascia}, {Morishita}, {Roberts-Borsani}, {Santini}, {Vanzella}, {Vulcani}, {Zakharova}, {Bakx}, {Dickinson}, {Grillo}, {Leethochawalit}, {Llerena}, {Merlin}, {Paris}, {Rojas-Ruiz}, {Rosati}, {Wang}, {Yoon}, \& {Zavala}}]{Napolitano2025}
{Napolitano}, L., {Castellano}, M., {Pentericci}, L., {et~al.} 2025, \aap, 693, A50, \dodoi{10.1051/0004-6361/202452090}

\bibitem[{{Nishijima} {et~al.}(2024){Nishijima}, {Hirano}, \& {Umeda}}]{Nishijima2024}
{Nishijima}, S., {Hirano}, S., \& {Umeda}, H. 2024, \apj, 965, 141, \dodoi{10.3847/1538-4357/ad2fc9}

\bibitem[{{Nomoto} {et~al.}(2013){Nomoto}, {Kobayashi}, \& {Tominaga}}]{Nomoto2013}
{Nomoto}, K., {Kobayashi}, C., \& {Tominaga}, N. 2013, \araa, 51, 457, \dodoi{10.1146/annurev-astro-082812-140956}

\bibitem[{{Nomoto} {et~al.}(2004){Nomoto}, {Maeda}, {Mazzali}, {Umeda}, {Deng}, \& {Iwamoto}}]{Nomoto2004}
{Nomoto}, K., {Maeda}, K., {Mazzali}, P.~A., {et~al.} 2004, in Astrophysics and Space Science Library, Vol. 302, Astrophysics and Space Science Library, ed. C.~L. {Fryer}, 277--325, \dodoi{10.1007/978-0-306-48599-2_10}

\bibitem[{{Nomoto} {et~al.}(1984){Nomoto}, {Thielemann}, \& {Yokoi}}]{Nomoto1984}
{Nomoto}, K., {Thielemann}, F.~K., \& {Yokoi}, K. 1984, \apj, 286, 644, \dodoi{10.1086/162639}

\bibitem[{{Oke} \& {Gunn}(1983)}]{Oke&Gunn1983}
{Oke}, J.~B., \& {Gunn}, J.~E. 1983, \apj, 266, 713, \dodoi{10.1086/160817}

\bibitem[{{Ono} {et~al.}(2025){Ono}, {Ouchi}, {Harikane}, {Yajima}, {Nakajima}, {Fujimoto}, {Nakane}, \& {Xu}}]{Ono2025}
{Ono}, Y., {Ouchi}, M., {Harikane}, Y., {et~al.} 2025, arXiv e-prints, arXiv:2502.08885.
\newblock \doarXiv{2502.08885}

\bibitem[{{Onoue} {et~al.}(2020){Onoue}, {Ba{\~n}ados}, {Mazzucchelli}, {Venemans}, {Schindler}, {Walter}, {Hennawi}, {Andika}, {Davies}, {Decarli}, {Farina}, {Jahnke}, {Nagao}, {Tominaga}, \& {Wang}}]{Onoue2020}
{Onoue}, M., {Ba{\~n}ados}, E., {Mazzucchelli}, C., {et~al.} 2020, \apj, 898, 105, \dodoi{10.3847/1538-4357/aba193}

\bibitem[{{Padovani} \& {Matteucci}(1993)}]{Padovani1993}
{Padovani}, P., \& {Matteucci}, F. 1993, \apj, 416, 26, \dodoi{10.1086/173212}

\bibitem[{{Pasquini} {et~al.}(2008){Pasquini}, {Ecuvillon}, {Bonifacio}, \& {Wolff}}]{Pasquini2008}
{Pasquini}, L., {Ecuvillon}, A., {Bonifacio}, P., \& {Wolff}, B. 2008, \aap, 489, 315, \dodoi{10.1051/0004-6361:200809963}

\bibitem[{{Pollock} {et~al.}(2025){Pollock}, {Gottumukkala}, {Heintz}, {Brammer}, {Roberts-Borsani}, {Oesch}, {Witstok}, {Arellano-C{\'o}rdova}, {Cullen}, {Scholte}, {Terp}, {Rowland}, {Sneppen}, {Ito}, {Valentino}, {Matthee}, {Watson}, \& {Toft}}]{Pollock2025}
{Pollock}, C.~L., {Gottumukkala}, R., {Heintz}, K.~E., {et~al.} 2025, arXiv e-prints, arXiv:2506.15779, \dodoi{10.48550/arXiv.2506.15779}

\bibitem[{{Price} {et~al.}(2024){Price}, {Bezanson}, {Labbe}, {Furtak}, {de Graaff}, {Greene}, {Kokorev}, {Setton}, {Suess}, {Brammer}, {Cutler}, {Leja}, {Pan}, {Wang}, {Weaver}, {Whitaker}, {Atek}, {Burgasser}, {Chemerynska}, {Dayal}, {Feldmann}, {F{\"o}rster Schreiber}, {Fudamoto}, {Fujimoto}, {Glazebrook}, {Goulding}, {Khullar}, {Kriek}, {Marchesini}, {Maseda}, {Miller}, {Muzzin}, {Nanayakkara}, {Nelson}, {Oesch}, {Shipley}, {Smit}, {Taylor}, {van Dokkum}, {Williams}, \& {Zitrin}}]{Price2024}
{Price}, S.~H., {Bezanson}, R., {Labbe}, I., {et~al.} 2024, arXiv e-prints, arXiv:2408.03920, \dodoi{10.48550/arXiv.2408.03920}

\bibitem[{{Rix} {et~al.}(2004){Rix}, {Pettini}, {Leitherer}, {Bresolin}, {Kudritzki}, \& {Steidel}}]{Rix2004}
{Rix}, S.~A., {Pettini}, M., {Leitherer}, C., {et~al.} 2004, \apj, 615, 98, \dodoi{10.1086/424031}

\bibitem[{{Rizzuti} {et~al.}(2024){Rizzuti}, {Matteucci}, {Molaro}, {Cescutti}, \& {Maiolino}}]{Rizzuti2024}
{Rizzuti}, F., {Matteucci}, F., {Molaro}, P., {Cescutti}, G., \& {Maiolino}, R. 2024, arXiv e-prints, arXiv:2412.05363, \dodoi{10.48550/arXiv.2412.05363}

\bibitem[{{Rodney} {et~al.}(2014){Rodney}, {Riess}, {Strolger}, {Dahlen}, {Graur}, {Casertano}, {Dickinson}, {Ferguson}, {Garnavich}, {Hayden}, {Jha}, {Jones}, {Kirshner}, {Koekemoer}, {McCully}, {Mobasher}, {Patel}, {Weiner}, {Cenko}, {Clubb}, {Cooper}, {Filippenko}, {Frederiksen}, {Hjorth}, {Leibundgut}, {Matheson}, {Nayyeri}, {Penner}, {Trump}, {Silverman}, {U}, {Azalee Bostroem}, {Challis}, {Rajan}, {Wolff}, {Faber}, {Grogin}, \& {Kocevski}}]{Rodney2014}
{Rodney}, S.~A., {Riess}, A.~G., {Strolger}, L.-G., {et~al.} 2014, \aj, 148, 13, \dodoi{10.1088/0004-6256/148/1/13}

\bibitem[{{Salpeter}(1955)}]{Salpeter1955}
{Salpeter}, E.~E. 1955, \apj, 121, 161, \dodoi{10.1086/145971}

\bibitem[{{Sameshima} {et~al.}(2017){Sameshima}, {Yoshii}, \& {Kawara}}]{Sameshima2017}
{Sameshima}, H., {Yoshii}, Y., \& {Kawara}, K. 2017, \apj, 834, 203, \dodoi{10.3847/1538-4357/834/2/203}

\bibitem[{{Sameshima} {et~al.}(2020){Sameshima}, {Yoshii}, {Matsunaga}, {Kobayashi}, {Ikeda}, {Kondo}, {Hamano}, {Mizumoto}, {Arai}, {Yasui}, {Fukue}, {Kawakita}, {Otsubo}, {Bono}, \& {Saviane}}]{Sameshima2020}
{Sameshima}, H., {Yoshii}, Y., {Matsunaga}, N., {et~al.} 2020, \apj, 904, 162, \dodoi{10.3847/1538-4357/abc33b}

\bibitem[{{Sanders} {et~al.}(2024){Sanders}, {Shapley}, {Topping}, {Reddy}, \& {Brammer}}]{Sanders2024}
{Sanders}, R.~L., {Shapley}, A.~E., {Topping}, M.~W., {Reddy}, N.~A., \& {Brammer}, G.~B. 2024, \apj, 962, 24, \dodoi{10.3847/1538-4357/ad15fc}

\bibitem[{{Shi} {et~al.}(2007){Shi}, {Zhao}, \& {Liang}}]{Shi2007}
{Shi}, F., {Zhao}, G., \& {Liang}, Y.~C. 2007, \aap, 475, 409, \dodoi{10.1051/0004-6361:20077183}

\bibitem[{{Shin} {et~al.}(2019){Shin}, {Nagao}, {Woo}, \& {Le}}]{Shin2019}
{Shin}, J., {Nagao}, T., {Woo}, J.-H., \& {Le}, H. A.~N. 2019, \apj, 874, 22, \dodoi{10.3847/1538-4357/ab05da}

\bibitem[{{Stanton} {et~al.}(2024){Stanton}, {Cullen}, {McLure}, {Shapley}, {Arellano-C{\'o}rdova}, {Begley}, {Amor{\'\i}n}, {Barrufet}, {Calabr{\`o}}, {Carnall}, {Cirasuolo}, {Dunlop}, {Donnan}, {Hamadouche}, {Liu}, {McLeod}, {Pentericci}, {Pozzetti}, {Sanders}, {Scholte}, \& {Topping}}]{Stanton2024}
{Stanton}, T.~M., {Cullen}, F., {McLure}, R.~J., {et~al.} 2024, \mnras, 532, 3102, \dodoi{10.1093/mnras/stae1705}

\bibitem[{{Stanton} {et~al.}(2025){Stanton}, {Cullen}, {Carnall}, {Scholte}, {Arellano-C{\'o}rdova}, {McLeod}, {Begley}, {Donnan}, {Dunlop}, {Hamadouche}, {McLure}, {Shapley}, {Bondestam}, \& {Stevenson}}]{Stanton2025}
{Stanton}, T.~M., {Cullen}, F., {Carnall}, A.~C., {et~al.} 2025, \mnras, 537, 1735, \dodoi{10.1093/mnras/staf106}

\bibitem[{{Stanway} \& {Eldridge}(2018)}]{Stanway2018}
{Stanway}, E.~R., \& {Eldridge}, J.~J. 2018, \mnras, 479, 75, \dodoi{10.1093/mnras/sty1353}

\bibitem[{{Steidel} {et~al.}(2016){Steidel}, {Strom}, {Pettini}, {Rudie}, {Reddy}, \& {Trainor}}]{Steidel2016}
{Steidel}, C.~C., {Strom}, A.~L., {Pettini}, M., {et~al.} 2016, \apj, 826, 159, \dodoi{10.3847/0004-637X/826/2/159}

\bibitem[{{Suzuki} \& {Maeda}(2018)}]{Suzuki2018}
{Suzuki}, A., \& {Maeda}, K. 2018, \apj, 852, 101, \dodoi{10.3847/1538-4357/aaa024}

\bibitem[{{Tacchella} {et~al.}(2023){Tacchella}, {Eisenstein}, {Hainline}, {Johnson}, {Baker}, {Helton}, {Robertson}, {Suess}, {Chen}, {Nelson}, {Pusk{\'a}s}, {Sun}, {Alberts}, {Egami}, {Hausen}, {Rieke}, {Rieke}, {Shivaei}, {Williams}, {Willmer}, {Bunker}, {Cameron}, {Carniani}, {Charlot}, {Curti}, {Curtis-Lake}, {Looser}, {Maiolino}, {Maseda}, {Rawle}, {Rix}, {Smit}, {{\"U}bler}, {Willott}, {Witstok}, {Baum}, {Bhatawdekar}, {Boyett}, {Danhaive}, {de Graaff}, {Endsley}, {Ji}, {Lyu}, {Sandles}, {Saxena}, {Scholtz}, {Topping}, \& {Whitler}}]{Tacchella2023}
{Tacchella}, S., {Eisenstein}, D.~J., {Hainline}, K., {et~al.} 2023, \apj, 952, 74, \dodoi{10.3847/1538-4357/acdbc6}

\bibitem[{{Tacchella} {et~al.}(2024){Tacchella}, {McClymont}, {Scholtz}, {Maiolino}, {Ji}, {Villanueva}, {Charlot}, {D'Eugenio}, {Helton}, {Williams}, {Witstok}, {Bhatawdekar}, {Carniani}, {Chevallard}, {Curti}, {Hainline}, {Ji}, {Johnson}, {Leja}, {Li}, {Maseda}, {Pusk{\'a}s}, {Rieke}, {Robertson}, {Shivaei}, {Silcock}, {Simmonds}, {{\"U}bler}, {Willmer}, \& {Willott}}]{Tacchella2024}
{Tacchella}, S., {McClymont}, W., {Scholtz}, J., {et~al.} 2024, arXiv e-prints, arXiv:2404.02194, \dodoi{10.48550/arXiv.2404.02194}

\bibitem[{{Takahashi} {et~al.}(2018){Takahashi}, {Yoshida}, \& {Umeda}}]{Takahashi2018}
{Takahashi}, K., {Yoshida}, T., \& {Umeda}, H. 2018, \apj, 857, 111, \dodoi{10.3847/1538-4357/aab95f}

\bibitem[{{Tang} {et~al.}(2023){Tang}, {Zhang}, {Yan}, {Zhang}, {Carigi}, \& {Fern{\'a}ndez-Trincado}}]{Tang2023}
{Tang}, B., {Zhang}, J., {Yan}, Z., {et~al.} 2023, \aap, 669, A125, \dodoi{10.1051/0004-6361/202244052}

\bibitem[{{Tang} {et~al.}(2024){Tang}, {Stark}, {Topping}, {Mason}, \& {Ellis}}]{Tang2024}
{Tang}, M., {Stark}, D.~P., {Topping}, M.~W., {Mason}, C., \& {Ellis}, R.~S. 2024, \apj, 975, 208, \dodoi{10.3847/1538-4357/ad7eb7}

\bibitem[{{Tegmark} {et~al.}(1997){Tegmark}, {Silk}, {Rees}, {Blanchard}, {Abel}, \& {Palla}}]{Tegmark1997}
{Tegmark}, M., {Silk}, J., {Rees}, M.~J., {et~al.} 1997, \apj, 474, 1, \dodoi{10.1086/303434}

\bibitem[{{Topping} {et~al.}(2020{\natexlab{a}}){Topping}, {Shapley}, {Reddy}, {Sanders}, {Coil}, {Kriek}, {Mobasher}, \& {Siana}}]{Topping2020a}
{Topping}, M.~W., {Shapley}, A.~E., {Reddy}, N.~A., {et~al.} 2020{\natexlab{a}}, \mnras, 495, 4430, \dodoi{10.1093/mnras/staa1410}

\bibitem[{{Topping} {et~al.}(2020{\natexlab{b}}){Topping}, {Shapley}, {Reddy}, {Sanders}, {Coil}, {Kriek}, {Mobasher}, \& {Siana}}]{Topping2020b}
---. 2020{\natexlab{b}}, \mnras, 499, 1652, \dodoi{10.1093/mnras/staa2941}

\bibitem[{{Topping} {et~al.}(2024{\natexlab{a}}){Topping}, {Stark}, {Senchyna}, {Plat}, {Zitrin}, {Endsley}, {Charlot}, {Furtak}, {Maseda}, {Smit}, {Mainali}, {Chevallard}, {Molyneux}, \& {Rigby}}]{Topping2024}
{Topping}, M.~W., {Stark}, D.~P., {Senchyna}, P., {et~al.} 2024{\natexlab{a}}, \mnras, 529, 3301, \dodoi{10.1093/mnras/stae682}

\bibitem[{{Topping} {et~al.}(2024{\natexlab{b}}){Topping}, {Stark}, {Senchyna}, {Chen}, {Zitrin}, {Endsley}, {Charlot}, {Furtak}, {Maseda}, {Plat}, {Smit}, {Mainali}, {Chevallard}, {Molyneux}, \& {Rigby}}]{Topping2025a}
---. 2024{\natexlab{b}}, arXiv e-prints, arXiv:2407.19009, \dodoi{10.48550/arXiv.2407.19009}

\bibitem[{{Topping} {et~al.}(2025){Topping}, {Sanders}, {Shapley}, {Pahl}, {Reddy}, {Stark}, {Berg}, {Clarke}, {Cullen}, {Dunlop}, {Ellis}, {F{\"o}rster Schreiber}, {Illingworth}, {Jones}, {Narayanan}, {Pettini}, \& {Schaerer}}]{Topping2025b}
{Topping}, M.~W., {Sanders}, R.~L., {Shapley}, A.~E., {et~al.} 2025, arXiv e-prints, arXiv:2502.08712, \dodoi{10.48550/arXiv.2502.08712}

\bibitem[{{Totani} {et~al.}(2008){Totani}, {Morokuma}, {Oda}, {Doi}, \& {Yasuda}}]{Totani2008}
{Totani}, T., {Morokuma}, T., {Oda}, T., {Doi}, M., \& {Yasuda}, N. 2008, \pasj, 60, 1327, \dodoi{10.1093/pasj/60.6.1327}

\bibitem[{{Tsuzuki} {et~al.}(2006){Tsuzuki}, {Kawara}, {Yoshii}, {Oyabu}, {Tanab{\'e}}, \& {Matsuoka}}]{Tsuzuki2006}
{Tsuzuki}, Y., {Kawara}, K., {Yoshii}, Y., {et~al.} 2006, \apj, 650, 57, \dodoi{10.1086/506376}

\bibitem[{{Umeda} \& {Nomoto}(2008)}]{Umeda2008}
{Umeda}, H., \& {Nomoto}, K. 2008, \apj, 673, 1014, \dodoi{10.1086/524767}

\bibitem[{{Valenti} {et~al.}(2011){Valenti}, {Origlia}, \& {Rich}}]{Valenti2011}
{Valenti}, E., {Origlia}, L., \& {Rich}, R.~M. 2011, \mnras, 414, 2690, \dodoi{10.1111/j.1365-2966.2011.18580.x}

\bibitem[{Virtanen {et~al.}(2020)Virtanen, Gommers, Oliphant, Haberland, Reddy, Cournapeau, Burovski, Peterson, Weckesser, Bright, {van der Walt}, Brett, Wilson, Millman, Mayorov, Nelson, Jones, Kern, Larson, Carey, Polat, Feng, Moore, {VanderPlas}, Laxalde, Perktold, Cimrman, Henriksen, Quintero, Harris, Archibald, Ribeiro, Pedregosa, {van Mulbregt}, \& {SciPy 1.0 Contributors}}]{Virtanen2020}
Virtanen, P., Gommers, R., Oliphant, T.~E., {et~al.} 2020, Nature Methods, 17, 261, \dodoi{10.1038/s41592-019-0686-2}

\bibitem[{{Watanabe} {et~al.}(2024){Watanabe}, {Ouchi}, {Nakajima}, {Isobe}, {Tominaga}, {Suzuki}, {Ishigaki}, {Nomoto}, {Takahashi}, {Harikane}, {Hatano}, {Kusakabe}, {Moriya}, {Nishigaki}, {Ono}, {Onodera}, \& {Sugahara}}]{Watanabe2024}
{Watanabe}, K., {Ouchi}, M., {Nakajima}, K., {et~al.} 2024, \apj, 962, 50, \dodoi{10.3847/1538-4357/ad13ff}

\bibitem[{Watanabe(2010)}]{Watanabe2010}
Watanabe, S. 2010, Journal of Machine Learning Research, 11, 3571.
\newblock \url{http://jmlr.org/papers/v11/watanabe10a.html}

\bibitem[{{Williams} {et~al.}(2023){Williams}, {Kelly}, {Chen}, {Brammer}, {Zitrin}, {Treu}, {Scarlata}, {Koekemoer}, {Oguri}, {Lin}, {Diego}, {Nonino}, {Hjorth}, {Langeroodi}, {Broadhurst}, {Rogers}, {Perez-Fournon}, {Foley}, {Jha}, {Filippenko}, {Strolger}, {Pierel}, {Poidevin}, \& {Yang}}]{Williams2023}
{Williams}, H., {Kelly}, P.~L., {Chen}, W., {et~al.} 2023, Science, 380, 416, \dodoi{10.1126/science.adf5307}

\bibitem[{{Yanagisawa} {et~al.}(2024){Yanagisawa}, {Ouchi}, {Nakajima}, {Harikane}, {Fujimoto}, {Ono}, {Umeda}, {Nakane}, {Yajima}, {Fukushima}, \& {Xu}}]{Yanagisawa2024}
{Yanagisawa}, H., {Ouchi}, M., {Nakajima}, K., {et~al.} 2024, arXiv e-prints, arXiv:2411.19893, \dodoi{10.48550/arXiv.2411.19893}

\bibitem[{{Yong} {et~al.}(2008){Yong}, {Grundahl}, {Johnson}, \& {Asplund}}]{Yong2008}
{Yong}, D., {Grundahl}, F., {Johnson}, J.~A., \& {Asplund}, M. 2008, \apj, 684, 1159, \dodoi{10.1086/590658}

\bibitem[{{Yong} {et~al.}(2005){Yong}, {Grundahl}, {Nissen}, {Jensen}, \& {Lambert}}]{Yong2005}
{Yong}, D., {Grundahl}, F., {Nissen}, P.~E., {Jensen}, H.~R., \& {Lambert}, D.~L. 2005, \aap, 438, 875, \dodoi{10.1051/0004-6361:20052916}

\bibitem[{{Yoshii} {et~al.}(2022){Yoshii}, {Sameshima}, {Tsujimoto}, {Shigeyama}, {Beers}, \& {Peterson}}]{Yoshii2022}
{Yoshii}, Y., {Sameshima}, H., {Tsujimoto}, T., {et~al.} 2022, \apj, 937, 61, \dodoi{10.3847/1538-4357/ac8163}

\bibitem[{{Zavala} {et~al.}(2024{\natexlab{a}}){Zavala}, {Castellano}, {Akins}, {Bakx}, {Burgarella}, {Casey}, {Ch{\~A}{\textexclamdown}vez Ortiz}, {Dickinson}, {Finkelstein}, {Mitsuhashi}, {Nakajima}, {P{\~A}{\textcopyright}rez-Gonz{\~A}{\textexclamdown}lez}, {Arrabal Haro}, {Bergamini}, {Buat}, {Backhaus}, {Calabr{\~A}{\texttwosuperior}}, {Cleri}, {Fern{\~A}{\textexclamdown}ndez-Arenas}, {Fontana}, {Franco}, {Grillo}, {Giavalisco}, {Grogin}, {Hathi}, {Hirschmann}, {Ikeda}, {Jung}, {Kartaltepe}, {Koekemoer}, {Larson}, {McKinney}, {Papovich}, {Rosati}, {Saito}, {Santini}, {Terlevich}, {Terlevich}, {Treu}, \& {Yung}}]{Zavala2024a}
{Zavala}, J.~A., {Castellano}, M., {Akins}, H.~B., {et~al.} 2024{\natexlab{a}}, Nature Astronomy, \dodoi{10.1038/s41550-024-02397-3}

\bibitem[{{Zavala} {et~al.}(2024{\natexlab{b}}){Zavala}, {Bakx}, {Mitsuhashi}, {Castellano}, {Calabro}, {Akins}, {Buat}, {Casey}, {Fernandez-Arenas}, {Franco}, {Fontana}, {Hatsukade}, {Ho}, {Ikeda}, {Kartaltepe}, {Koekemoer}, {McKinney}, {Napolitano}, {Perez-Gonzalez}, {Santini}, {Serjeant}, {Terlevich}, {Terlevich}, \& {Yung}}]{Zavala2024b}
{Zavala}, J.~A., {Bakx}, T., {Mitsuhashi}, I., {et~al.} 2024{\natexlab{b}}, arXiv e-prints, arXiv:2411.03593, \dodoi{10.48550/arXiv.2411.03593}

\bibitem[{{Zhao} {et~al.}(2016){Zhao}, {Mashonkina}, {Yan}, {Alexeeva}, {Kobayashi}, {Pakhomov}, {Shi}, {Sitnova}, {Tan}, {Zhang}, {Zhang}, {Zhou}, {Bolte}, {Chen}, {Li}, {Liu}, \& {Zhai}}]{Zhao2016}
{Zhao}, G., {Mashonkina}, L., {Yan}, H.~L., {et~al.} 2016, \apj, 833, 225, \dodoi{10.3847/1538-4357/833/2/225}

\end{thebibliography}



\clearpage

\end{document}